\documentstyle[10pt,epsf,epsfig,dp_delphititle,oldlfont,lineno,amssymb,axodraw]{dp_delphi}
\makeindex
\def\DpPaperGroup{EP}
\def\DpPaperRef{2002-076}
\def\DpDate{8 August 2002}
\def\DpAuthors{DELPHI Collaboration}
\def\DpTitle{{\boldmath
 Search for an LSP gluino at LEP with the DELPHI detector}}
\def\DpSubmit{(Eur. Phys. J. C26 (2003) 505)}
\def\DpComment{
}
\def\DpEMail{
}

\newcommand{\into}{{\ifmmode \rightarrow \else $\rightarrow$\fi}}
\newcommand{\epem}{{\ifmmode e^+e^- \else $e^+e^-$\fi}}
\newcommand{\zz}{{\ifmmode Z^0~ \else $Z^0~$\fi}}
\newcommand{\sqs}{{\ifmmode \sqrt s~ \else $\sqrt s~$\fi}}
\newcommand{\de}{{\ifmmode ^{\circ} \else ${^\circ}$\fi}}
\newcommand{\q}{{\ifmmode q~ \else $q~$\fi}}
\newcommand{\qb}{{\ifmmode \bar{q}~ \else $\bar{q}~$\fi}}
\newcommand{\sx}{{\ifmmode s~ \else $s~$\fi}}
\newcommand{\sxb}{{\ifmmode \bar{s}~ \else $\bar{s}~$\fi}}
\newcommand{\sxx}{{\ifmmode s \else $s$\fi}}
\newcommand{\sxbx}{{\ifmmode \bar{s} \else $\bar{s}$\fi}}
\newcommand{\Kp}{{\ifmmode K^{+}~ \else $K^{+}~$\fi}}
\newcommand{\Kpx}{{\ifmmode K^{+} \else $K^{+}$\fi}}
\newcommand{\Km}{{\ifmmode K^{-}~ \else $K^{-}~$\fi}}
\newcommand{\Kmx}{{\ifmmode K^{-} \else $K^{-}$\fi}}
\newcommand{\p}{{\ifmmode p~ \else $p~$\fi}}
\newcommand{\px}{{\ifmmode p \else $p$\fi}}
\newcommand{\bp}{{\ifmmode \bar{p}~ \else $\bar{p}~$\fi}}
\newcommand{\bpx}{{\ifmmode \bar{p} \else $\bar{p}$\fi}}

\newcommand{\qqb}{{\ifmmode q\,\bar{q}~ \else $q\,\bar{q}~$\fi}}
\newcommand{\ssb}{{\ifmmode s\,\bar{s}~ \else $s\,\bar{s}~$\fi}}
\newcommand{\pipi}{{\ifmmode \pi^+\,\pi^-~ \else $\pi^+\,\pi^-~$\fi}}
\newcommand{\pipix}{{\ifmmode \pi^+\,\pi^- \else $\pi^+\,\pi^-$\fi}}
\newcommand{\KK}{{\ifmmode K^+K^-~ \else $K^+K^-~$\fi}}
\newcommand{\KKx}{{\ifmmode K^+K^- \else $K^+K^-$\fi}}
\newcommand{\pp}{{\ifmmode p\,\bar{p}~ \else $p\,\bar{p}~$\fi}}
\newcommand{\ppx}{{\ifmmode p\,\bar{p} \else $p\,\bar{p}$\fi}}
\newcommand{\dy}{{\ifmmode {\mit\Delta y}~ \else ${\mit\Delta y}~$\fi}}
\newcommand{\dyx}{{\ifmmode {\mit\Delta y} \else ${\mit\Delta y}$\fi}}
\newcommand{\PPP}{{\ifmmode \cal{P}~ \else $\cal{P}~$\fi}}
\newcommand{\PPPx}{{\ifmmode \cal{P} \else $\cal{P}$\fi}}

\newcommand{\pT}{{$p_T~$}}

\newcommand{\dpT}{{\ifmmode \Delta\pT~ \else $\Delta\pT~$\fi}}
\newcommand{\dpTx}{{\ifmmode \Delta\pT \else $\Delta\pT$\fi}}
\newcommand{\Nin}{{\ifmmode N_{in}~ \else $N_{in}~$\fi}}
\newcommand{\Ninx}{{\ifmmode N_{in} \else $N_{in}$\fi}}
\newcommand{\Nout}{{\ifmmode N_{out}~ \else $N_{out}~$\fi}}
\newcommand{\Noutx}{{\ifmmode N_{out} \else $N_{out}$\fi}}
\newcommand{\dphi}{{\ifmmode {\mit\Delta}\phi~ \else ${\mit\Delta}\phi~$\fi}}
\newcommand{\dphix}{{\ifmmode {\mit\Delta}\phi \else ${\mit\Delta}\phi $\fi}}

\newcommand{\glui}    {\mbox{$\widetilde{{\mathrm g}}                    $}}

\newcommand{\mglui}   {\mbox{$m_{\tilde{{\mathrm g}}}  		         $}}

\newcommand{\xoi}     {\mbox{$\widetilde{{\mathrm \chi}}_1^0             $}}

\newcommand{\Sti}     {\mbox{$\widetilde{{\mathrm t}}_1                  $}}

\newcommand{\msti}    {\mbox{$m_{\tilde{{\mathrm t}}_1}                  $}}

\newcommand{\Sbi}     {\mbox{$\widetilde{{\mathrm b}}_1                  $}}

\newcommand{\msbi}    {\mbox{$m_{\tilde{{\mathrm b}}_1}                  $}}

\newcommand{\Sqi}     {\mbox{$\widetilde{{\mathrm q}}_1                  $}}

\newcommand{\Sql}     {\mbox{$\widetilde{{\mathrm q}}_L                  $}}
\newcommand{\Sqr}     {\mbox{$\widetilde{{\mathrm q}}_R                  $}}
\newcommand{\msq}     {\mbox{$m_{\tilde{{\mathrm q}}}                    $}}
\newcommand{\msqi}    {\mbox{$m_{\tilde{{\mathrm q}}_1}                  $}}

\newcommand{\Zo}      {\mbox{${\mathrm Z}^0                              $}}

\newcommand{\glu}     {\mbox{${\mathrm g}                                $}}

\newcommand{\qu}      {\mbox{${\mathrm u}                                $}}

\newcommand{\qc}      {\mbox{${\mathrm c}                                $}}

\newcommand{\au}      {\mbox{$\bar{{\mathrm u}}                          $}}
\newcommand{\ad}      {\mbox{$\bar{{\mathrm d}}                          $}}
\newcommand{\ac}      {\mbox{$\bar{{\mathrm c}}                          $}}

\newcommand{\ab}      {\mbox{$\bar{{\mathrm b}}                          $}}

\newcommand{\elep}    {\mbox{${\mathrm e}^+				 $}}
\newcommand{\elem}    {\mbox{${\mathrm e}^-				 $}}

\newcommand{\muop}    {\mbox{${\mathrm \mu}^+				 $}}
\newcommand{\muom}    {\mbox{${\mathrm \mu}^-				 $}}

\newcommand{\tap}     {\mbox{${\mathrm \tau}^+				 $}}
\newcommand{\tam}     {\mbox{${\mathrm \tau}^-				 $}}

\newcommand{\neu}     {\mbox{${\mathrm \nu}				 $}}

\newcommand{\qqbar}	{\mbox{$\mathrm{q\bar{q}}		       $}}
\newcommand{\ro}  	{\mbox{$\mathrm{R^{\circ}}		       $}}
\newcommand{\rpm}  	{\mbox{$\mathrm{R^{\pm}}		       $}}
\newcommand{\roro}  	{\mbox{$\mathrm{R^{\circ}R^{\circ}}	       $}}
\newcommand{\rpmrpm}  	{\mbox{$\mathrm{R^{\pm}R^{\pm}}		       $}}
\newcommand{\rpmro}  	{\mbox{$\mathrm{R^{\pm}R^{\circ}}	       $}}

\newcommand{\MeVc}    {\mbox{$\mathrm{MeV}/\mathrm{c}		       $}}

\newcommand{\GeV}     {\mbox{$\mathrm{GeV}			       $}}
\newcommand{\GeVc}    {\mbox{$\mathrm{GeV}/\mathrm{c}		       $}}
\newcommand{\GeVcc}   {\mbox{$\mathrm{GeV}/\mathrm{c}^2 	       $}}

\newcommand{\pb}      {\mbox{$\mathrm{pb}^{-1}			       $}}
\newcommand{\ee}      {\mbox{$\mathrm{e}^+\mathrm{e}^-		       $}}
\newcommand{\dgr}     {\mbox{$ ^\circ                                  $}}

\def\vev#1{\langle #1 \rangle}

\def\PLB#1#2#3{{\rm Phys.~Lett.}             {\bf{B#1}} (#2) #3}
\def\PRD#1#2#3{{\rm Phys.~Rev.}              {\bf{D#1}} (#2) #3}
\def\PRL#1#2#3{{\rm Phys.~Rev.~Lett.}        {\bf{#1}}  (#2) #3}
\def\ZPC#1#2#3{{\rm Z.~Phys.}                {\bf C#1}  (#2) #3}

\def\NIM#1#2#3{{\rm Nucl.~Instr.~and~Meth.}  {\bf{#1}} (#2) #3} 
\def\NIMA#1#2#3{{\rm Nucl.~Instr.~and~Meth.} {\bf{A#1}} (#2) #3} 
\def\CPC#1#2#3{{\rm Comp.~Phys.~Comm.}       {\bf#1}    (#2) #3}
\def\EJC#1#2#3{{\rm E.~Phys.~J.}             {\bf{C#1}} (#2) #3}

\newcommand{\etal}{{\it et al.}}

\newcommand{\aSti}    {\mbox{$\bar{\tilde{{\mathrm t}}}_1            $}}
\newcommand{\aSbi}    {\mbox{$\bar{\tilde{{\mathrm b}}}_1            $}}
\newcommand{\aSqi}    {\mbox{$\bar{\tilde{{\mathrm q}}}_1            $}}

\def\vev#1{\langle #1 \rangle}

\begin{document}
\makeatletter
\newcount\@tempcntc
\def\@citex[#1]#2{\if@filesw\immediate\write\@auxout{\string\citation{#2}}\fi
  \@tempcnta\z@\@tempcntb\m@ne\def\@citea{}\@cite{\@for\@citeb:=#2\do
    {\@ifundefined
       {b@\@citeb}{\@citeo\@tempcntb\m@ne\@citea\def\@citea{,}{\bf ?}\@warning
       {Citation `\@citeb' on page \thepage \space undefined}}%
    {\setbox\z@\hbox{\global\@tempcntc0\csname b@\@citeb\endcsname\relax}%
     \ifnum\@tempcntc=\z@ \@citeo\@tempcntb\m@ne
       \@citea\def\@citea{,}\hbox{\csname b@\@citeb\endcsname}%
     \else
      \advance\@tempcntb\@ne
      \ifnum\@tempcntb=\@tempcntc
      \else\advance\@tempcntb\m@ne\@citeo
      \@tempcnta\@tempcntc\@tempcntb\@tempcntc\fi\fi}}\@citeo}{#1}}
\def\@citeo{\ifnum\@tempcnta>\@tempcntb\else\@citea\def\@citea{,}%
  \ifnum\@tempcnta=\@tempcntb\the\@tempcnta\else
   {\advance\@tempcnta\@ne\ifnum\@tempcnta=\@tempcntb \else \def\@citea{--}\fi
    \advance\@tempcnta\m@ne\the\@tempcnta\@citea\the\@tempcntb}\fi\fi}
 
\makeatother
\begin{titlepage}
\pagenumbering{roman}
\CERNpreprint{\DpPaperGroup}{\DpPaperRef} 
\date{{\small\DpDate}} 
\title{\DpTitle} 
\address{\DpAuthors} 
\begin{shortabs} 
\noindent
In some supersymmetric models, the gluino (\glui) is predicted to be light 
and stable. In that case, it would hadronize to form R-hadrons.
In these models, the missing energy signature of the lightest
supersymmetric particle is no longer valid, even if R-parity is conserved.
Therefore, such a gluino is not constrained by hadron collider results, 
which looked for the decay $\glui\to\qqbar\xoi$.\\
Data collected by the DELPHI detector in 1994 at 91.2~GeV have been analysed to
search for $\qqbar\glui\glui$ events. No deviation from Standard Model
predictions is observed and a gluino mass between 2 and 18~\GeVcc\ is
excluded at the 95\% confidence level in these models. Then, R-hadrons produced
in the squark decays were searched for in the data collected by DELPHI
at the centre-of-mass energies of 189 to 208 GeV, corresponding to an 
overall integrated luminosity of 609~\pb. 
The observed number of events is in agreement with the Standard 
Model predictions. Limits at 95\% confidence level are derived on the squark
masses from the excluded regions in the plane (\msqi,\mglui):
\begin{itemize}
\item[\ ]
$\msti>90~\GeVcc$, and $\msbi>96~\GeVcc$ for purely left squarks.
\item[\ ]
$\msti>87~\GeVcc$, and $\msbi>82~\GeVcc$ independent of the mixing angle.
\end{itemize}
\end{shortabs}
\vfill
\begin{center}
\DpSubmit \ \\ 
\DpComment \ \\
\DpEMail \ \\
\end{center}
\vfill
\clearpage
\headsep 10.0pt
\addtolength{\textheight}{10mm}
\addtolength{\footskip}{-5mm}
\begingroup
%
\newcommand{\DpName}[2]{\hbox{#1$^{\ref{#2}}$},\hfill}
\newcommand{\DpNameTwo}[3]{\hbox{#1$^{\ref{#2},\ref{#3}}$},\hfill}
\newcommand{\DpNameThree}[4]{\hbox{#1$^{\ref{#2},\ref{#3},\ref{#4}}$},\hfill}
\newskip\Bigfill \Bigfill = 0pt plus 1000fill
\newcommand{\DpNameLast}[2]{\hbox{#1$^{\ref{#2}}$}\hspace{\Bigfill}}
%
\footnotesize
\noindent
\DpName{J.Abdallah}{LPNHE}
\DpName{P.Abreu}{LIP}
\DpName{W.Adam}{VIENNA}
\DpName{P.Adzic}{DEMOKRITOS}
\DpName{T.Albrecht}{KARLSRUHE}
\DpName{T.Alderweireld}{AIM}
\DpName{R.Alemany-Fernandez}{CERN}
\DpName{T.Allmendinger}{KARLSRUHE}
\DpName{P.P.Allport}{LIVERPOOL}
\DpName{U.Amaldi}{MILANO2}
% ...
\normalsize
\endgroup
\titlefoot{Department of Physics and Astronomy, Iowa State
     University, Ames IA 50011-3160, USA
    \label{AMES}}
\titlefoot{Physics Department, Universiteit Antwerpen,
     Universiteitsplein 1, B-2610 Antwerpen, Belgium \\
     \indent~~and IIHE, ULB-VUB,
     Pleinlaan 2, B-1050 Brussels, Belgium \\
     \indent~~and Facult\'e des Sciences,
     Univ. de l'Etat Mons, Av. Maistriau 19, B-7000 Mons, Belgium
    \label{AIM}}
\titlefoot{Physics Laboratory, University of Athens, Solonos Str.
     104, GR-10680 Athens, Greece
    \label{ATHENS}}
\titlefoot{Department of Physics, University of Bergen,
     All\'egaten 55, NO-5007 Bergen, Norway
    \label{BERGEN}}
\titlefoot{Dipartimento di Fisica, Universit\`a di Bologna and INFN,
     Via Irnerio 46, IT-40126 Bologna, Italy
    \label{BOLOGNA}}
\titlefoot{Centro Brasileiro de Pesquisas F\'{\i}sicas, rua Xavier Sigaud 150,
     BR-22290 Rio de Janeiro, Brazil \\
     \indent~~and Depto. de F\'{\i}sica, Pont. Univ. Cat\'olica,
     C.P. 38071 BR-22453 Rio de Janeiro, Brazil \\
     \indent~~and Inst. de F\'{\i}sica, Univ. Estadual do Rio de Janeiro,
     rua S\~{a}o Francisco Xavier 524, Rio de Janeiro, Brazil
    \label{BRASIL}}
\titlefoot{Coll\`ege de France, Lab. de Physique Corpusculaire, IN2P3-CNRS,
     FR-75231 Paris Cedex 05, France
    \label{CDF}}
\titlefoot{CERN, CH-1211 Geneva 23, Switzerland
    \label{CERN}}
\titlefoot{Institut de Recherches Subatomiques, IN2P3 - CNRS/ULP - BP20,
     FR-67037 Strasbourg Cedex, France
    \label{CRN}}
\titlefoot{Now at DESY-Zeuthen, Platanenallee 6, D-15735 Zeuthen, Germany
    \label{DESY}}
\titlefoot{Institute of Nuclear Physics, N.C.S.R. Demokritos,
     P.O. Box 60228, GR-15310 Athens, Greece
    \label{DEMOKRITOS}}
\titlefoot{FZU, Inst. of Phys. of the C.A.S. High Energy Physics Division,
     Na Slovance 2, CZ-180 40, Praha 8, Czech Republic
    \label{FZU}}
\titlefoot{Dipartimento di Fisica, Universit\`a di Genova and INFN,
     Via Dodecaneso 33, IT-16146 Genova, Italy
    \label{GENOVA}}
\titlefoot{Institut des Sciences Nucl\'eaires, IN2P3-CNRS, Universit\'e
     de Grenoble 1, FR-38026 Grenoble Cedex, France
    \label{GRENOBLE}}
\titlefoot{Helsinki Institute of Physics, HIP,
     P.O. Box 9, FI-00014 Helsinki, Finland
    \label{HELSINKI}}
\titlefoot{Joint Institute for Nuclear Research, Dubna, Head Post
     Office, P.O. Box 79, RU-101 000 Moscow, Russian Federation
    \label{JINR}}
\titlefoot{Institut f\"ur Experimentelle Kernphysik,
     Universit\"at Karlsruhe, Postfach 6980, DE-76128 Karlsruhe,
     Germany
    \label{KARLSRUHE}}
\titlefoot{Institute of Nuclear Physics,Ul. Kawiory 26a,
     PL-30055 Krakow, Poland
    \label{KRAKOW1}}
\titlefoot{Faculty of Physics and Nuclear Techniques, University of Mining
     and Metallurgy, PL-30055 Krakow, Poland
    \label{KRAKOW2}}
\titlefoot{Universit\'e de Paris-Sud, Lab. de l'Acc\'el\'erateur
     Lin\'eaire, IN2P3-CNRS, B\^{a}t. 200, FR-91405 Orsay Cedex, France
    \label{LAL}}
\titlefoot{School of Physics and Chemistry, University of Lancaster,
     Lancaster LA1 4YB, UK
    \label{LANCASTER}}
\titlefoot{LIP, IST, FCUL - Av. Elias Garcia, 14-$1^{o}$,
     PT-1000 Lisboa Codex, Portugal
    \label{LIP}}
\titlefoot{Department of Physics, University of Liverpool, P.O.
     Box 147, Liverpool L69 3BX, UK
    \label{LIVERPOOL}}
\titlefoot{LPNHE, IN2P3-CNRS, Univ.~Paris VI et VII, Tour 33 (RdC),
     4 place Jussieu, FR-75252 Paris Cedex 05, France
    \label{LPNHE}}
\titlefoot{Department of Physics, University of Lund,
     S\"olvegatan 14, SE-223 63 Lund, Sweden
    \label{LUND}}
\titlefoot{Universit\'e Claude Bernard de Lyon, IPNL, IN2P3-CNRS,
     FR-69622 Villeurbanne Cedex, France
    \label{LYON}}
\titlefoot{Dipartimento di Fisica, Universit\`a di Milano and INFN-MILANO,
     Via Celoria 16, IT-20133 Milan, Italy
    \label{MILANO}}
\titlefoot{Dipartimento di Fisica, Univ. di Milano-Bicocca and
     INFN-MILANO, Piazza della Scienza 2, IT-20126 Milan, Italy
    \label{MILANO2}}
\titlefoot{IPNP of MFF, Charles Univ., Areal MFF,
     V Holesovickach 2, CZ-180 00, Praha 8, Czech Republic
    \label{NC}}
\titlefoot{NIKHEF, Postbus 41882, NL-1009 DB
     Amsterdam, The Netherlands
    \label{NIKHEF}}
\titlefoot{National Technical University, Physics Department,
     Zografou Campus, GR-15773 Athens, Greece
    \label{NTU-ATHENS}}
\titlefoot{Physics Department, University of Oslo, Blindern,
     NO-0316 Oslo, Norway
    \label{OSLO}}
\titlefoot{Dpto. Fisica, Univ. Oviedo, Avda. Calvo Sotelo
     s/n, ES-33007 Oviedo, Spain
    \label{OVIEDO}}
\titlefoot{Department of Physics, University of Oxford,
     Keble Road, Oxford OX1 3RH, UK
    \label{OXFORD}}
\titlefoot{Dipartimento di Fisica, Universit\`a di Padova and
     INFN, Via Marzolo 8, IT-35131 Padua, Italy
    \label{PADOVA}}
\titlefoot{Rutherford Appleton Laboratory, Chilton, Didcot
     OX11 OQX, UK
    \label{RAL}}
\titlefoot{Dipartimento di Fisica, Universit\`a di Roma II and
     INFN, Tor Vergata, IT-00173 Rome, Italy
    \label{ROMA2}}
\titlefoot{Dipartimento di Fisica, Universit\`a di Roma III and
     INFN, Via della Vasca Navale 84, IT-00146 Rome, Italy
    \label{ROMA3}}
\titlefoot{DAPNIA/Service de Physique des Particules,
     CEA-Saclay, FR-91191 Gif-sur-Yvette Cedex, France
    \label{SACLAY}}
\titlefoot{Instituto de Fisica de Cantabria (CSIC-UC), Avda.
     los Castros s/n, ES-39006 Santander, Spain
    \label{SANTANDER}}
\titlefoot{Inst. for High Energy Physics, Serpukov
     P.O. Box 35, Protvino, (Moscow Region), Russian Federation
    \label{SERPUKHOV}}
\titlefoot{J. Stefan Institute, Jamova 39, SI-1000 Ljubljana, Slovenia
     and Laboratory for Astroparticle Physics,\\
     \indent~~Nova Gorica Polytechnic, Kostanjeviska 16a, SI-5000 Nova Gorica, Slovenia, \\
     \indent~~and Department of Physics, University of Ljubljana,
     SI-1000 Ljubljana, Slovenia
    \label{SLOVENIJA}}
\titlefoot{Fysikum, Stockholm University,
     Box 6730, SE-113 85 Stockholm, Sweden
    \label{STOCKHOLM}}
\titlefoot{Dipartimento di Fisica Sperimentale, Universit\`a di
     Torino and INFN, Via P. Giuria 1, IT-10125 Turin, Italy
    \label{TORINO}}
\titlefoot{INFN,Sezione di Torino, and Dipartimento di Fisica Teorica,
     Universit\`a di Torino, Via P. Giuria 1,\\
     \indent~~IT-10125 Turin, Italy
    \label{TORINOTH}}
\titlefoot{Dipartimento di Fisica, Universit\`a di Trieste and
     INFN, Via A. Valerio 2, IT-34127 Trieste, Italy \\
     \indent~~and Istituto di Fisica, Universit\`a di Udine,
     IT-33100 Udine, Italy
    \label{TU}}
\titlefoot{Univ. Federal do Rio de Janeiro, C.P. 68528
     Cidade Univ., Ilha do Fund\~ao
     BR-21945-970 Rio de Janeiro, Brazil
    \label{UFRJ}}
\titlefoot{Department of Radiation Sciences, University of
     Uppsala, P.O. Box 535, SE-751 21 Uppsala, Sweden
    \label{UPPSALA}}
\titlefoot{IFIC, Valencia-CSIC, and D.F.A.M.N., U. de Valencia,
     Avda. Dr. Moliner 50, ES-46100 Burjassot (Valencia), Spain
    \label{VALENCIA}}
\titlefoot{Institut f\"ur Hochenergiephysik, \"Osterr. Akad.
     d. Wissensch., Nikolsdorfergasse 18, AT-1050 Vienna, Austria
    \label{VIENNA}}
\titlefoot{Inst. Nuclear Studies and University of Warsaw, Ul.
     Hoza 69, PL-00681 Warsaw, Poland
    \label{WARSZAWA}}
\titlefoot{Fachbereich Physik, University of Wuppertal, Postfach
     100 127, DE-42097 Wuppertal, Germany \\
\noindent
{$^\dagger$~deceased}
    \label{WUPPERTAL}}

\addtolength{\textheight}{-10mm}
\addtolength{\footskip}{5mm}
\clearpage
\headsep 30.0pt
\end{titlepage}
%
\pagenumbering{arabic} 
\setcounter{footnote}{0} %
\large
\section{Introduction}

\begin{figure}[!t]
\begin{center}
\begin{tabular}{cc}
(a) & (b) \\
\begin{picture}(175,200)(0,0)
\ArrowLine(0,50)(25,100)
\ArrowLine(25,100)(0,150)
\Photon(25,100)(50,100){4}{4}
\Vertex(25,100){2}
\ArrowLine(50,100)(100,175)
\ArrowLine(100,25)(50,100)
\Vertex(50,100){2}
\Vertex(60,115){2}
\Gluon(60,115)(75,100){4}{2}
\Vertex(75,100){2}
\DashArrowLine(75,100)(100,125){4}{}
\DashArrowLine(75,100)(100,75){4}{}
\Text(15,145)[]{$\mathrm{e^+}$}
\Text(15,55)[]{$\mathrm{e^-}$}
\Text(38,115)[]{$\gamma$,\Zo}
\Text(90,175)[]{$\mathrm{q}$}
\Text(90,25)[]{$\mathrm{\bar{q}}$}
\Text(75,115)[]{g}
\Text(90,130)[]{\glui}
\Text(90,70)[]{\glui}
\end{picture}
&
\includegraphics[width=6.5cm]{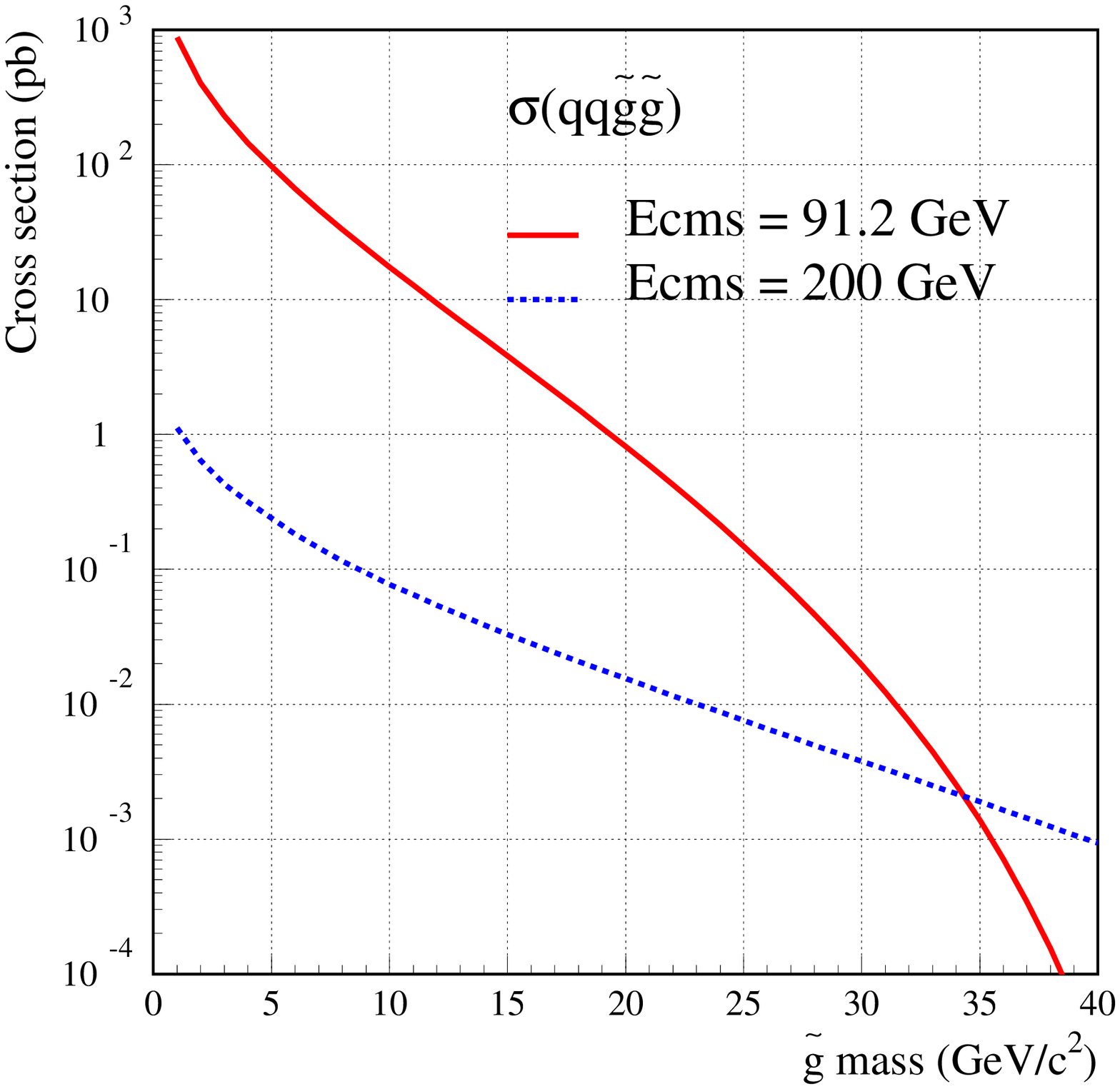} \\
\end{tabular}
\caption{(a) Gluon splitting into a pair of gluinos. 
(b) Comparison of the cross-section (pb) of this process at centre-of-mass
energies of 91.2~GeV (LEP1) and of 200~GeV (LEP2).}
\label{fi:qqgg}
\end{center}
\end{figure}

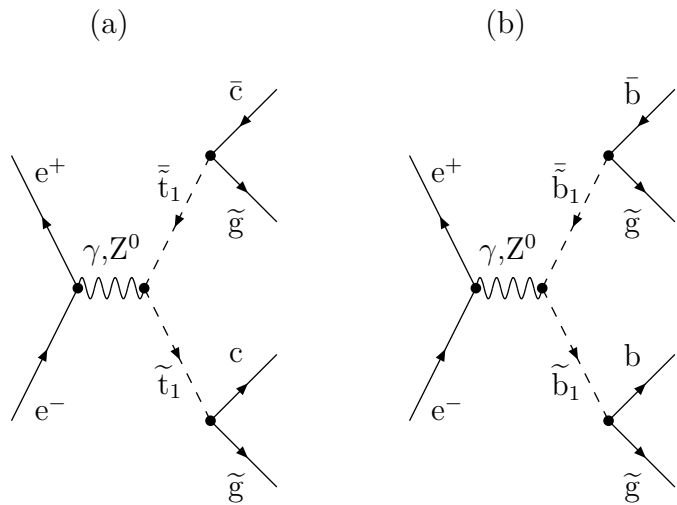
\begin{figure}[!thb]
\begin{center}
\begin{picture}(250,200)(0,0)
\ArrowLine(0,50)(25,100)
\ArrowLine(25,100)(0,150)
\Photon(25,100)(50,100){4}{4}
\Vertex(25,100){2}
\DashArrowLine(75,150)(50,100){4}{}
\DashArrowLine(50,100)(75,50){4}{}
\Vertex(50,100){2}
\ArrowLine(100,175)(75,150)
\ArrowLine(75,150)(100,125)
\Vertex(75,150){2}
\ArrowLine(75,50)(100,25)
\ArrowLine(75,50)(100,75)
\Vertex(75,50){2}
\Text(15,145)[]{$\mathrm{e^+}$}
\Text(15,55)[]{$\mathrm{e^-}$}
\Text(38,115)[]{$\gamma$,\Zo}
\Text(60,140)[]{\aSti}
\Text(60,65)[]{\Sti}
\Text(85,175)[]{\ac}
\Text(85,125)[]{\glui}
\Text(85,75)[]{\qc}
\Text(85,25)[]{\glui}
\ArrowLine(150,50)(175,100)
\ArrowLine(175,100)(150,150)
\Photon(175,100)(200,100){4}{4}
\Vertex(175,100){2}
\DashArrowLine(225,150)(200,100){4}{}
\DashArrowLine(200,100)(225,50){4}{}
\Vertex(200,100){2}
\ArrowLine(250,175)(225,150)
\ArrowLine(225,150)(250,125)
\Vertex(225,150){2}
\ArrowLine(225,50)(250,25)
\ArrowLine(225,50)(250,75)
\Vertex(225,50){2}
\Text(165,145)[]{$\mathrm{e^+}$}
\Text(165,55)[]{$\mathrm{e^-}$}
\Text(188,115)[]{$\gamma$,\Zo}
\Text(210,140)[]{\aSbi}
\Text(210,65)[]{\Sbi}
\Text(235,175)[]{\ab}
\Text(235,125)[]{\glui}
\Text(235,75)[]{\mbox{${\mathrm b}$}}
\Text(235,25)[]{\glui}
\Text(37,200)[]{(a)}
\Text(187,200)[]{(b)}
\end{picture}
\caption{Stop (a) and sbottom (b) production and decay at LEP2.}
\label{fi:sqdecay}
\end{center}
\end{figure}

In minimal supergravity supersymmetry models (mSUGRA), the gaugino masses
($M_i$) are usually supposed to evolve from a common value $m_{1/2}$ at the GUT 
scale. In such models, the $M_i$ are proportional to the corresponding coupling
constants ($g_i$) and the gluino is naturally heavier than the other
gauginos at the electroweak scale. 
\begin{eqnarray*}
\frac{M_1}{g^2_1} = \frac{M_2}{g^2_2} = \frac{M_3}{g^2_3} & \mbox{ that is, } & 
M_1:M_2:M_3 \, \sim \, 1:2:7 
\end{eqnarray*}
Nevertheless, models exist where the $M_i$ do not follow this relation. 
$M_3$ could be lighter than the other gaugino masses~\cite{GUNION}, and in 
this case, the gluino is the Lightest 
Supersymmetric Particle (LSP). For example there is a particular Gauge Mediated 
Supersymmetry Breaking model (GMSB)~\cite{RABY} where the gluino can either be 
the LSP or the next to lightest  supersymmetric particle (NLSP) with a 
gravitino LSP. In the latter case, the lifetime of the
gluino would ever be sufficiently large to consider the gluino as a stable
particle for collider physic. If R-parity is assumed, 
the gluino is stable in all these models and it should 
hadronize to form R-hadrons because of color confinement.\\
\indent
The gluino has been 
intensively searched for in hadron collisions in various decay 
channels~\cite{TEVATRON}.
However, the limit obtained ($m_{\tilde{g}}>$ 173~\GeVcc\ for 
$m_{\tilde{g}}=m_{\tilde{q}}$) does not apply to 
a stable gluino. For this model, it has been shown that CDF run I data 
could not constrain a stable gluino with mass lower than 
35~\GeVcc~\cite{GUNION,RABY}. On the other hand, the gluino mass could
be much larger than it is in the so-called light gluino scenario~\cite{FARRAR}, 
which seems to be excluded by the measurement of the triple gluon 
coupling and of the four-jet rates at LEP~\cite{LEPALPHAS}.\\ 
\indent
A pair of gluinos can be produced in
the splitting of a gluon. Figure~\ref{fi:qqgg} shows the Feynman diagram of this
process and the corresponding cross-sections at centre-of-mass energies of
91.2~GeV (LEP1) and of 200~GeV (LEP2). The production rate is too low at LEP2, 
and LEP1 data must be analysed to be sensitive to this process.
In this channel, the gluino does not originate
from the decay of another sparticle, so it can be produced even if the other
supersymmetric particles are not accessible.\\
\indent
Gluinos can also be produced from other sparticle decays.
\Sql\ and \Sqr\ are the supersymmetric partners of the left-handed and right handed 
quarks. 
With large Yukawa coupling running and important off-diagonal terms 
the supersymmetric partners of top and bottom quarks are the most probable candidates 
for the charged lightest supersymmetric particle.
The squark mass eigenstates are parametrised by a mixing angle 
$\mathrm{\theta_{\tilde{q}}}$. The lighter squark is given by
$\Sqi = \Sql \sin \theta_{\tilde{q}} + \Sqr \cos \theta_{\tilde{q}}$.
The stop (\Sti) and the sbottom (\Sbi) could be pair-produced at LEP 
via \ee\ annihilation into $\Zo/\gamma$. A squark mixing angle equal to zero 
leads to the maximal squark pair production cross-section, while 
the minimal cross-sections are obtained for a mixing angle of 
$56^\circ$ for the stop and of $68^\circ$ for the sbottom.
For these particular angles, the $\Zo-\Sqi-\aSqi$ coupling is suppressed.
This paper also describes the search for R-hadrons from stop and sbottom decays
at LEP2. The dominant decay of the 
stop and of the sbottom are $\Sti~\to~\qc\glui$ and 
$\Sbi~\to~{\mathrm b}\glui$~\cite{KOBAY} when the gluino is lighter than
the squarks, as in the stable gluino scenario. The branching ratios
of these decay channels were taken to be 100\%.
Figure~\ref{fi:sqdecay} shows the squark production and decay diagrams.\\
\indent
DELPHI data collected in 1994 at a centre-of-mass energy of 91.2~GeV were 
used to search for the process $\ee\to\qqbar\glui\glui$. 
Then, stop and sbottom squarks 
were searched for in DELPHI data collected from 1998 to 2000 at centre-of-mass 
energies ranging from 189 to 208~GeV.
The three analyses presented in this paper are therefore searches for:
\begin{eqnarray*}
\left\{
\begin{array}{lclcl}
\ee & \to & \qqbar\glu  & \to & \qqbar\glui\glui \\
\ee & \to & \Sti\aSti   & \to & {\mathrm c}\glui\bar{{\mathrm c}}\glui \\
\ee & \to & \Sbi\aSbi   & \to & {\mathrm b}\glui\bar{{\mathrm b}}\glui \\
\end{array}
\right.
\end{eqnarray*}
all giving the same topology of two standard jets plus two gluino jets.
The gluino could either fragment to neutral \ro\ states (\glui\glu, 
\glui\qu\au, ...) or to charged \rpm\ states (\glui\qu\ad,...). 
If $P$ is the probability that a gluino fragments to a charged R-hadron, then
for $P=1$, R-hadrons are identified by an anomalous ionizing 
energy loss in the tracking chambers, and for $P=0$, the gluino hadronizes into
neutral states which reach the calorimeters where they deposit a part of their 
energy. The missing energy carried away by the LSP is reduced
compared with usual SUSY models with R-parity conservation.
For neutral R-hadrons, the estimate of the experimental sensitivity depends on 
the model used to calculate the energy loss in the calorimeters.

\section{The DELPHI detector}

The description of the DELPHI detector and its performance can be found in
references~\cite{DELPHI1,DELPHI2}. We only summarize here the parts
relevant to the analysis.\\
\indent
Charged particles are reconstructed in a 1.2 T magnetic field by a system of
cylindrical tracking detectors. The closest to the beam is the Vertex
Detector (VD) which consists of three cylindrical layers of silicon detectors 
at radii 6.3~cm, 9.0~cm and 11.0~cm. They measure 
coordinates in the $R\phi$ plane~\footnote{The DELPHI coordinate system is 
defined with $z$ along the e$^-$ beam direction; $\theta$ and $\phi$ are the 
polar and azimuthal angles and $R$ is the radial distance from the $z$ axis.}.
In addition, the inner and the outer layers are  
double-sided giving also a $z$ measurement. VD is the barrel part of the
Silicon Tracker (ST), which extends the polar angle acceptance down to 10 
degrees. The Inner Detector (ID) is a 
drift chamber with inner radius 12~cm and outer radius 22~cm covering polar 
angles between $15\dgr$ 
and $165\dgr$. The principal tracking detector of DELPHI is the Time
Projection Chamber (TPC). It is a cylinder of 30 cm inner radius, 122 cm 
outer radius and 2.7~m length. Each end-plate is divided into 6 sectors, with
192 sense wires to allow the dE/dx measurement, and with 16 circular pad rows
which provide 3-dimensional track reconstruction. The TPC covers polar angles
from $20\dgr$ to $160\dgr$. Finally, the Outer Detector (OD) consists of 
drift cells at radii between 192~cm and 208~cm, covering polar angles between 
$43\dgr$ and $137\dgr$. In addition, two planes of drift chambers 
perpendicular to the  beam axis (Forward Chambers A and B) are installed in
the endcaps covering polar angles $11\dgr<\theta<33\dgr$ and 
$147\dgr<\theta<169\dgr$.\\
\indent
The electromagnetic calorimeters are the High density Projection
Chamber~(HPC) in the barrel region ($40\dgr<\theta<140\dgr$) and the Forward 
Electromagnetic Calorimeter~(FEMC) in the endcaps ($11\dgr<\theta<36\dgr$ and
$144\dgr<\theta<169\dgr$). In the forward and backward regions, the 
Scintillator TIle Calorimeter (STIC) extends the coverage down to 1.66$\dgr$ 
from the beam axis. The number of radiation lengths are respectively 18, 20 
and 27 in the HPC, the FEMC and the STIC.
In the gap between the HPC and the FEMC, 
hermeticity taggers made of single layer scintillator-lead counters are used
to veto events with electromagnetic particles which would otherwise escape 
detection. Between the HPC modules, gaps at $\theta=90\dgr$ and 
gaps in $\phi$ are also instrumented with such taggers.
Finally, the hadron calorimeter (HCAL) covers polar angle between 
$11\dgr<\theta<169\dgr$. The iron thickness in the HCAL is 110~cm 
which corresponds to 6.6 nuclear interaction lengths.

\section{Data and Monte Carlo samples}

The total integrated luminosity collected by the DELPHI detector in 1994 at the
\Zo\ peak ($\sqrt{s}=$~91.2~GeV) was 46~\pb. It corresponded to
around 1.6 million hadronic \Zo\ events. The Standard Model hadronic background 
was estimated
with the JETSET~7.3~\cite{JETSET} program tuned to reproduce LEP1 
data~\cite{DELTUNE}. The program described in~\cite{GUNION} was
used to simulate the $\ee\to\qqbar\glui\glui$ signal.\\
\indent
At LEP2, the total integrated luminosity collected by the DELPHI detector at 
centre-of-mass energies from 189 to 208~\GeV\ was 609~\pb. 
In September 2000, one of the twelve TPC sectors (sector 6) stopped functioning.
About 60~\pb\ of data were collected without that sector until the end of the
data taking. The reconstruction programs were modified to allow the analysis of
the data taken with the DELPHI TPC not fully operational. After this
modification, there was only a small degradation of the performance of the 
tracking. The data collected in year 2000 were divided in 
centre-of-mass energy windows to optimize the analysis sensitivity.
Table~\ref{tab:ecmvslumi} summarizes the LEP2 data samples used in the 
analysis.\\

\begin{table}[!htb]
\begin{center}
\begin{tabular}{|l||c|c|c|}
\hline
 Year          & $<\sqrt{s}>$ (GeV) & $\sqrt{s}$ (GeV) & Integrated luminosity  \\
               &      Data          & Simulated MC     & ($pb^{-1}$) \\
\hline \hline
1998 & 188.6 & 189    & 158.0 \\  \hline
1999 & 191.6 & 192    &  25.9 \\
     & 195.6 & 196   &  76.4  \\
     & 199.6 & 200   &  83.4  \\
     & 201.6 & 202   &  40.6  \\  \hline
2000 & 204.8 & 204   &  78.1  \\
     & 206.6 & 206   &  78.5  \\
     & 208.1 & 208   &   7.3  \\ \hline
2000(*) & 206.5 & 206.7 &  60.6  \\ \hline
\end{tabular}
\caption{Total integrated luminosity as a function of the centre-of-mass
energy of the LEP2 analysed data samples. The third column shows the centre-of-mass
energy of the simulated events. (*) indicates the data collected by DELPHI in
2000 without the sector 6 of the TPC.}
\label{tab:ecmvslumi}
\end{center}
\end{table}

\begin{figure}[t!]
\begin{center}
\includegraphics[width=6cm]{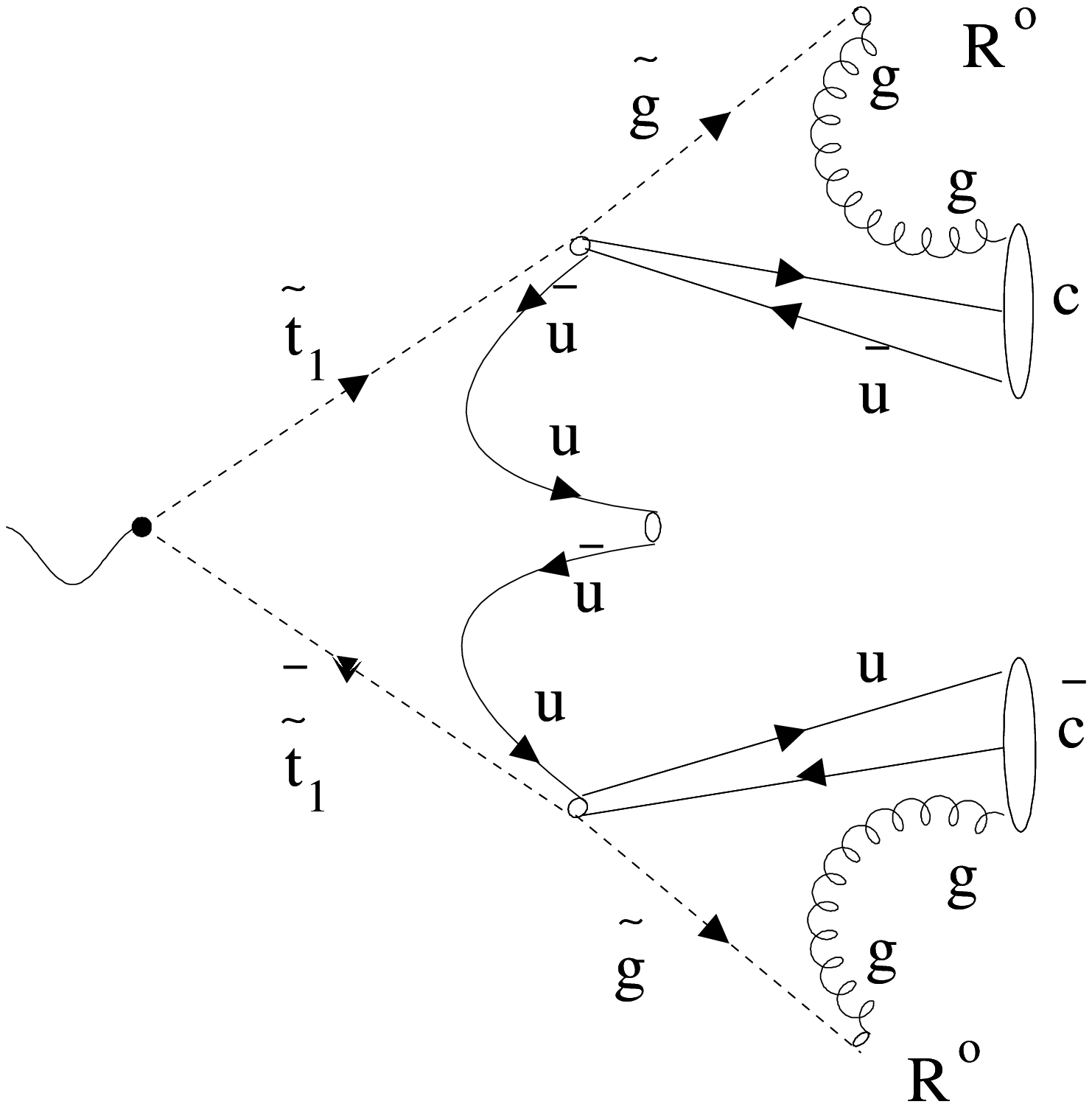}
\includegraphics[width=6cm]{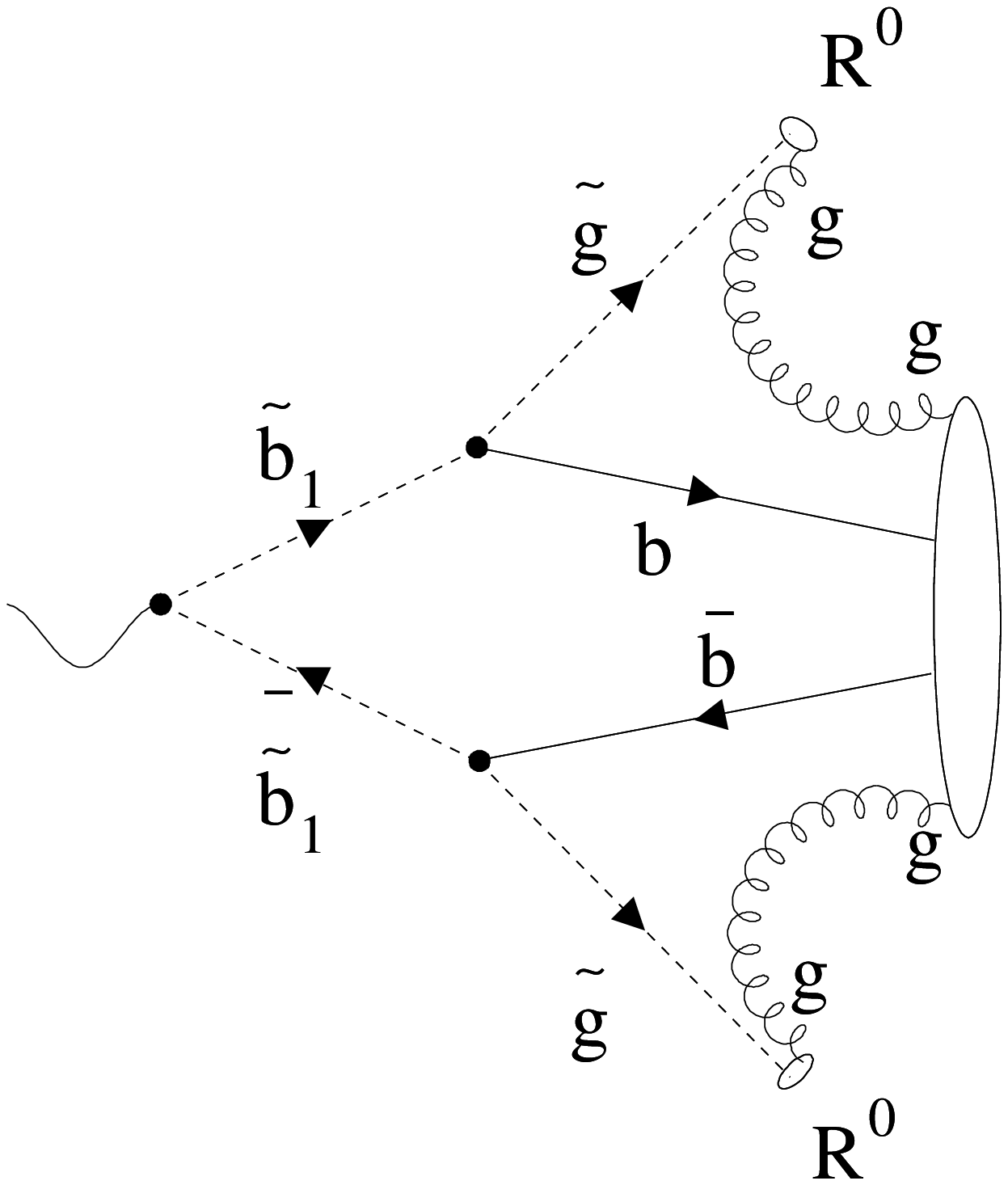}
\caption{Production and decay of the stop and sbottom squarks. Ellipses indicate
the color singlet and the color string stretched between the
partons.}
\label{fi:sqgluprod}
\end{center}
\end{figure}

The \ee\ interactions leading to four-fermion final states were generated using 
EXCALIBUR~\cite{EXCALIBUR}. GRC4F~\cite{GRC4F} was used to simulate 
the processes \ee~$\to$~$\mathrm{e\nu}\qqbar$ and 
\ee~$\to$~$Z^o \mathrm{ee}$ with electrons 
emitted at polar angles lower than the cut imposed
in EXCALIBUR. The two-fermion final states were 
generated with PYTHIA~\cite{JETSET} for \ee~$\to$~$\qqbar(n\gamma)$, 
KORALZ~\cite{KORALZ} for \ee~$\to$~$\tap\tam(\gamma)$, 
\ee~$\to$~$\muop\muom(\gamma)$, \ee~$\to$~$\neu\neu(\gamma)$, 
and BHWIDE~\cite{BHWIDE} for \ee~$\to$~$\elep\elem(\gamma)$. 
PYTHIA~6.143~\cite{PYTHIA} was used to simulate $\gamma\gamma$ interactions 
leading to hadronic final states. 
BDKRC~\cite{BDKRC} was used for $\gamma\gamma$ interactions leading to leptonic
final states.
In all cases, the final hadronization of the particles was performed with 
JETSET~\cite{JETSET}.\\
\indent
The flavour changing decay \Sti\ $\to$\ \qc\glui\ goes through one-loop 
diagrams. Therefore, \Sti\ is expected to be long-lived and to hadronize before 
its decay. A modified version of the SUSYGEN~\cite{SUSYGEN} generator was 
used to simulate this process. Special care was taken to introduce hard 
gluon radiation off the scalar stop at the matrix-element level and to treat 
the stop hadronization as a non-perturbative strong interaction effect. 
A detailed description of this hadronization model can be found 
in~\cite{GENDREES}. Such a model, based on the Peterson 
function~\cite{PETERSON}, was also used to perform the gluino 
hadronization into R-hadrons. SUSYGEN has also been modified to perform the
sbottom decay into $\mathrm{b}\glui$. Figure~\ref{fi:sqgluprod} 
summarizes the stop and 
sbottom production and the fragmentation steps. The final hadronization was 
performed using JETSET~\cite{JETSET}.\\
\indent
The Monte Carlo samples used to simulate the Standard Model processes and 
the supersymmetric signals were passed through DELSIM~\cite{DELSIM}, the  
program simulating the full DELPHI detector response. They were subsequently 
processed with 
the same reconstruction program as the real data. The number of generated 
events was always several times higher than the number expected for the
integrated luminosity collected.\\

\section{R-hadron simulation}

\begin{figure}[ht!]
\begin{center}
\includegraphics[width=8cm]{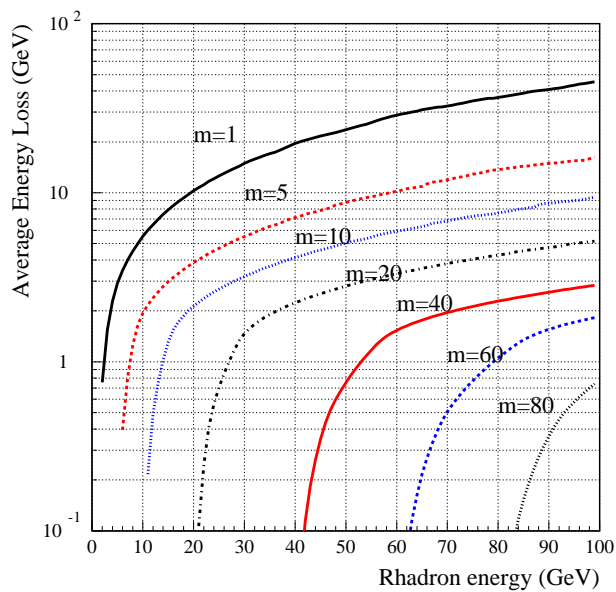}
\caption{Average energy loss by neutral R-hadrons in the DELPHI calorimeters 
as a function of their initial energy for different mass cases.}
\label{ELOSS}
\end{center}
\end{figure}

In the analysis, two generic R-hadron states were considered:
one charged denoted \rpm\ and one neutral, \ro, which corresponds 
to the glueballino, a \glui\glu\ state.
It is important to understand how an \ro\ would
manifest itself in the detector. We refer to the results of 
reference~\cite{GUNION}. The energy loss in the scattering on a nucleon
$\ro\mathrm{N} \rightarrow \ro\mathrm{X}$ is given by:
\begin{eqnarray*}
\Delta E={m_X^2-m_N^2+|t|\over 2 m_N}
\end{eqnarray*}
where $|t|$ is the usual momentum transfer invariant for the \ro\ 
and $m_X$ is the mass of the system produced in the 
$\ro\mathrm{N} \rightarrow \ro\mathrm{X}$ collision. 
The average energy loss in the reaction 
$\ro\mathrm{N} \rightarrow \ro\mathrm{X}$ is then
given by:
\begin{eqnarray*}
\vev{\Delta E}= {\int_{m_N}^{\sqrt{s}-m_{R^{\circ}}} dm_X
\int_{|t|_{\rm min}(m_X)}^{|t|_{\rm max}(m_X)} d|t| 
\,\Delta E{d\sigma\over d|t|dm_X}
\over 
\int_{m_N}^{\sqrt{s}-m_{R^{\circ}}} dm_X
\int_{|t|_{\rm min}(m_X)}^{|t|_{\rm max}(m_X)} d|t| {d\sigma\over d|t|dm_X}}
\end{eqnarray*}
where $s=M_{R^{\circ}}^2+m_N^2+2\gamma M_{R^{\circ}} m_N$ describes the collision energy.
The following functions are defined:
\begin{eqnarray*}
\left\{
\begin{array}{lcl}
\gamma                 & = & (1-\beta^2)^{-1/2}                        \\
|t|_{\rm min,max}(m_X) & = & 2[E(m_N)E(m_X)\mp p(m_N)p(m_X)-M_{R^{\circ}}^2] \\
E(m)                   & = & (s+M_{R^{\circ}}^2-m^2)/(2\sqrt{s})             \\
p(m)                   & = & \lambda^{1/2}(s,M_{R^{\circ}}^2,m^2)/(2\sqrt{s})\\
\lambda(a,b,c)         & = & a^2+b^2+c^2-2(ab+ac+bc)
\end{array}
\right.
\end{eqnarray*}
From the results of studies in~\cite{GUNION}, the differential
cross-section ${d\sigma\over d|t|dm_X}$ is taken as: 
\begin{eqnarray*}
{d\sigma\over d|t|dm_X}\propto
\left\{ 
\begin{array}{l}
1,\ {\mathrm if}\ |t|\leq 1\ {\mathrm GeV}^2 \\
0,\ {\mathrm if}\ |t|>    1\ {\mathrm GeV}^2 \\
\end{array}
\right.
\end{eqnarray*}
\indent
The average number of collisions of an \ro\ particle in the calorimeters is 
given by the depth of the calorimeter in units of equivalent iron 
interaction lengths, $\lambda_T$. In DELPHI, the electromagnetic 
calorimeter's thickness represents around 1~$\lambda_T$ while this value is 
6.6~$\lambda_T$ for the hadronic calorimeter. We have adopted a correction 
factor for the interaction length of 9/16 as suggested in 
reference~\cite{GUNION}:
a factor $C_A/C_F=9/4$ comes from the colour octet nature
of the \ro\ constituents increasing the $\sigma_{R^{\circ} N}$ cross-section as 
compared to $\sigma_{\pi N}$, while a factor 
$\vev{r^2_{R^{\circ}}}/\vev{r^2_{\pi}}=1/4$ takes into account the
relative size of the R-hadrons as compared to standard hadrons. On
average, neutral R-hadrons should undergo 4.3 collisions in DELPHI calorimeters.
Figure~\ref{ELOSS} shows the total energy loss by an \ro\ after 4 collisions
in iron.\\
\indent
The difficulty in separating the signal of a neutral R-hadron from the
background increases with the amount of energy lost in the calorimeters. The
choice of the interaction model made here is conservative in this respect.
The \ro\ scatters were subsequently treated in the DELPHI detector simulation 
as $\mathrm{K^0_L}$ with the energy that the \ro\ should deposit in four
collisions according to the above formula.\\
\indent
The charged R-hadrons were treated as heavy muons to 
reproduce the anomalous dE/dx signature. In this case, only the tracking
information was used to calculate the R-hadron momentum.

\section{Particle Identification and analysis method}

\subsection{Particle Identification and event preselection}

Particle selection was identical for LEP1 and LEP2 data. 
Reconstructed charged 
particles were required to have momenta above 100~\MeVc\ with $\Delta p / p < 1$, 
where $\Delta p$ is the momentum error, and impact parameter below 5~cm in the 
transverse plane and below 10~cm$/\sin\theta$ in the beam 
direction. More stringent cuts were applied for tracks without TPC information. 
A cluster in the calorimeters was selected as a neutral particle if not
associated to a charged particle and if the cluster energy was greater than 
500~MeV in the HPC, 400~MeV in the FEMC, 300~MeV in the STIC or 900~MeV in the 
HAC. Particles were then clustered into jets with the DURHAM
algorithm~\cite{DURHAM}. b-quarks were tagged using a probabilistic method 
based on the impact parameters of tracks with respect to the main vertex. A
combined b-tagging variable was defined by including the properties of secondary
vertices~\cite{BTAG}.\\
\indent
Events were then kept if there were at least two charged particles, and at
least one with a transverse momentum above 1.5~\GeVc, and if the transverse
energy~\footnote{The transverse energy is defined as the sum of 
$\sqrt{p_T^2 + m^2}$ over all particles; $p_T$ is the transverse momentum.}
exceeded 4~GeV.

\subsection{Neural networks}

A neural network allows one discriminating variable to be constructed from the
set of variables given as input. The form used here contains
three layers of nodes: the input layer where each neuron corresponds to a 
discriminating variable, the hidden layer, and the output layer which is the
response of the neural network. 
The layers were connected in a ``feed forward'' architecture.
The back-propagation algorithm was 
used to train the network with simulated events. This entails minimising 
a $\chi^2$ to adjust the neurons' weights and connections. An independent 
validation sample was also used not to overtrain the network. 
The outputs of neural networks were used to isolate events containing two 
neutral R-hadrons at LEP2.

\subsection{General description of the analyses}

The LEP1 and LEP2 data analyses have all the same final states which consists of
two jets and two gluino jets. Depending on the probability $P$ that the gluino
hadronizes into a charged R-hadron, three topologies are possible:
\begin{itemize}
\item{charged:} $\ee\to\qqbar\rpmrpm$
\item{mixed:}   $\ee\to\qqbar\rpmro$
\item{neutral:} $\ee\to\qqbar\roro$
\end{itemize}

\indent
For LEP1 data, the search analyses corresponding to the charged and the mixed
topologies were identical, while they were optimised separately
for LEP2 data. For these two topologies, the LEP1 and LEP2 
analyses were only based on anomalous ionizing energy loss, and no quark tagging
was applied.\\
\indent
For the neutral topology, LEP1 data were
analyzed using sequential cuts without trying to tag the quark flavor. Since
there are b-quarks in the final state of the sbottom decay, the 
b-quark tagging improves the isolation of this signal from the Standard Model 
backgrounds. Therefore, the b-tagging variable was used in the stop and 
sbottom search analyses at LEP2 in the neutral topology. This variable was
included into the neural networks which were used to isolate the stop and the 
sbottom signals.

\section{Search for a stable gluino at LEP1}

\subsection{Search for \qqbar\rpmrpm\ and \qqbar\rpmro\ events}

\begin{figure}[p!]
\begin{center}
\begin{tabular}{cc}
(a) & (b) \\
\includegraphics[width=8cm]{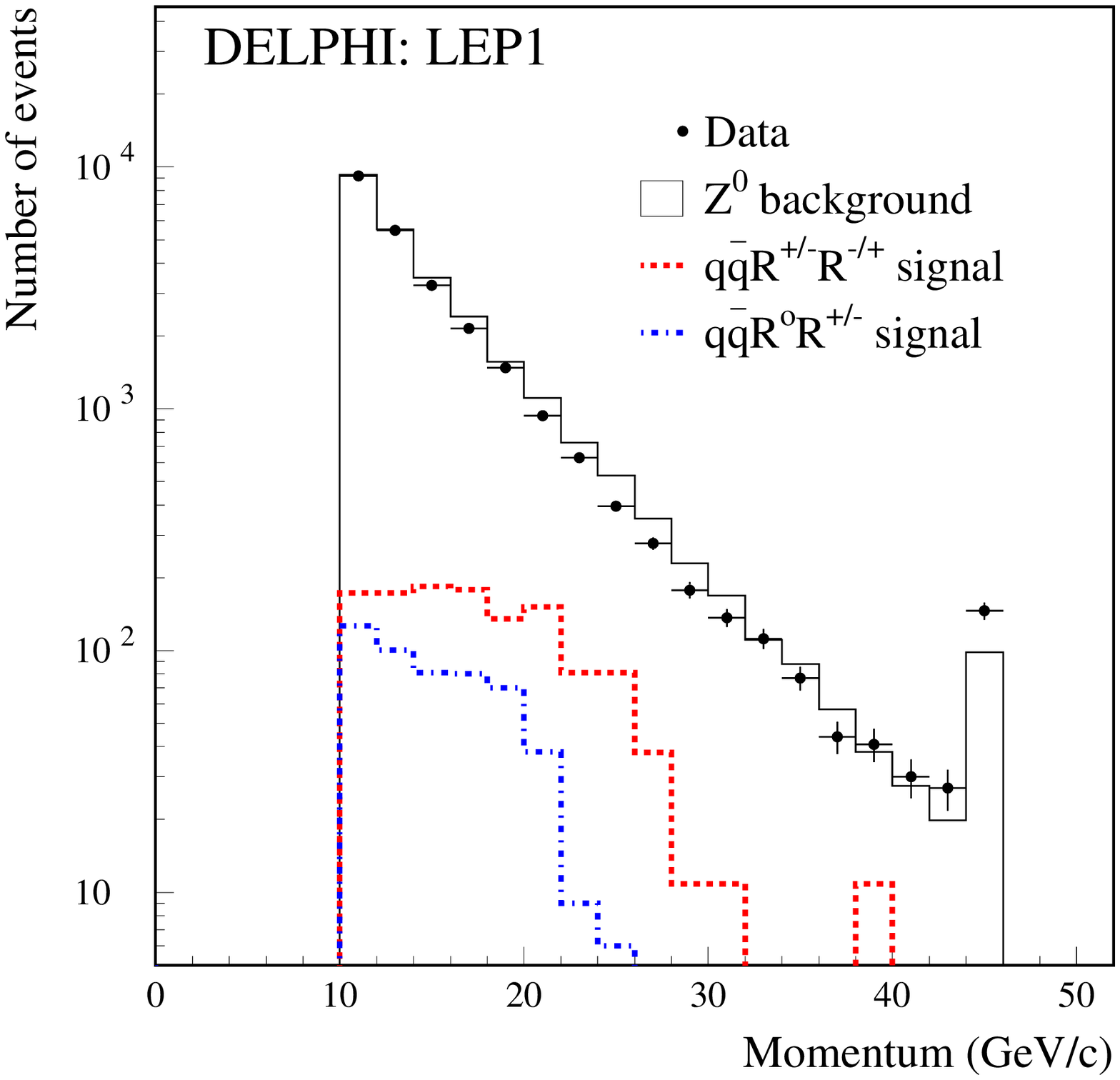} & 
\includegraphics[width=8cm]{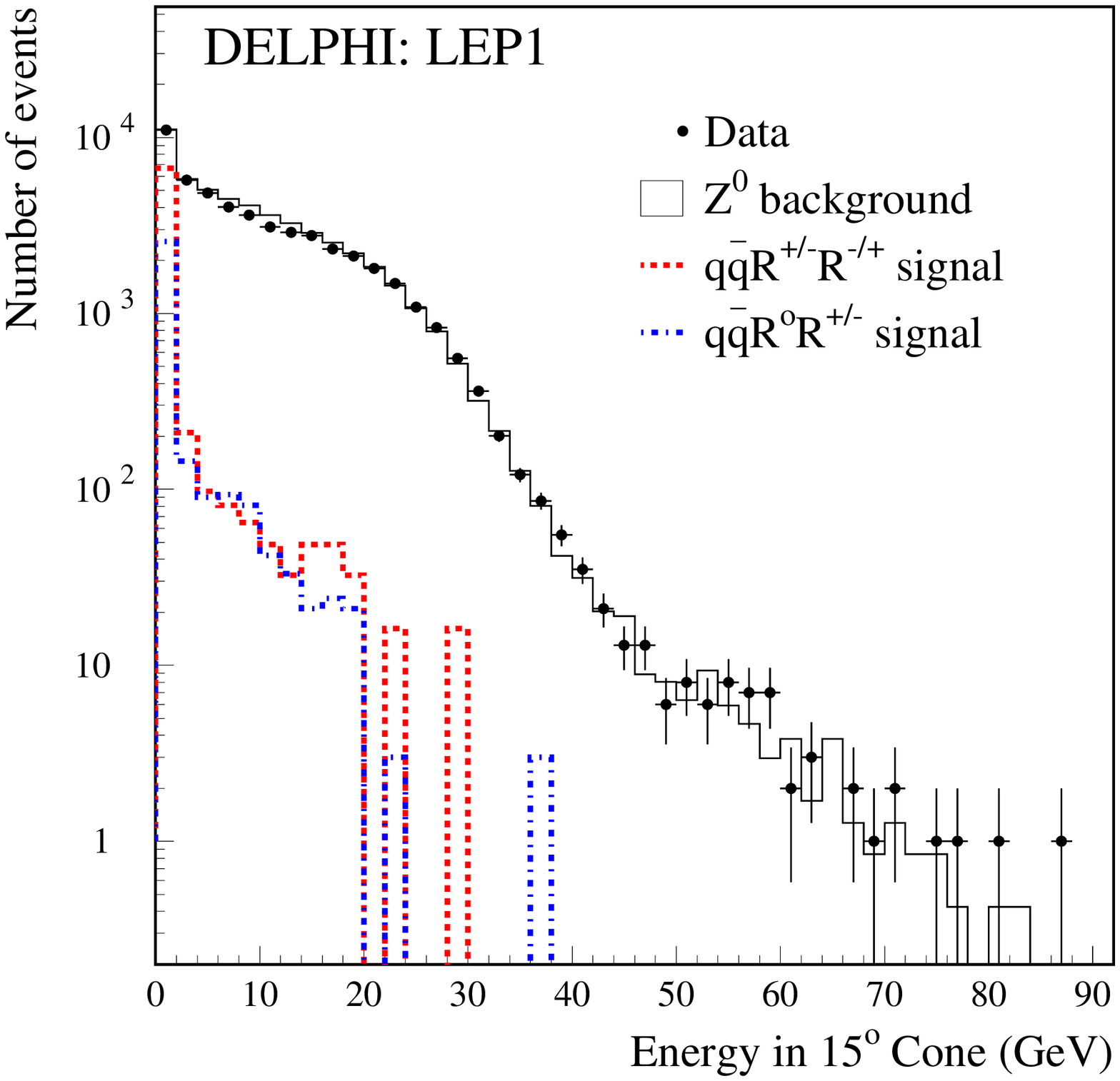} \\
(c) & (d) \\
\includegraphics[width=8cm]{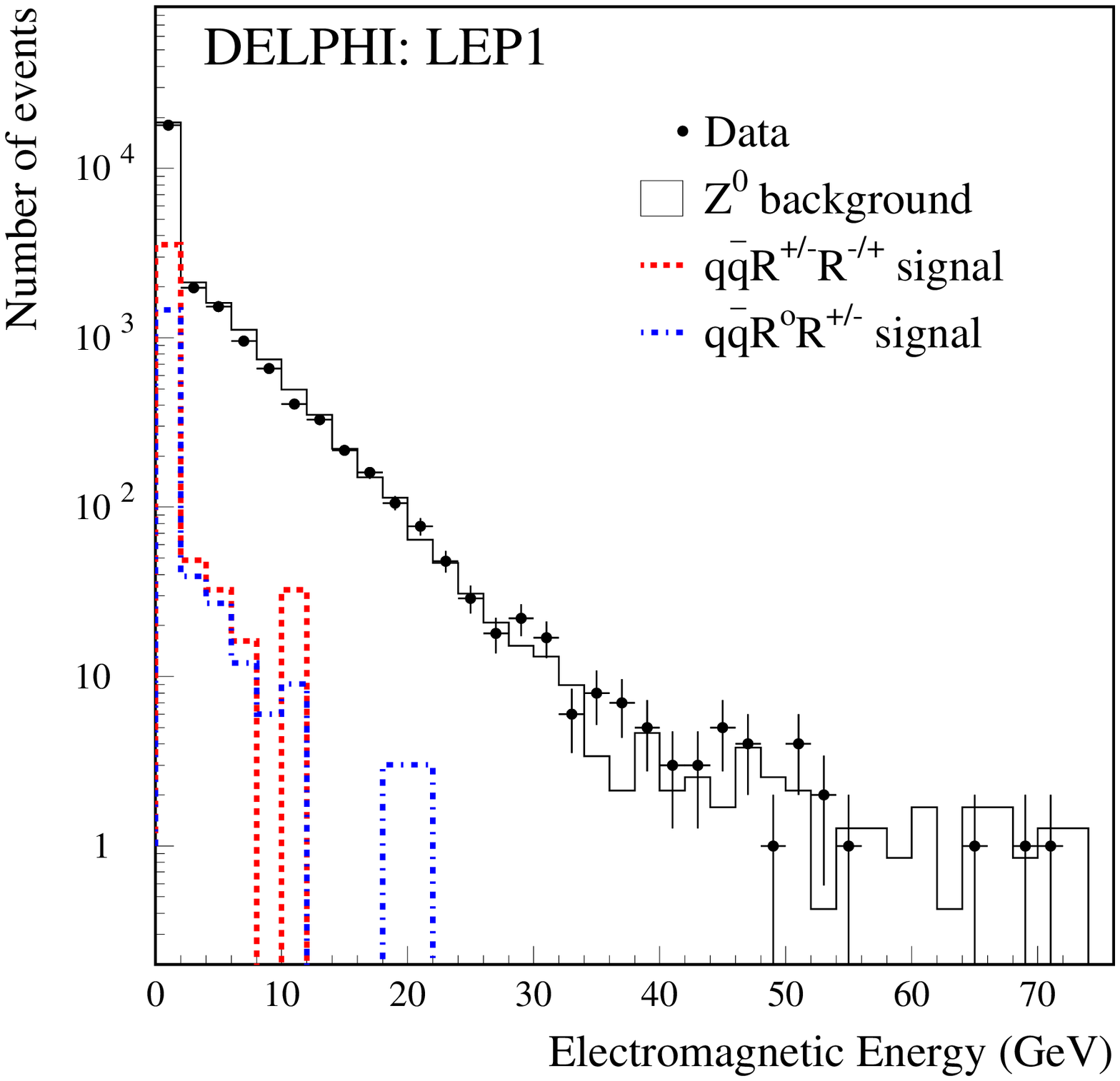} &
\includegraphics[width=8cm]{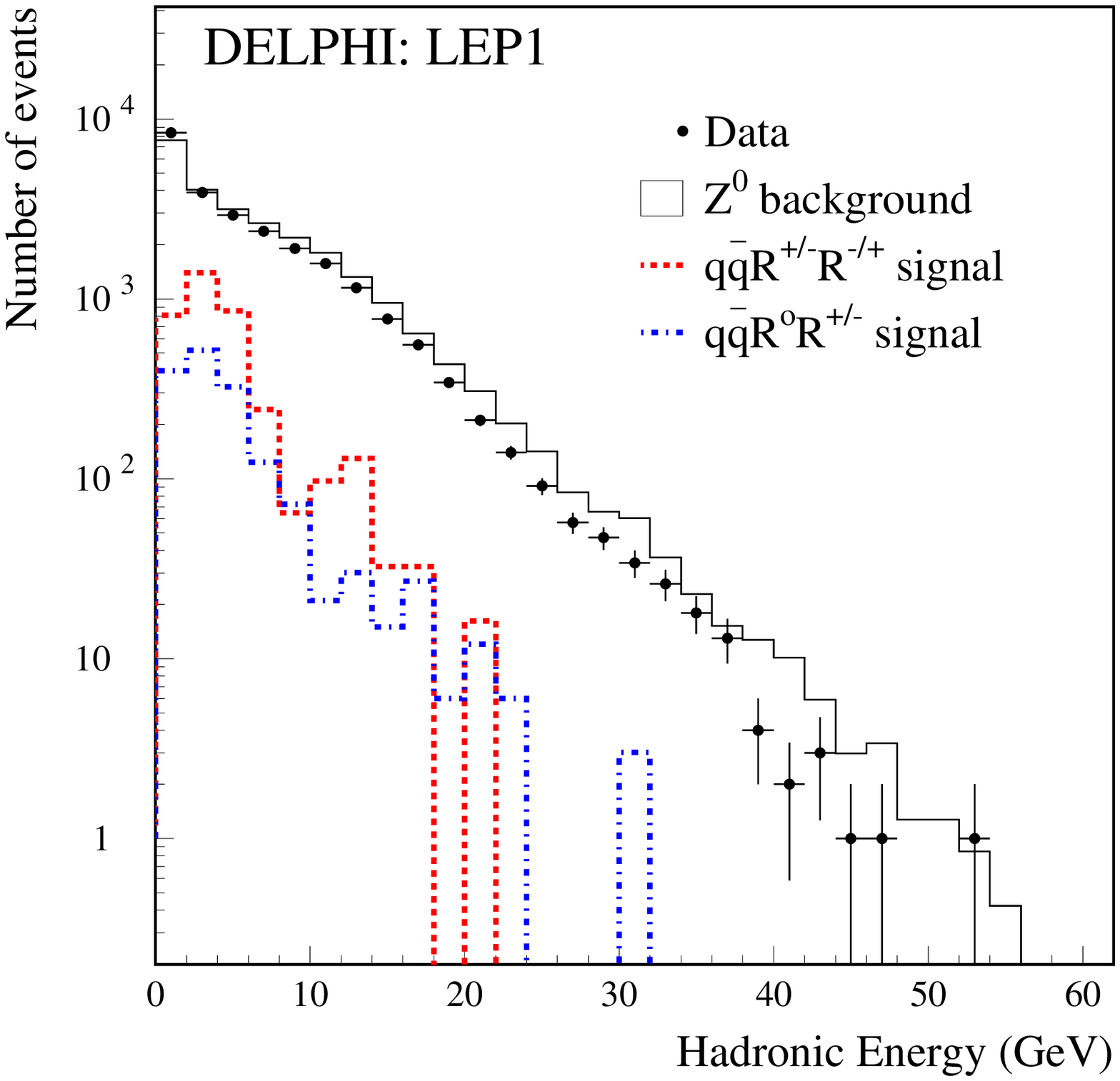} \\
\end{tabular}
\caption{Comparison between data and simulation in the search for charged
R-hadrons at LEP1. The plots show characteristic distributions before the
selection of the charged R-hadron candidates: (a) the momentum, (b) the total 
energy in a $15^\circ$ degree cone around the particle, (c) its electromagnetic
and (d) hadronic energy. Dotted lines show the \qqbar\rpmrpm\ and \qqbar\rpmro\ 
signal distributions with arbitrary normalization when all simulated samples are
added together.}
\label{fi:prerclep1}
\end{center}
\end{figure}

The same analysis based on the dE/dx measurement was performed to identify
\qqbar\rpmrpm\ and \qqbar\rpmro\ events.\\
\indent
In the preselection step, events were required to contain at least 5 charged
particles. At least one of these had to satisfy the following conditions.
The track was required to be reconstructed including a TPC track element and to
have a momentum above 10~\GeVc. At least 80 wires of the TPC were required to
have been included in the dE/dx measurement. The dE/dx had to be either greater
than 1.8 mip (units of energy loss for a minimum ionizing particle), or less 
than the
dE/dx expected for a particle of mass equal to 1~\GeVcc. The $Y_{23}$ variable is
the $y_{cut}$ value in the DURHAM algorithm for which the number of jets changes
between two and three. \qqbar\rpmrpm\ and \qqbar\rpmro\
events contain three or four jets. Thus, $Y_{23}$ was required to 
be less than 0.01. 
Figure~\ref{fi:prerclep1} shows a comparison between simulated and real data
at this level.\\
\indent
A \rpm\ candidate had to satisfy the following conditions: it had to be
reconstructed with the VD, the ID and the TPC detectors, and the dE/dx 
measurement had to be based on at least
80 wires of the TPC. In addition, the energy of the other particles in a $15^\circ$ cone 
around the 
\rpm\ candidate had to be less than 2~GeV. Finally, its associated
electromagnetic energy had to be less than 5~GeV.\\
\indent
The final selection was performed by cuts in the plane (P,dE/dx).
Figure~\ref{fi:dedx} shows the expected dE/dx as a function of the particle momentum.  
The analysis was separated into two mass windows:
\begin{itemize}
\item{$\mglui < 14~\GeVcc$:}\\
Here, charged R-hadrons were identified by low dE/dx values. The \rpm\ 
candidates were selected if their momentum was greater
than 15~\GeVc, and if their dE/dx was less than the dE/dx expected for a
particle of mass equal to 3~\GeVcc. 
\item{$\mglui \geq 14~\GeVcc$:}\\
In this mass window, R-hadrons were identified by high dE/dx values. The \rpm\
candidates were selected if their dE/dx was greater than 2~mip. 
\end{itemize}

\indent
The final selection was performed by requiring at least one charged R-hadron
candidate in either mass window. Table~\ref{tab:prelep1rhadc} contains
the number of events selected after each cut of this analysis.
For $\mglui < 14~\GeVcc$, 5 events were selected when 4.2 were expected. These
numbers are 12 and 13.5 in the $\mglui \geq 14~\GeVcc$ mass window.
Unlike the expected signal, all selected candidates in the data have only one
particle with anomalous dE/dx.
Figure~\ref{fi:effirlep1} shows the signal detection efficiencies. For 
\qqbar\rpmrpm, they ranged from a few percent for gluino masses 
close to 2~\GeVcc\ to around 50\% for gluino masses of the order of 25~\GeVcc.  
\qqbar\rpmro\ efficiencies were about half of the 
\qqbar\rpmrpm\ ones.

\subsection{Search for \qqbar\roro events}

\begin{figure}[p!]
\begin{center}
\includegraphics[width=14cm]{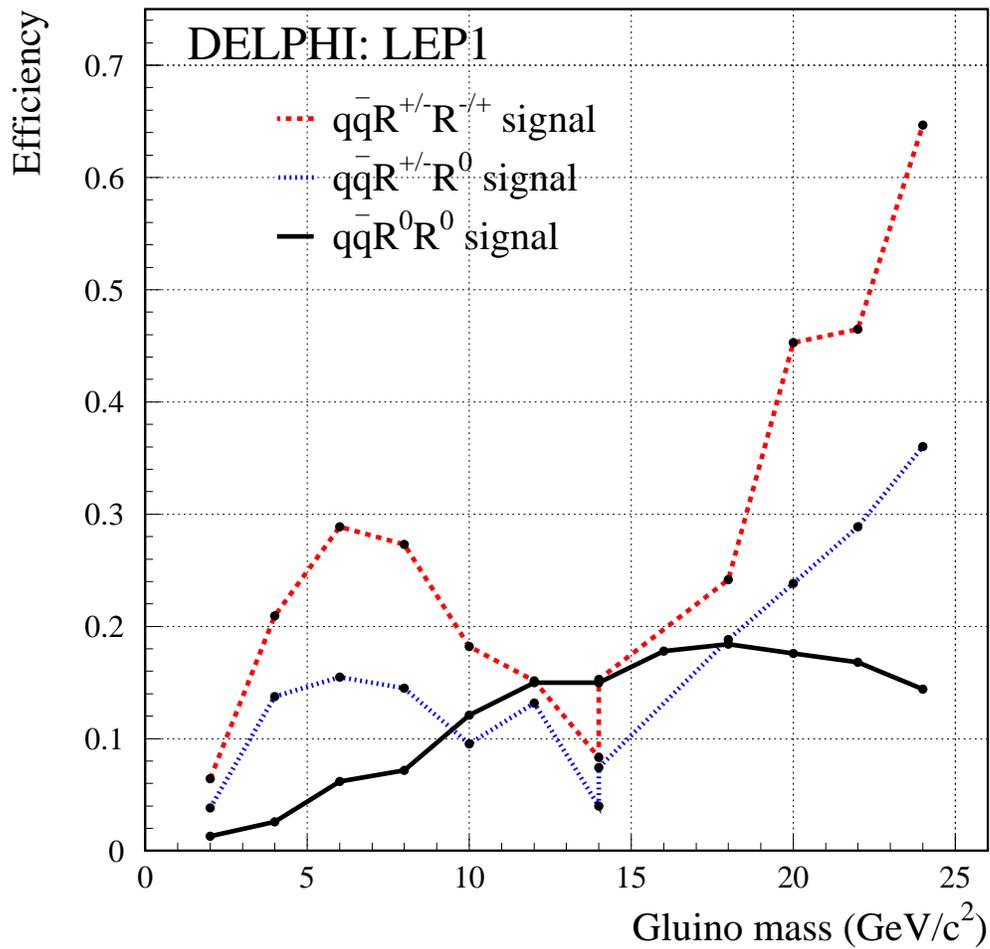}
\caption{Signal detection efficiencies in the LEP1 data analyses as a 
function of the gluino mass for the three signal topologies 
\qqbar\rpmrpm, \qqbar\rpmro\ and 
\qqbar\roro. The search for charged R-hadrons was optimized separately for low
gluino masses ($<14~\GeVcc$) and for high gluino masses ($\geq 14~\GeVcc$).}
\label{fi:effirlep1}
\end{center}
\end{figure}

\begin{figure}[p!]
\begin{center}
\begin{tabular}{cc}
(a) & (b) \\
\includegraphics[width=8cm]{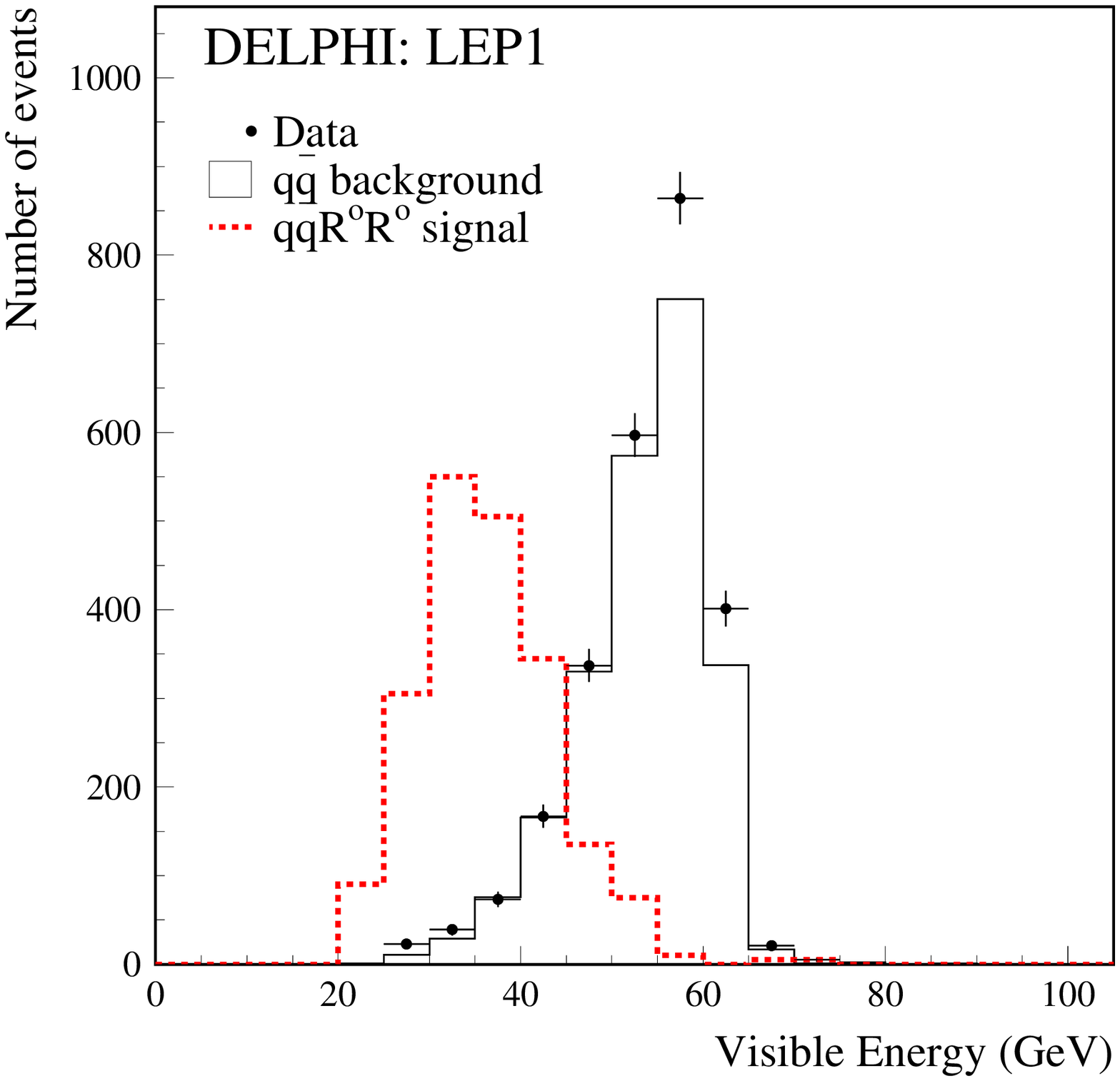} & 
\includegraphics[width=8cm]{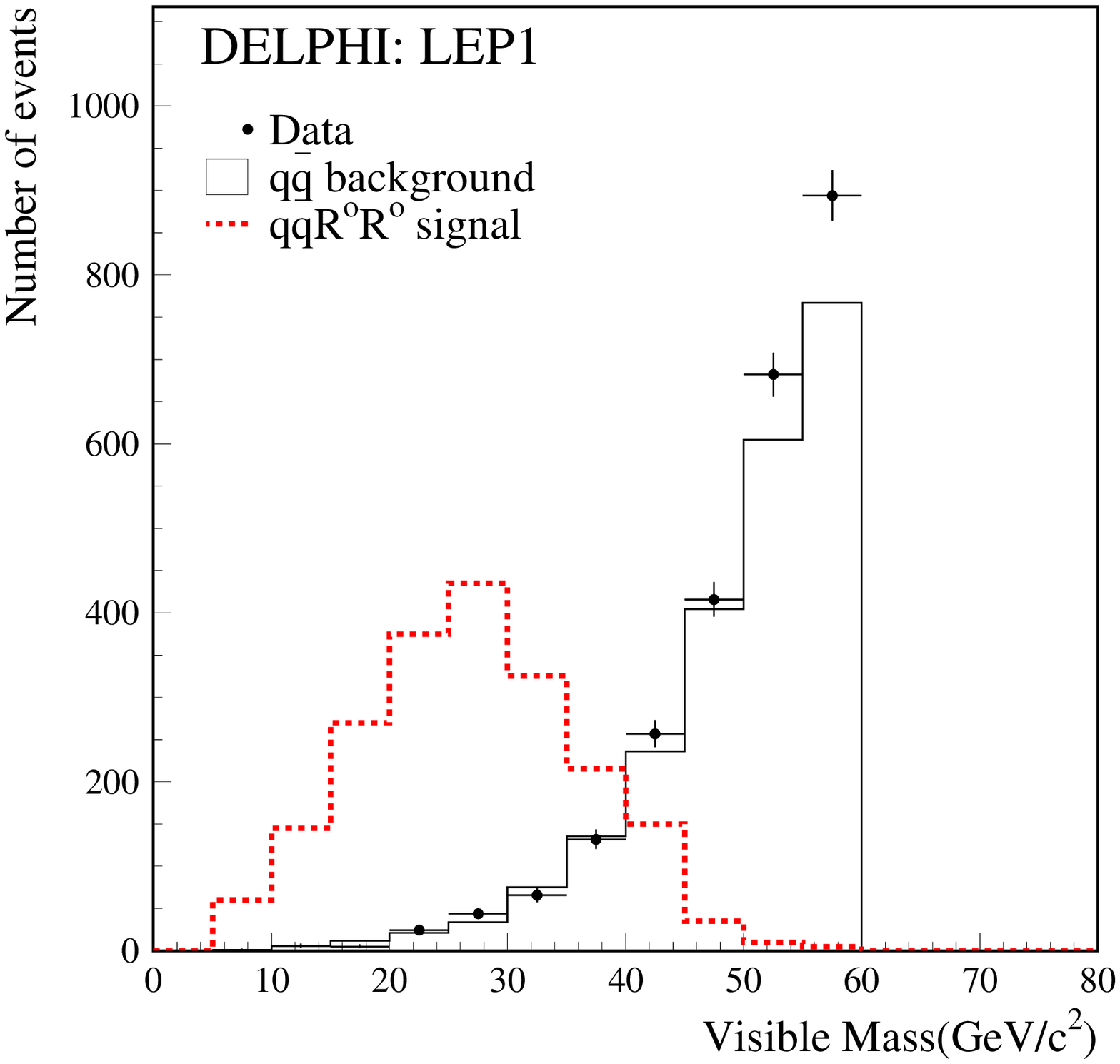} \\
(c) & (d) \\
\includegraphics[width=8cm]{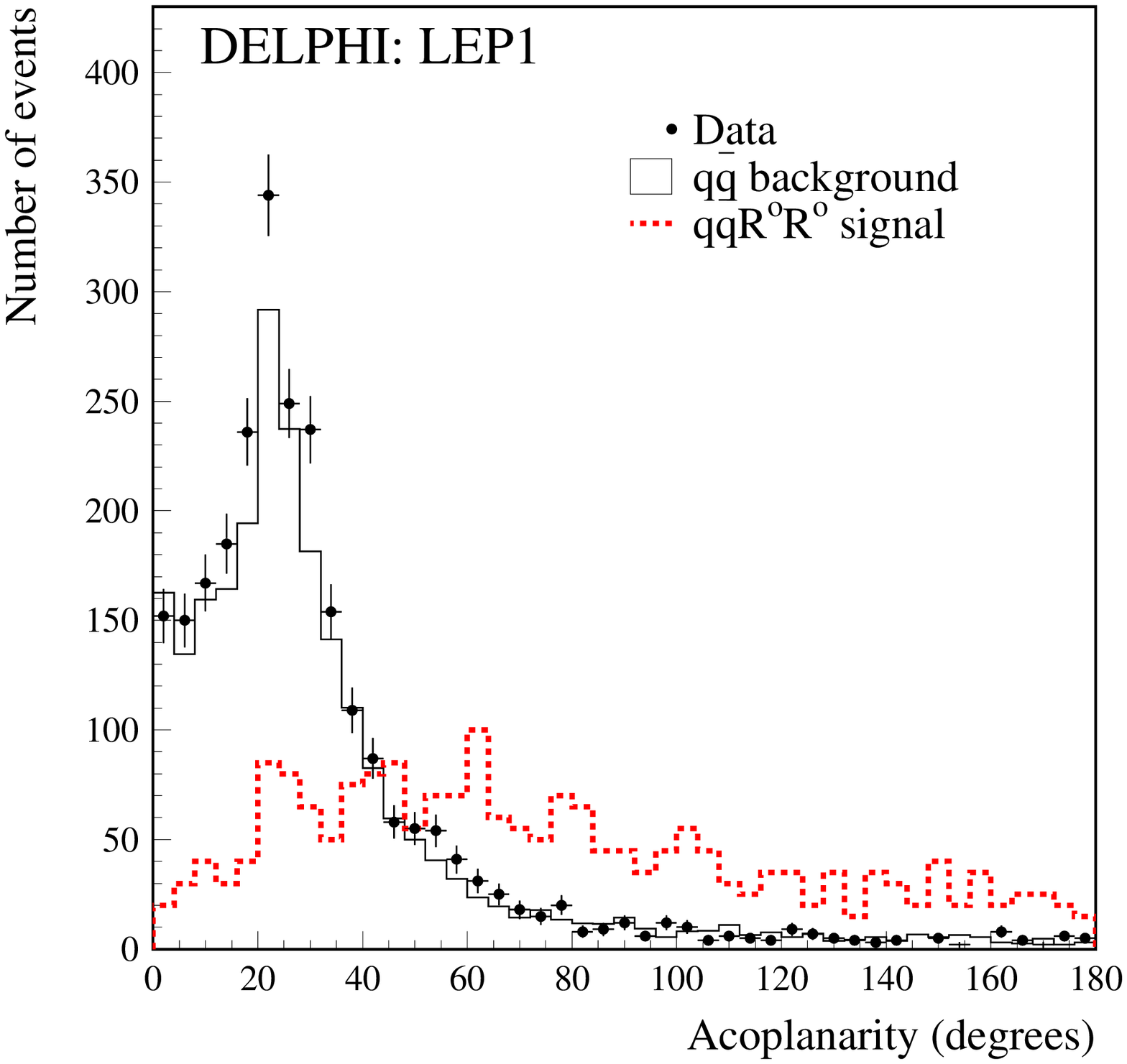} &
\includegraphics[width=8cm]{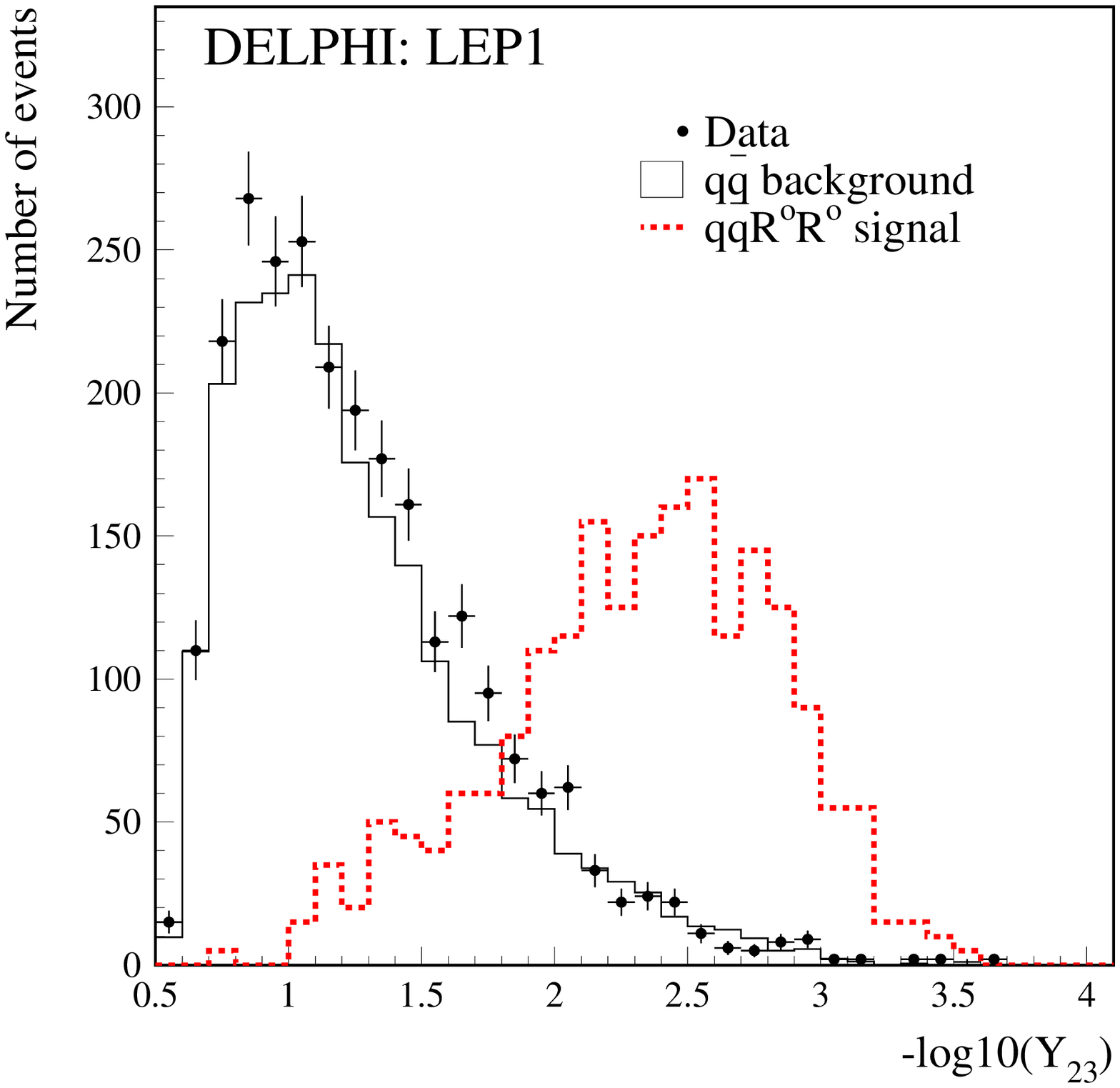} \\
\end{tabular}
\caption{Comparison between data and simulation in the \qqbar\roro\
analysis at LEP1. (a) visible energy, (b) visible mass, 
(c) acoplanarity (d) DURHAM distance $Y_{23}$.
Dotted lines show the \qqbar\roro\ 
signal distributions with arbitrary normalization when all simulated samples are
added together.}
\label{fi:prernlep1}
\end{center}
\end{figure}

The search for \qqbar\roro\ events was performed at LEP1 with a sequential cut
analysis. It was based on the search for the small part of missing energy 
carried away by the neutral R-hadrons.
Hadronic events were first selected by requiring 
at least 5 charged particles. After forcing the events into two jets,
the acollinearity~\footnote{The acollinearity of two jets is defined as the 
complement of the angle between their directions.}
was required to be greater than $20^\circ$ to reduce the huge number
of background Standard Model $\Zo\to\qqbar$ events.\\
\indent
The following cuts were applied to reduce the number
of hadronic $\gamma\gamma$ interactions. The number of tracks reconstructed 
with the TPC had to be greater than 4, and the energy of the particles with
tracks reconstructed using only the VD and ID detectors had to be less 
than 20\% of the total 
energy. The energies in $40^\circ$ and $20^\circ$ cones around the beam axis were 
required to be less than 40\% and 10\% of the total energy respectively. The 
transverse energy had to be greater than 20~GeV.\\
\indent
Hadronic events with missing
energy were then selected in the barrel region of the detector. The visible mass
was required to be less than 60~\GeVcc. The thrust axis and the missing momentum 
had to point in the polar regions $[37^\circ,143^\circ]$ and 
$[45^\circ,135^\circ]$ respectively. Figure~\ref{fi:prernlep1} shows a comparison
between data and simulation at this level of the selection.\\
\indent
The $Y_{23}$ quantity was then required to be less than 0.01 and 
events had to contain less than 20 charged particles. In order to reduce the
number of events with two back-to-back jets, the acoplanarity~\footnote{The 
acoplanarity of two jets is defined as the complement of the angle between
their directions projected onto the plane perpendicular to $z$.} was 
required to be greater than $10^\circ$ and the thrust to be less than 0.95.
The final cut was bi-dimensional. The value of the variable 
$M_{jet1}/E_{jet1}+M_{jet2}/E_{jet2}$
was calculated from the two jets reconstructed with the DURHAM algorithm.
Events were rejected if this variable was greater than 0.45 and if the 
acollinearity was less than $50^\circ$. Table~\ref{tab:prelep1rhadn} shows the number
of events after each cut of the \qqbar\roro\ 
analysis. 12 events were
selected in the data while 10.6 were expected in the hadronic background. Signal
efficiencies as a function of the gluino mass are shown in 
figure~\ref{fi:effirlep1}. They ranged from a few percent for low gluino
masses to around 20\% for $m_{\tilde{g}}=$~18~\GeVcc.

\section{Search for a stable gluino at LEP2}

\subsection{Preselection}

A common preselection for the charged and neutral R-hadron analyses 
was applied to reduce the background coming from soft $\gamma\gamma$ 
interactions. The cuts are the same for the stop and sbottom analysis at all
centre-of-mass energies ranging from 189 to 208~GeV.\\
\indent
First, events were forced into two jets.
To select hadronic events,
the number of charged particles reconstructed with the TPC was required to be
greater than three, and  the energy in the STIC to be less than 70\% of the total
visible energy. The polar angle of the thrust axis had to be in the interval 
$[20^\circ,160^\circ]$. Then, quality cuts were applied. The fraction of 
good tracks was defined as the ratio between the number of charged 
particles remaining after the track selection divided by this number before the
selection.
It had to be greater than 35\%. In addition, the scalar sum of particle momenta
reconstructed with the TPC was required to be greater than 55\% of the total 
reconstructed energy and the
number of charged particles to be greater than six. To remove radiative events,
the energy of the most energetic neutral particle had to be less than 40~\GeV.
Table~\ref{tab:sqpresel} contains the number of events after each of these
cuts.\\
\indent
For the \qqbar\rpmrpm\ and \qqbar\rpmro\ analyses, charged
R-hadron candidates were defined at this level. They had to be reconstructed
with the VD, ID and TPC detectors and their momentum was required to be greater
than 10~\GeVc. At least 80 sense wires of the TPC were required to have
contributed to the measurement of their dE/dx.
Their associated electromagnetic energy was required to be less than 5~GeV, and
the energy of the other charged particles in a $15^\circ$ cone around a candidate 
had to be less than  
5~GeV. In 2000 , the dE/dx could not be used in sector
6 of the TPC for almost any of the data. For this sample, charged R-hadron 
candidates in this sector were removed.

\subsection{Search for \qqbar\rpmrpm\ events}

\begin{figure}[p!]
\begin{center}
\begin{tabular}{cc}
 & \\
\includegraphics[width=8cm]{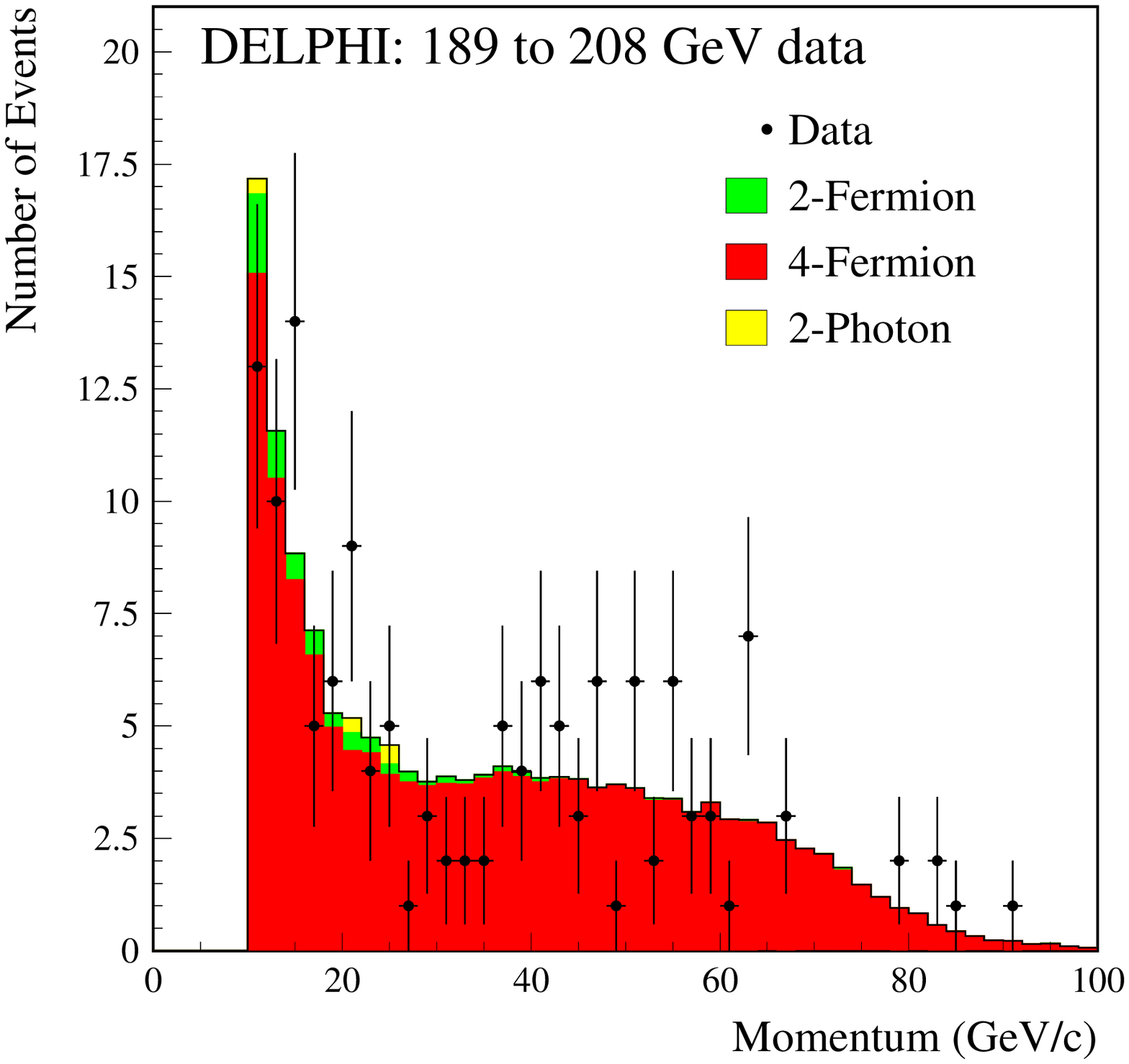} &
\includegraphics[width=8cm]{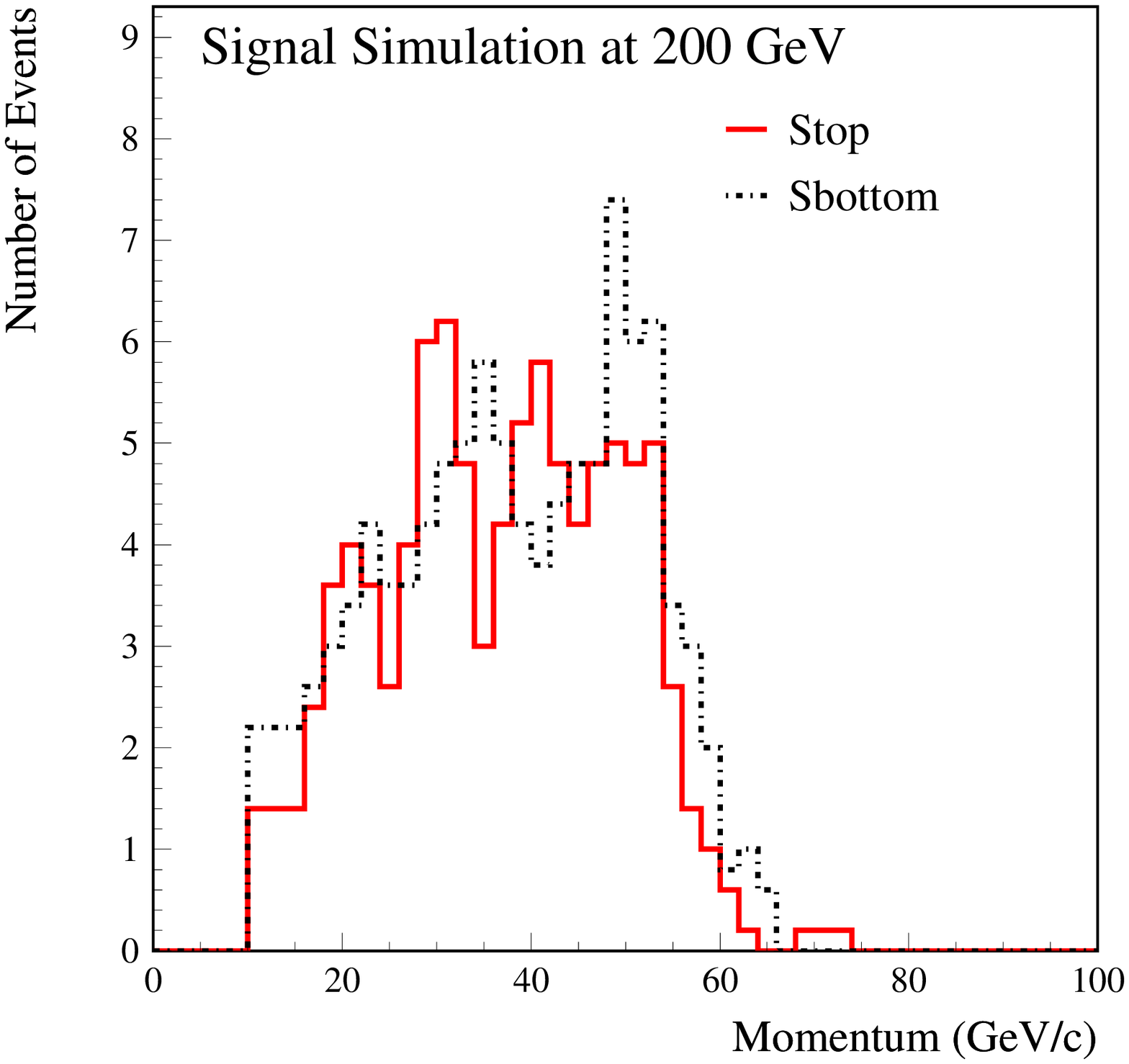} \\
 & \\
\includegraphics[width=8cm]{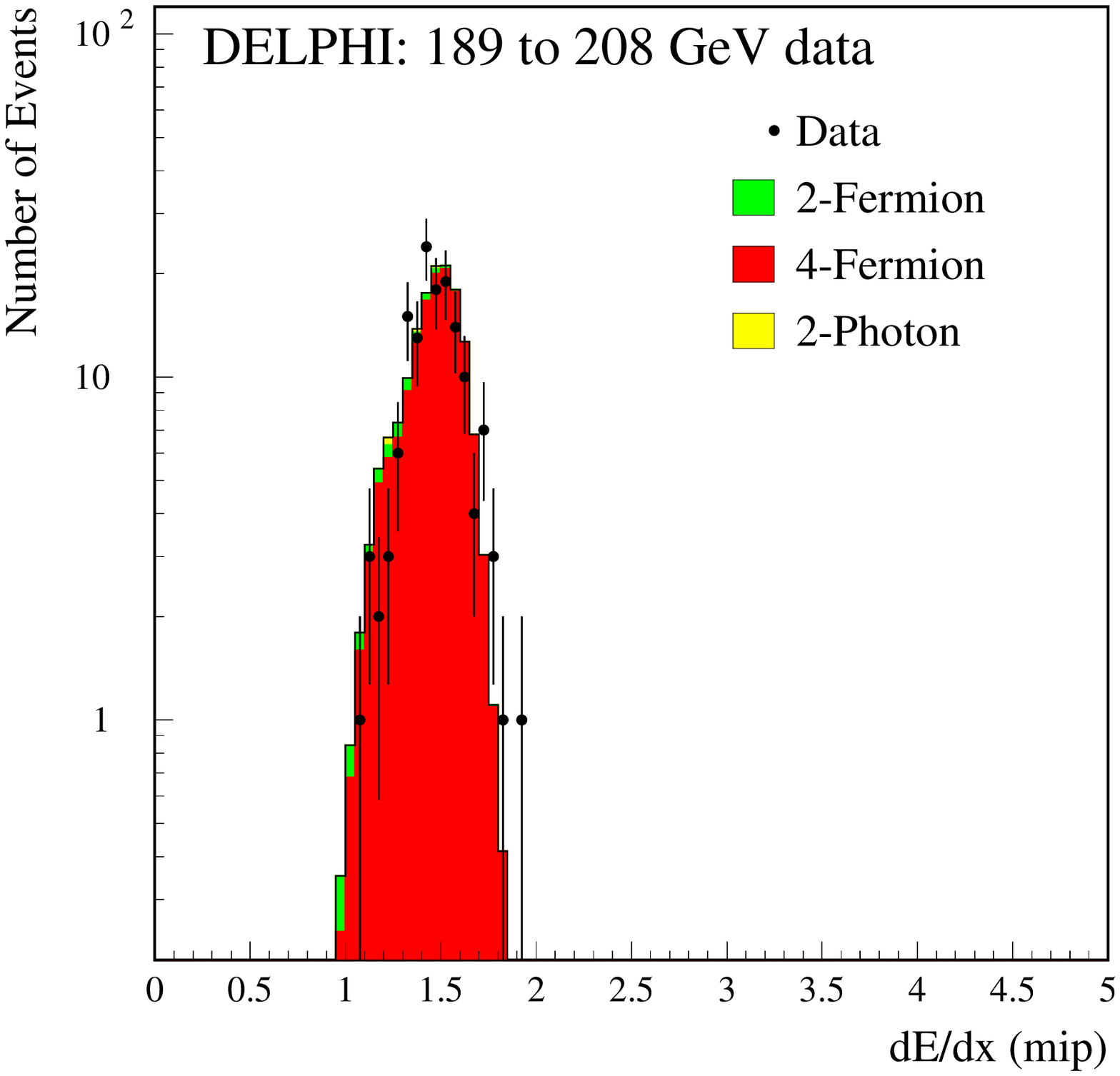} &
\includegraphics[width=8cm]{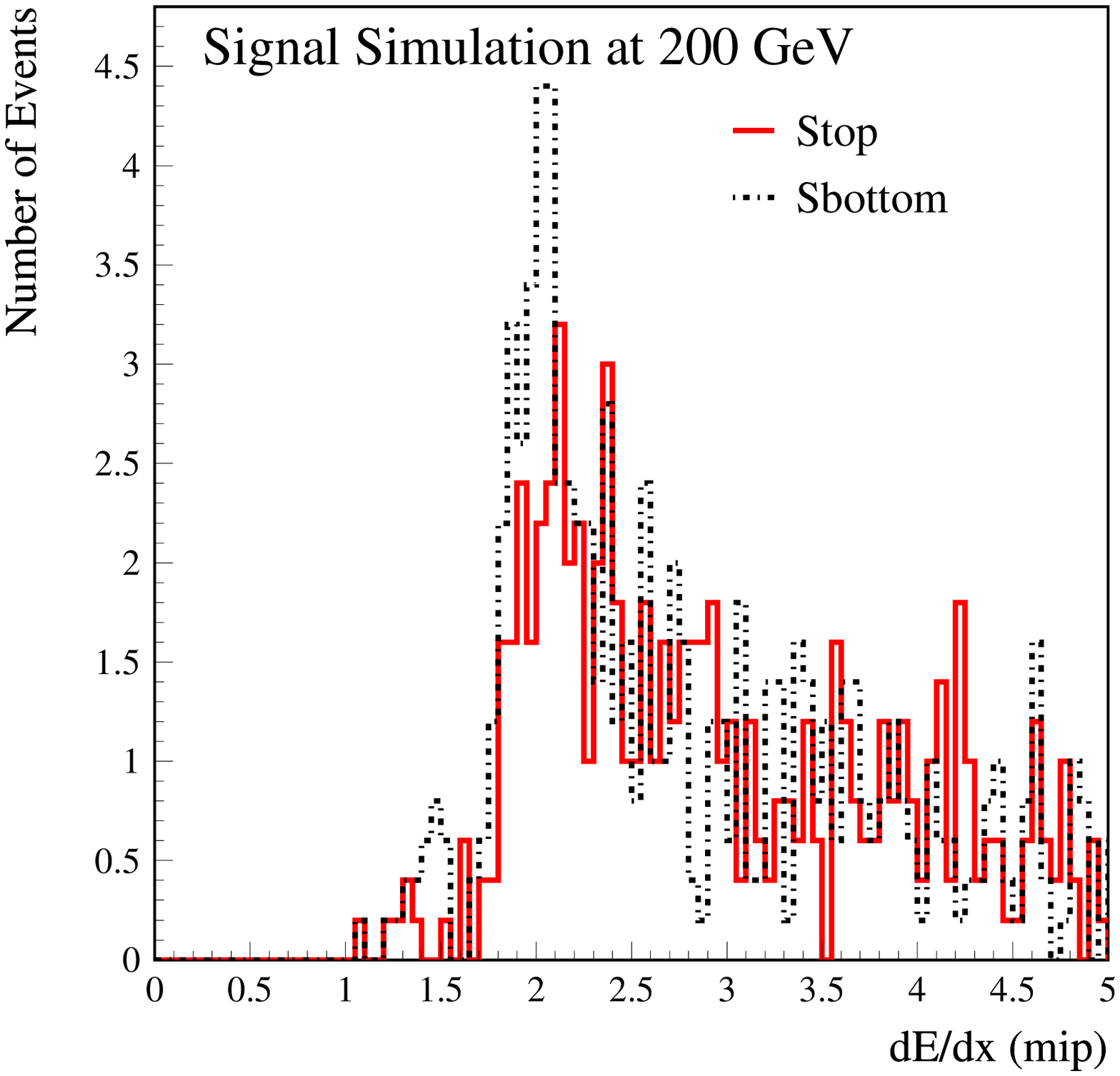} \\
\end{tabular}
\caption{Momentum and dE/dx of the charged R-hadron candidates selected by the   
\qqbar\rpmrpm\ analysis at LEP2. Data taken in the centre-of-mass 
energy range between 189 and 208~GeV were included. 
Right-hand side histograms show the expected distributions for one possible 
stop and sbottom signal ($\msqi=90~\GeVcc$, $\mglui=60~\GeVcc$) 
at $\sqrt{s}=200~\GeV$ with arbitrary normalization.}
\label{fi:rcpresel}
\end{center}
\end{figure}

The search for \qqbar\rpmrpm\ events was exactly the same for the
stop and the sbottom analyses. Events were selected if they contained at least two
charged R-hadron candidates. Figure~\ref{fi:rcpresel} shows the momentum
and the dE/dx distribution of the selected \rpm\ candidates. 
Table~\ref{tab:sqpresel} shows the number of selected events.\\
\indent
The analysis 
was then separated into three windows in 
gluino mass, and cuts in the plane (P,dE/dx) were applied:
\begin{itemize}
\item{$\mglui \leq 30~\GeVcc$:}\\
events had to contain at least one charged R-hadron candidate with momentum
greater than 20~\GeVc, and with dE/dx less than the dE/dx expected for a
particle of mass equal to 3~\GeVcc. 
\item{$30~\GeVcc < \mglui < 60~\GeVcc$:}\\
events were selected if they contained at least two charged R-hadron candidates 
with dE/dx both greater than the dE/dx expected for a particle of mass equal to
30~\GeVcc, and less than the dE/dx expected for a particle of mass equal to
60~\GeVcc. Moreover, this dE/dx had also to be either less than the dE/dx expected for a
particle of mass equal to 1~\GeVcc, or greater than 1.8~mip.
\item{$\mglui \geq 60~\GeVcc$:}\\
events were kept if they contained at least two charged R-hadron candidates with
dE/dx greater than the dE/dx expected for a particle of mass equal to 60~\GeVcc.
\end{itemize}

\begin{figure}[p!]
\begin{center}
\begin{tabular}{c}
(a) \\
\includegraphics[width=9cm]{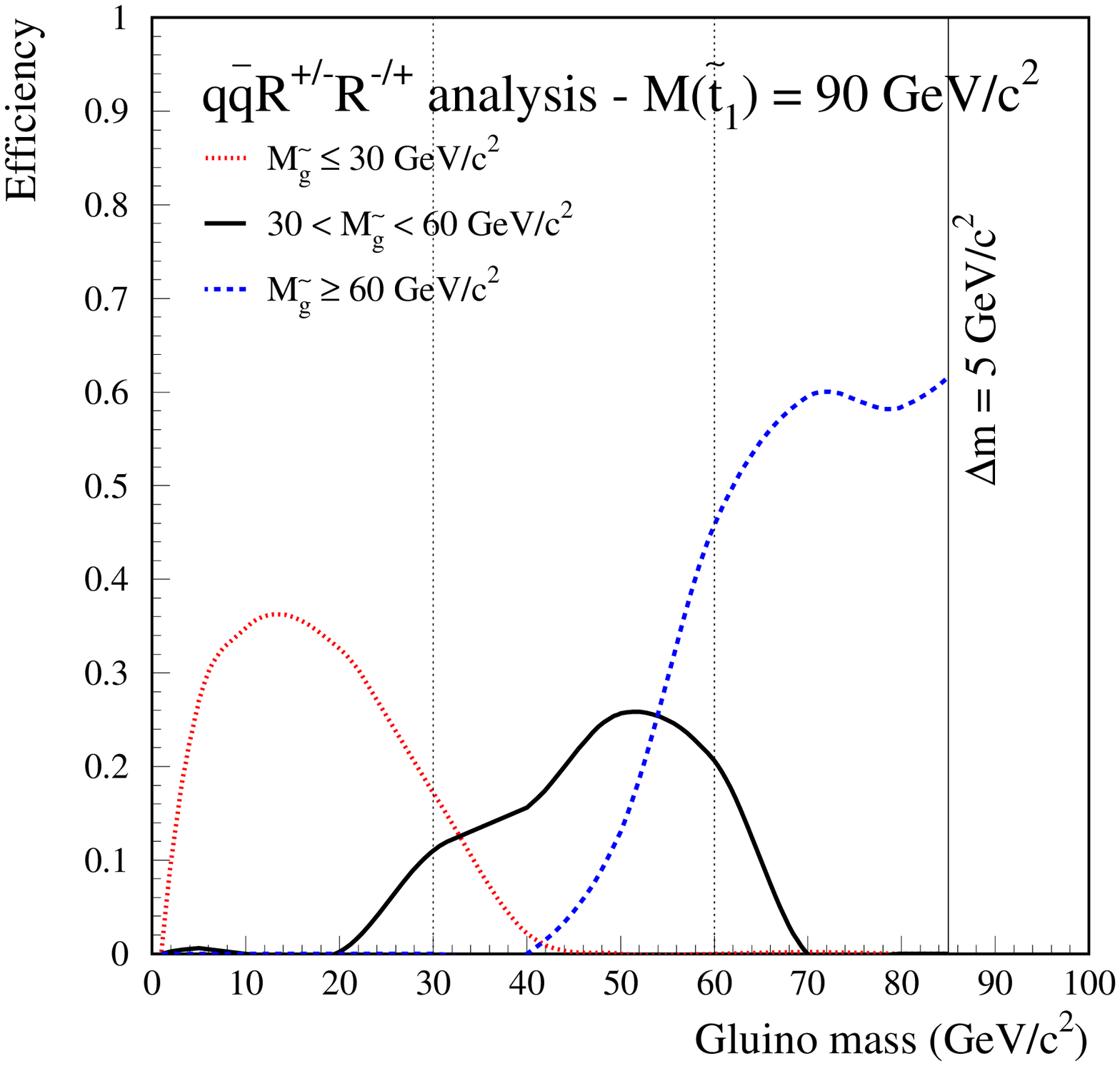} \\
(b) \\
\includegraphics[width=9cm]{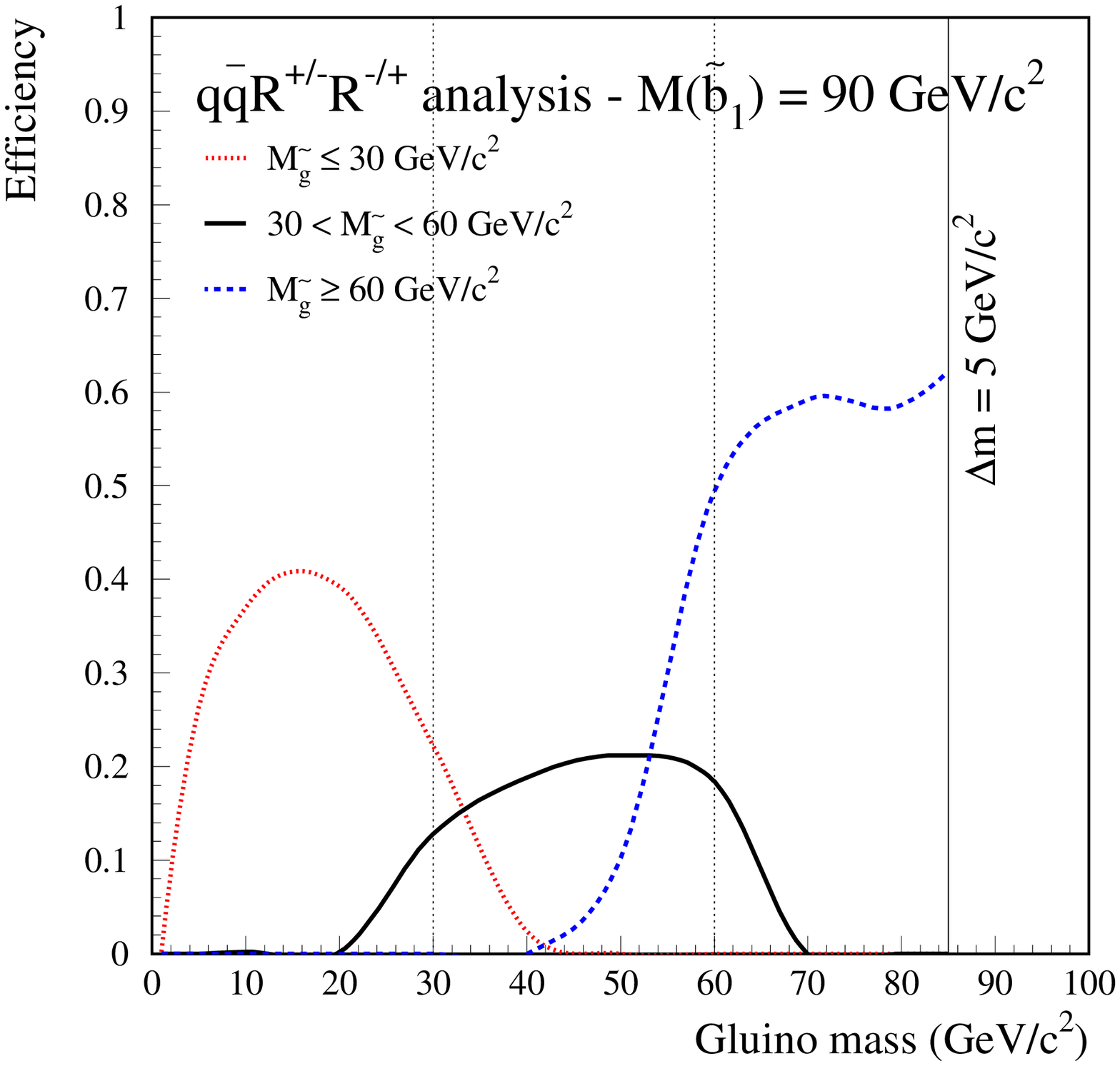} \\
\end{tabular}
\caption{Signal detection efficiencies at $\sqrt{s}=$200~GeV for the 
stop (a) and sbottom (b) \qqbar\rpmrpm\ analysis as a function of the gluino
mass ($\msqi=~90~\GeVcc$). $\Delta M$ is the mass difference between the squark
and the gluino. Vertical lines show the limits of the mass analysis window. The
last one ends at $\Delta M=5~\GeVcc$ which corresponds the last simulated 
signal points.}
\label{fi:rceffi}
\end{center}
\end{figure}

\indent
In all LEP2 data which were analysed, no events were selected in any of these
windows. The number of expected Standard Model background events were 0.115, 0.009 
and 0.011 in the analyses for $\mglui \leq 30~\GeVcc$, 
$30~\GeVcc < \mglui < 60~\GeVcc$ and 
$\mglui \geq 60~\GeVcc$ respectively. Table~\ref{tab:resrhadc} contains the 
number of events expected for the different centre-of-mass energies. 
Figure~\ref{fi:rceffi} shows the signal detection efficiencies near the
kinematical limit ($\msqi=90~\GeVcc$). The difference between stop and 
sbottom efficiencies is not large.
The highest efficiencies were always obtained for high gluino
masses, where the dE/dx is very high.  

\subsection{Search for \qqbar\rpmro\ events}

\begin{figure}[p!]
\begin{center}
\begin{tabular}{cc}
 & \\
\includegraphics[width=8cm]{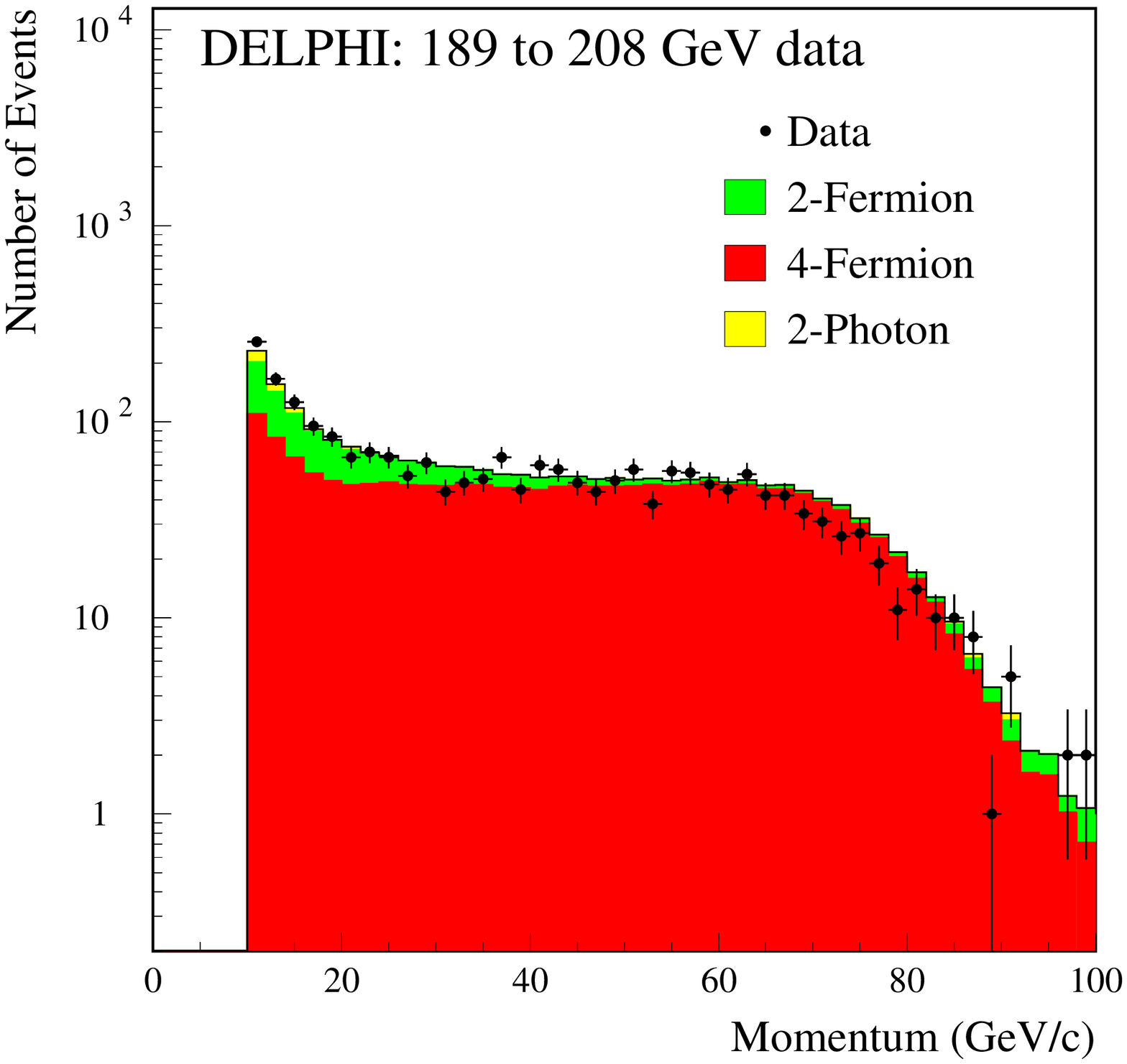} &
\includegraphics[width=8cm]{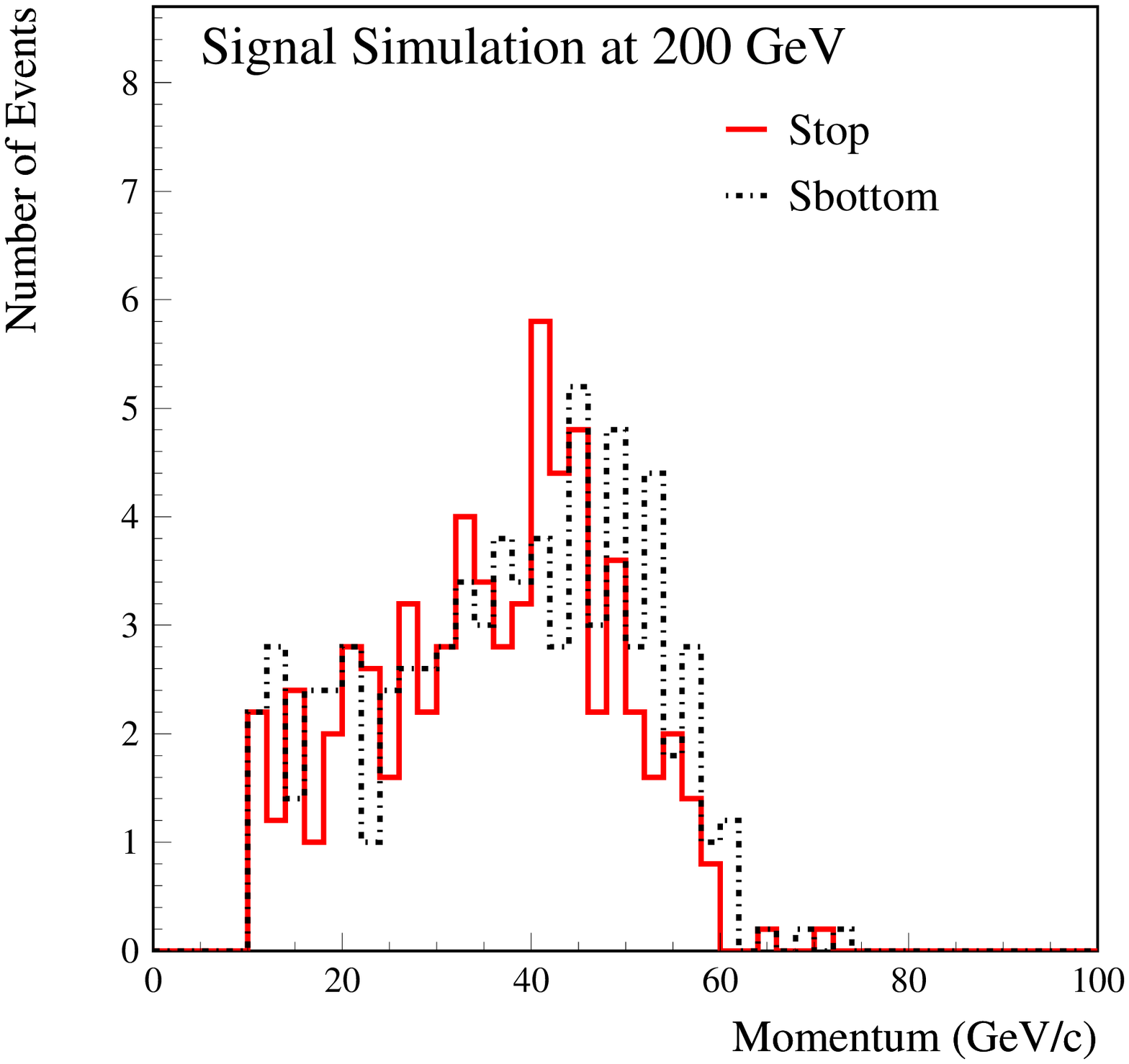} \\
 & \\
\includegraphics[width=8cm]{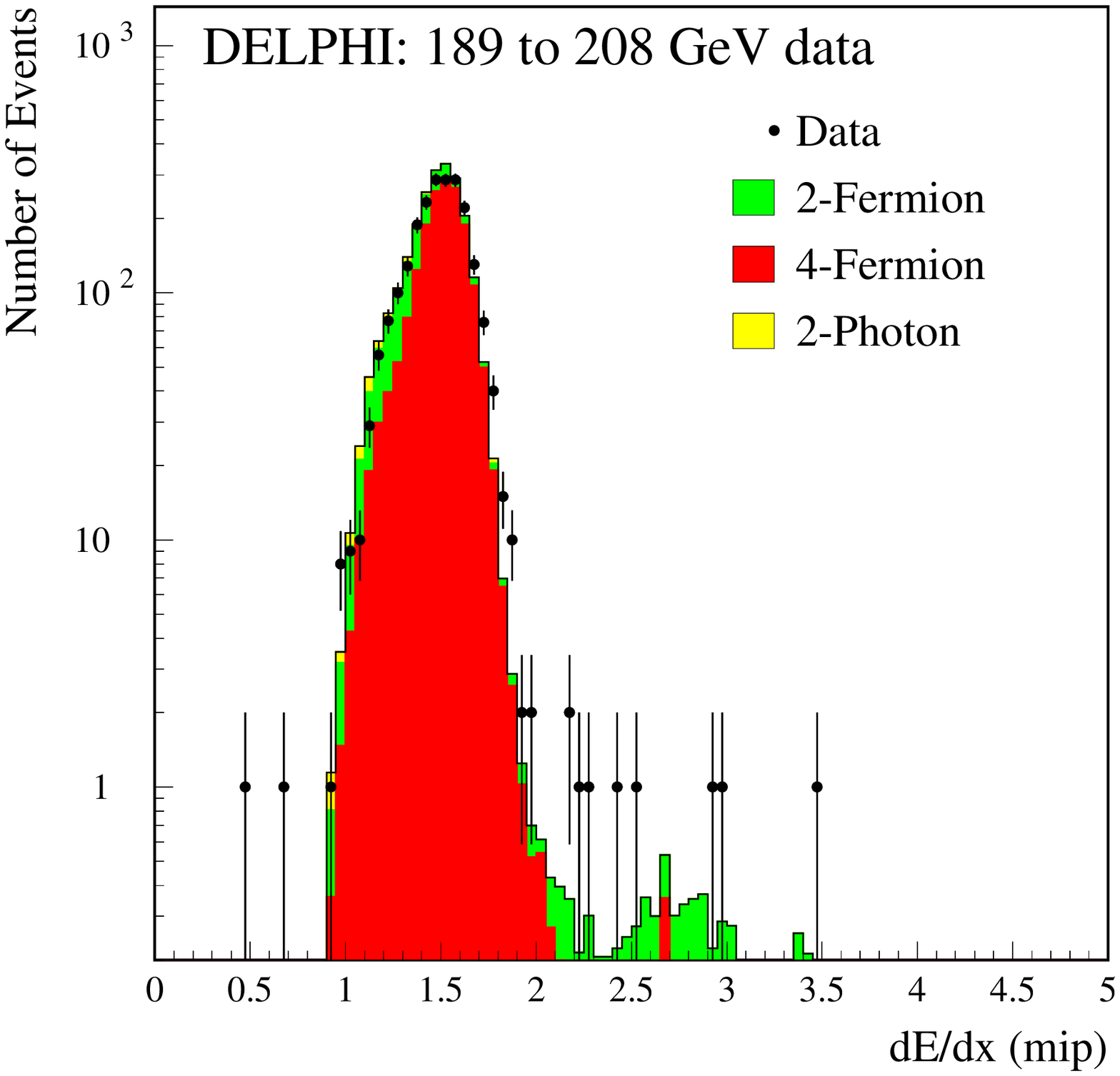} &
\includegraphics[width=8cm]{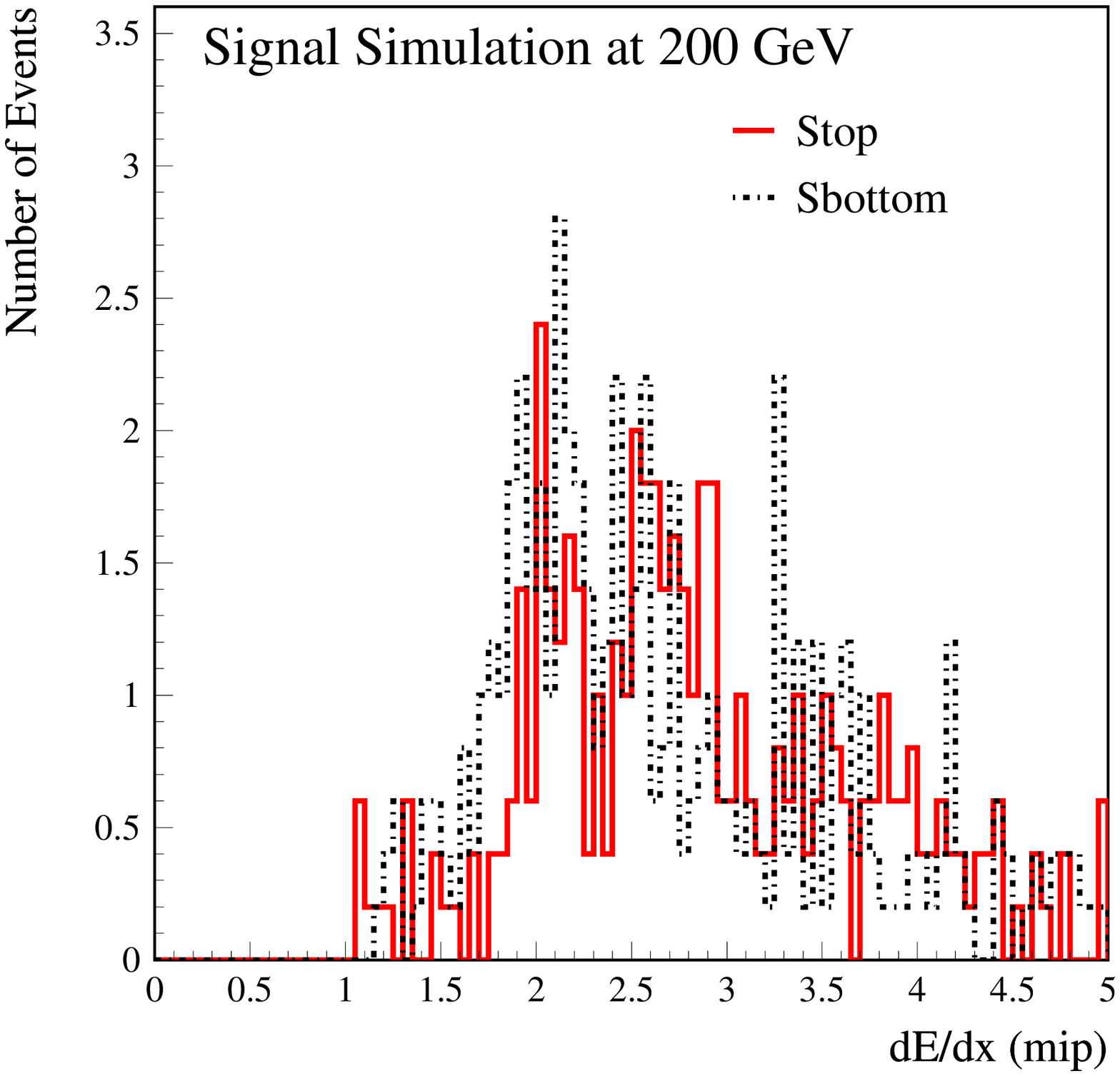} \\
\end{tabular}
\caption{Momentum and dE/dx of the charged R-hadron candidates selected by the   
\qqbar\rpmro\ analysis at LEP2. Data taken in the centre-of-mass
energy range between 189 and 208~GeV were included. 
Right-hand side histograms show the expected distributions for one possible 
stop and sbottom signal ($\msqi=90~\GeVcc$, $\mglui=60~\GeVcc$) 
at $\sqrt{s}=200~\GeV$ with arbitrary normalization.}
\label{fi:rmpresel}
\end{center}
\end{figure}

The search for \qqbar\rpmro\ events was also the same for the
stop and sbottom analyses. Events were selected if they contained at least one
charged R-hadron candidate. Figure~\ref{fi:rmpresel} shows the momentum
and the dE/dx distribution of the selected \rpm\ candidates.
Table~\ref{tab:sqpresel} shows the number of selected events.\\
\indent
The analysis 
was then separated into three gluino mass windows, and cuts in the plane 
(P,dE/dx) were applied:
\begin{itemize}
\item{$\mglui \leq 30~\GeVcc$:}\\
events had to contain at least one charged R-hadron candidate with momentum
greater than 20~\GeVc, and with dE/dx less than the dE/dx expected for a
particle of mass equal to 3~\GeVcc. 
\item{$\mglui \geq 60~\GeVcc$:}\\
events were kept if they contained at least one charged R-hadron candidate with
dE/dx greater than the dE/dx expected for a particle of mass equal to 60~\GeVcc,
and greater than 2~mip.
\item{$30~\GeVcc < \mglui < 60~\GeVcc$:}\\
events selected in either of the above windows (higher or lower 
$m_{\tilde{g}}$) were accepted.
\end{itemize}

\begin{figure}[p!]
\begin{center}
\begin{tabular}{c}
(a) \\
\includegraphics[width=9cm]{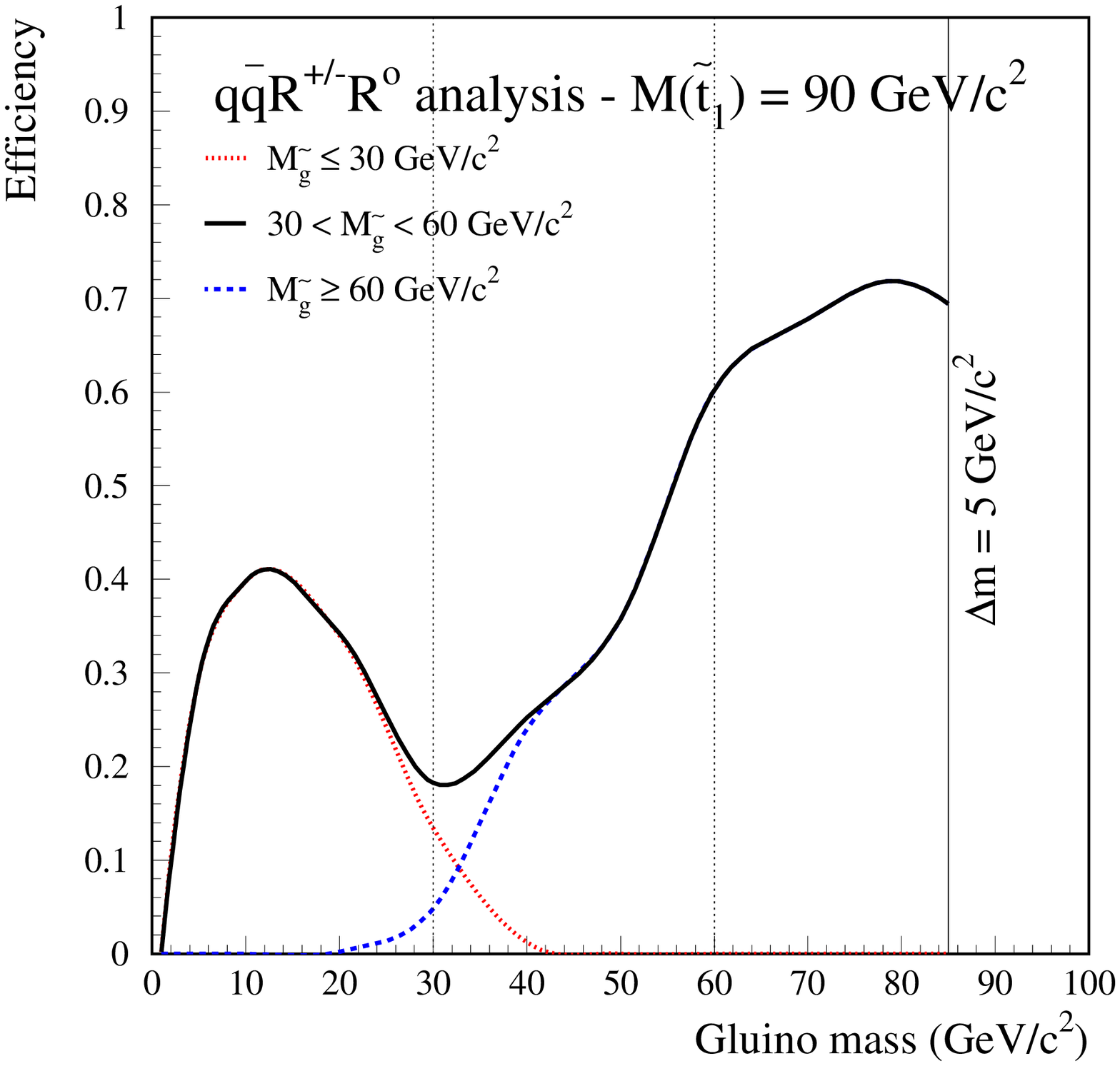} \\
(b) \\
\includegraphics[width=9cm]{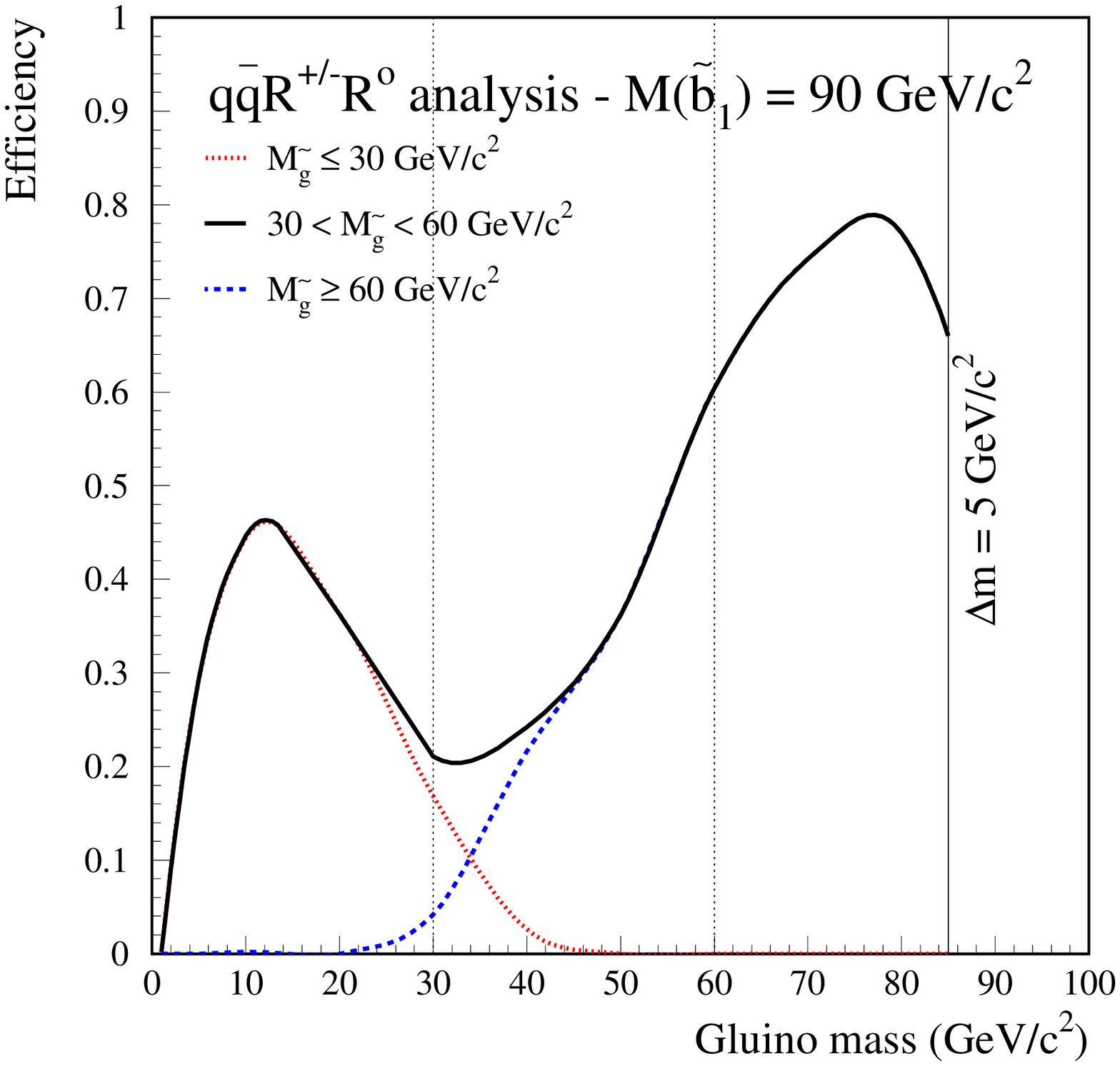} \\
\end{tabular}
\caption{Signal detection efficiencies at $\sqrt{s}=$200~GeV for the 
stop (a) and sbottom (b) \qqbar\rpmro\ analysis as a function of the gluino
mass ($\msqi=~90~\GeVcc$).$\Delta M$ is the mass difference between the squark
and the gluino. Vertical lines show the limits of the mass analysis window. The
last one ends at $\Delta M=5~\GeVcc$ which corresponds the last simulated 
signal points.}
\label{fi:rmeffi}
\end{center}
\end{figure}

\indent
Three, nine and six events were selected in the mass windows
$\mglui \leq 30~\GeVcc$, $30~\GeVcc < \mglui < 60~\GeVcc$ and 
$\mglui \geq 60~\GeVcc$ respectively. The number of expected background events 
were 1.6, 8.2 and 6.6.
All selected events in the data are more likely Standard Model instead than 
signal like. In particular, they do not follow any 
mass iso-curve in the (P,dE/dx) plane.
Table~\ref{tab:resrhadm} contains the number of selected
events as a function of the centre-of-mass energy. 
Figure~\ref{fi:rmeffi} shows the signal detection efficiencies near the
kinematical limit ($\msqi=90~\GeVcc$).
The highest efficiencies were obtained 
for high gluino masses where the dE/dx is very high.

\subsection{Search for \qqbar\roro\ events}

\begin{figure}[p!]
\begin{center}
\begin{tabular}{cc}
 & \\
\includegraphics[width=8cm]{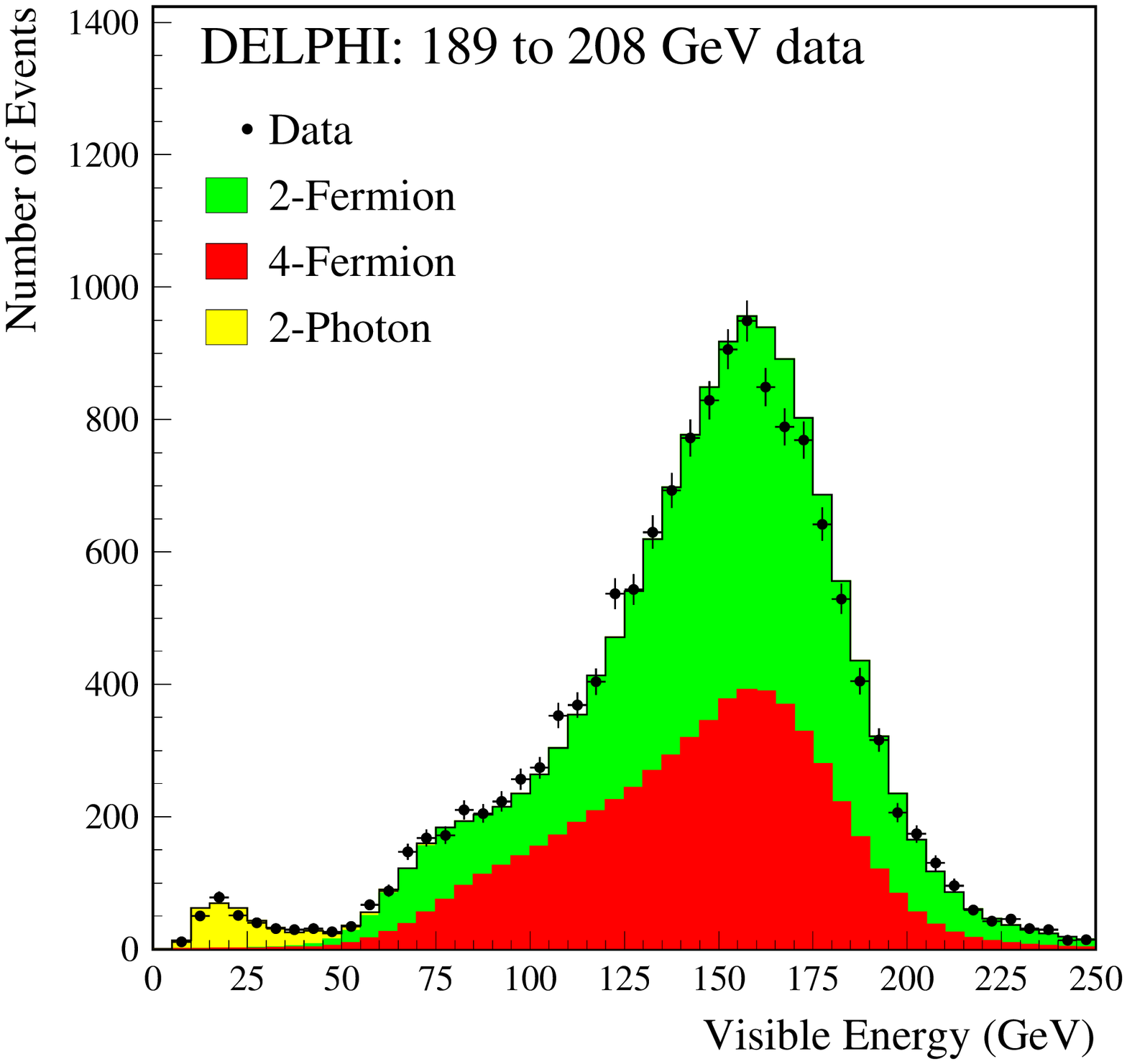} &
\includegraphics[width=8cm]{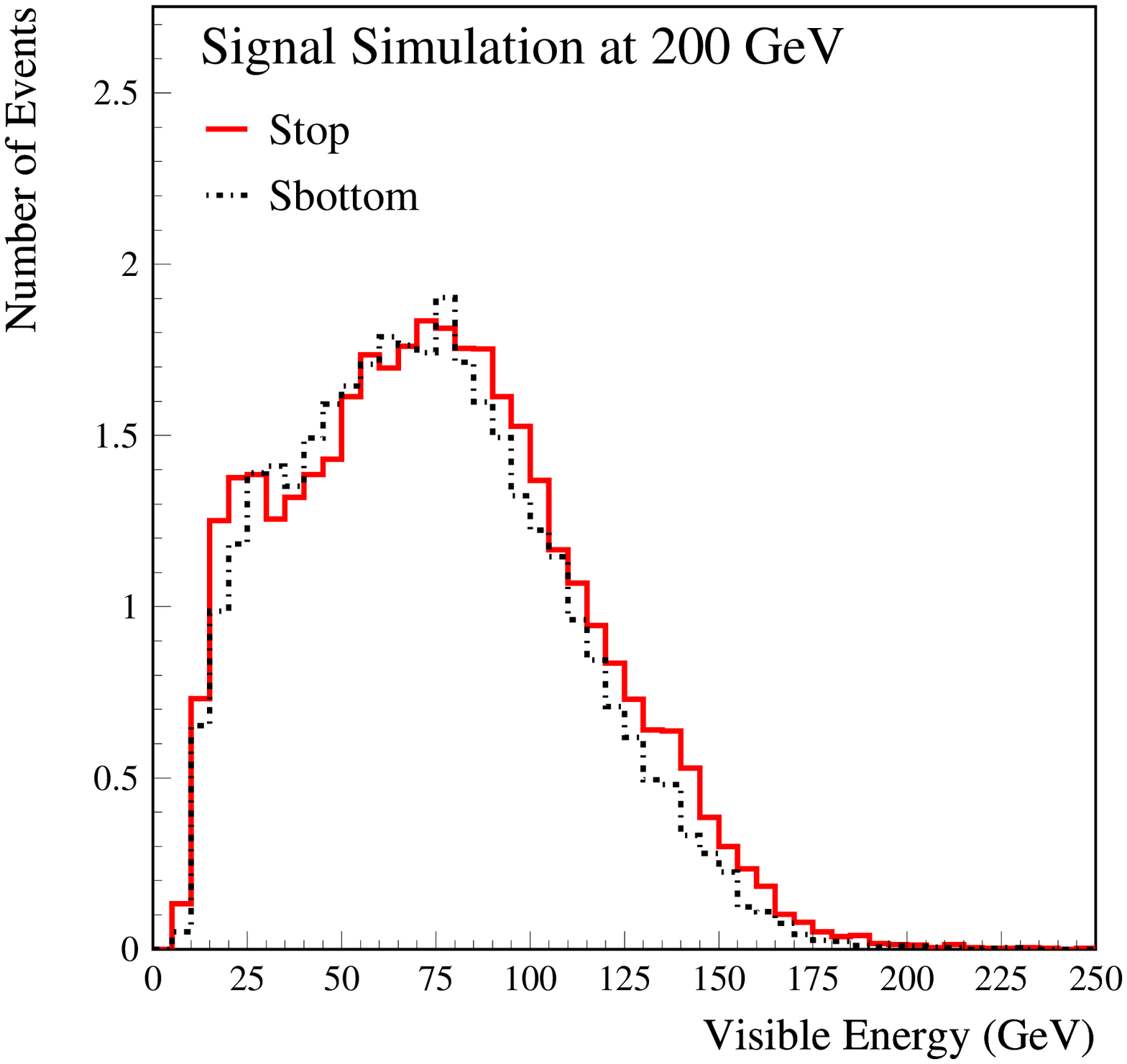} \\
 & \\
 &  \\
\includegraphics[width=8cm]{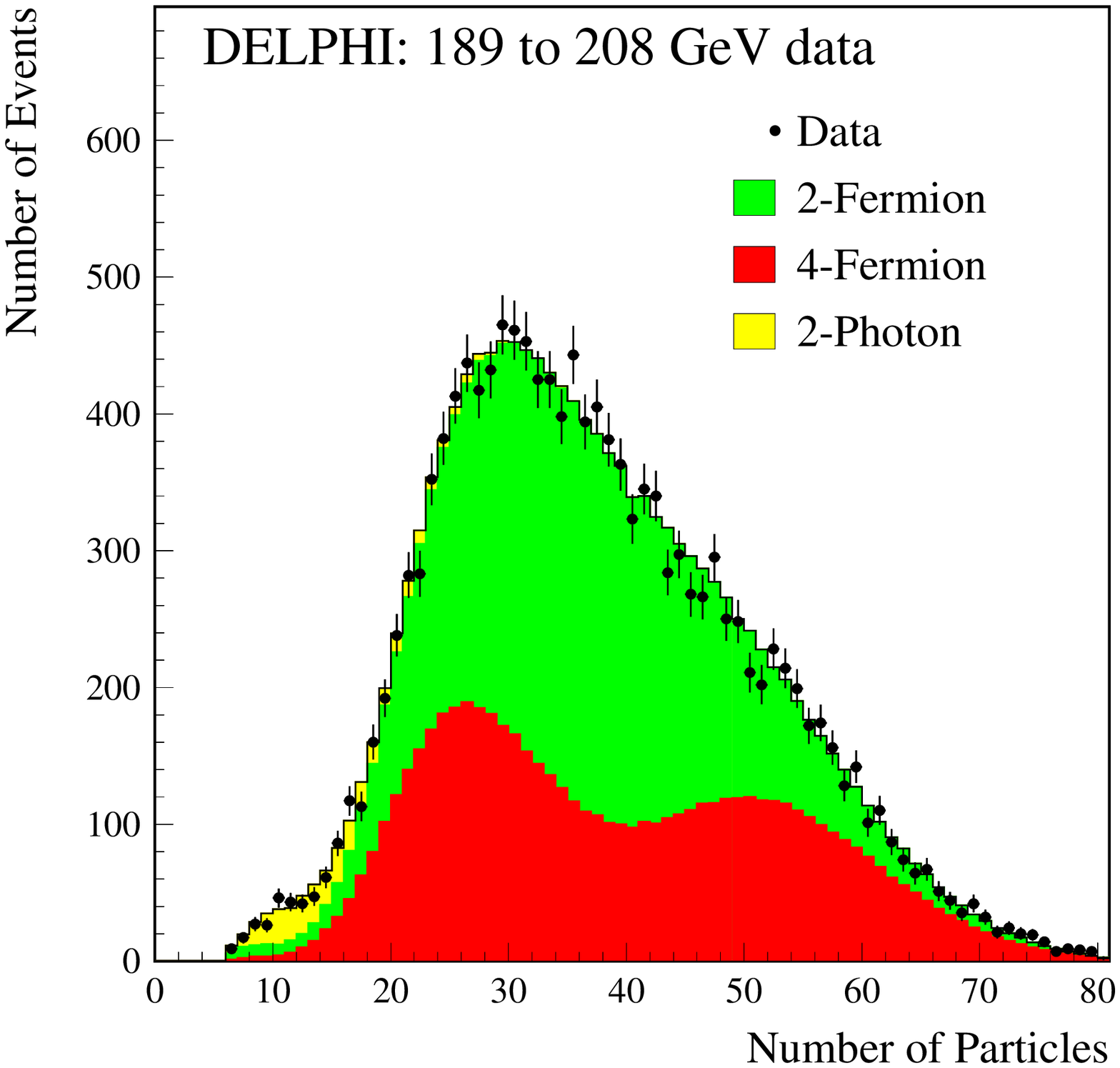} &
\includegraphics[width=8cm]{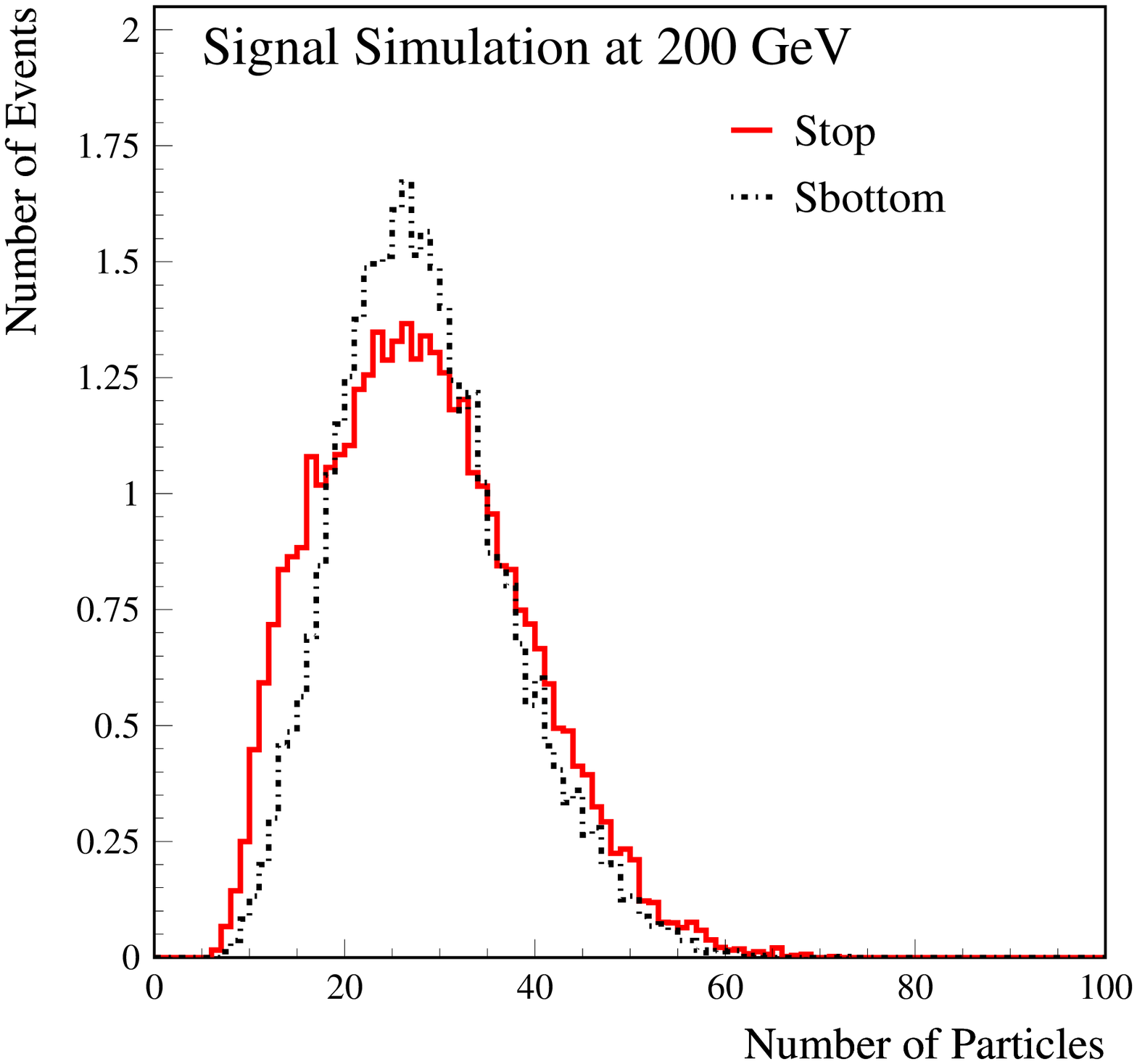} \\
\end{tabular}
\caption{Data-simulation comparison at the preselection level of the LEP2 
\qqbar\roro\ analysis. Data taken in the centre-of-mass energy 
range between 189 and 208~GeV were included. Right-hand side histograms 
show the expected distributions with arbitrary normalization  for the stop and
the sbottom signal at 200~GeV when all simulated samples are added together.}
\label{fi:rnpresel}
\end{center}
\end{figure}

\begin{figure}[p!]
\begin{center}
\begin{tabular}{cc}
 &  \\
\includegraphics[width=8cm]{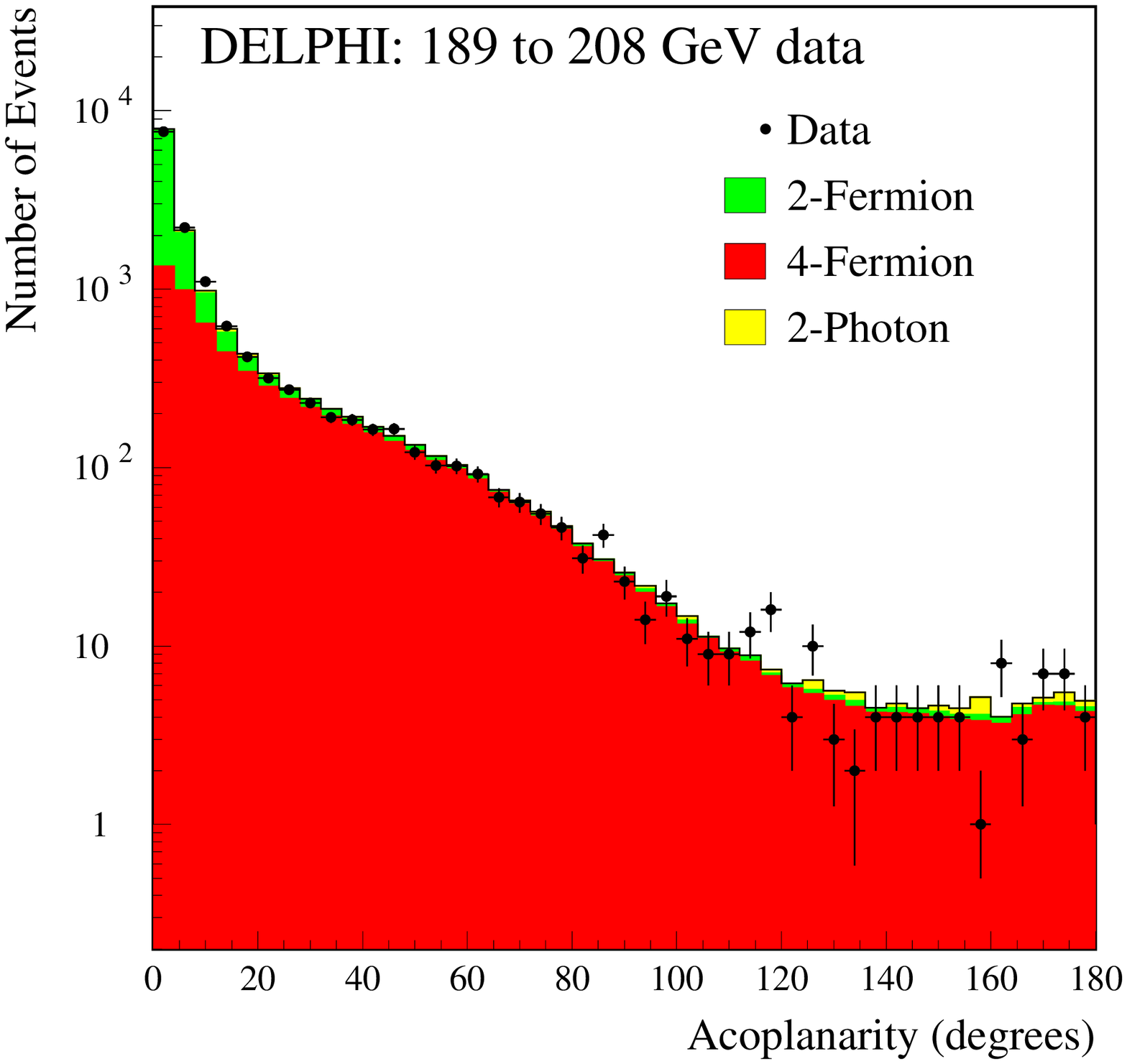} &
\includegraphics[width=8cm]{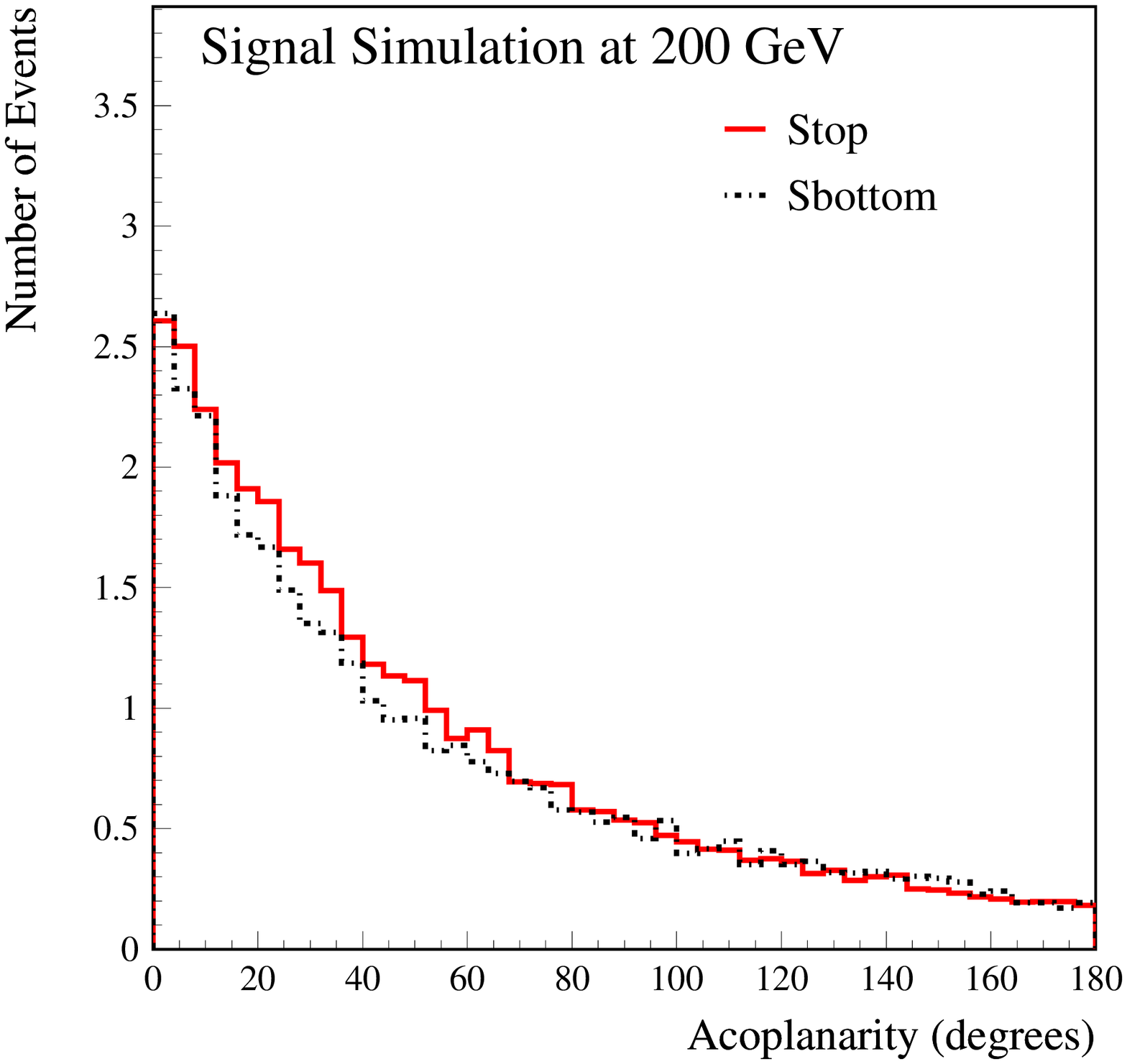} \\
 & \\
 &  \\
\includegraphics[width=8cm]{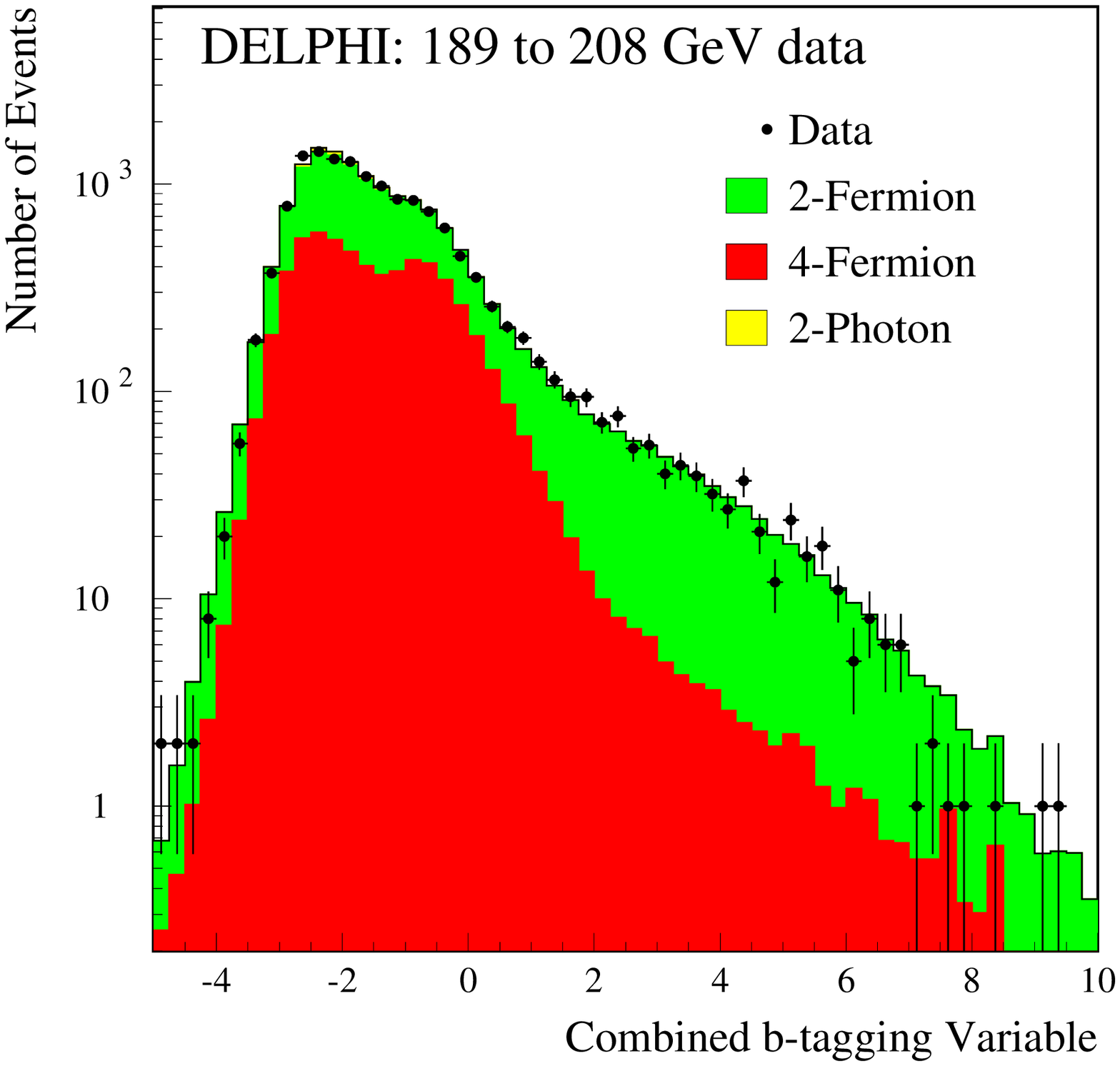} &
\includegraphics[width=8cm]{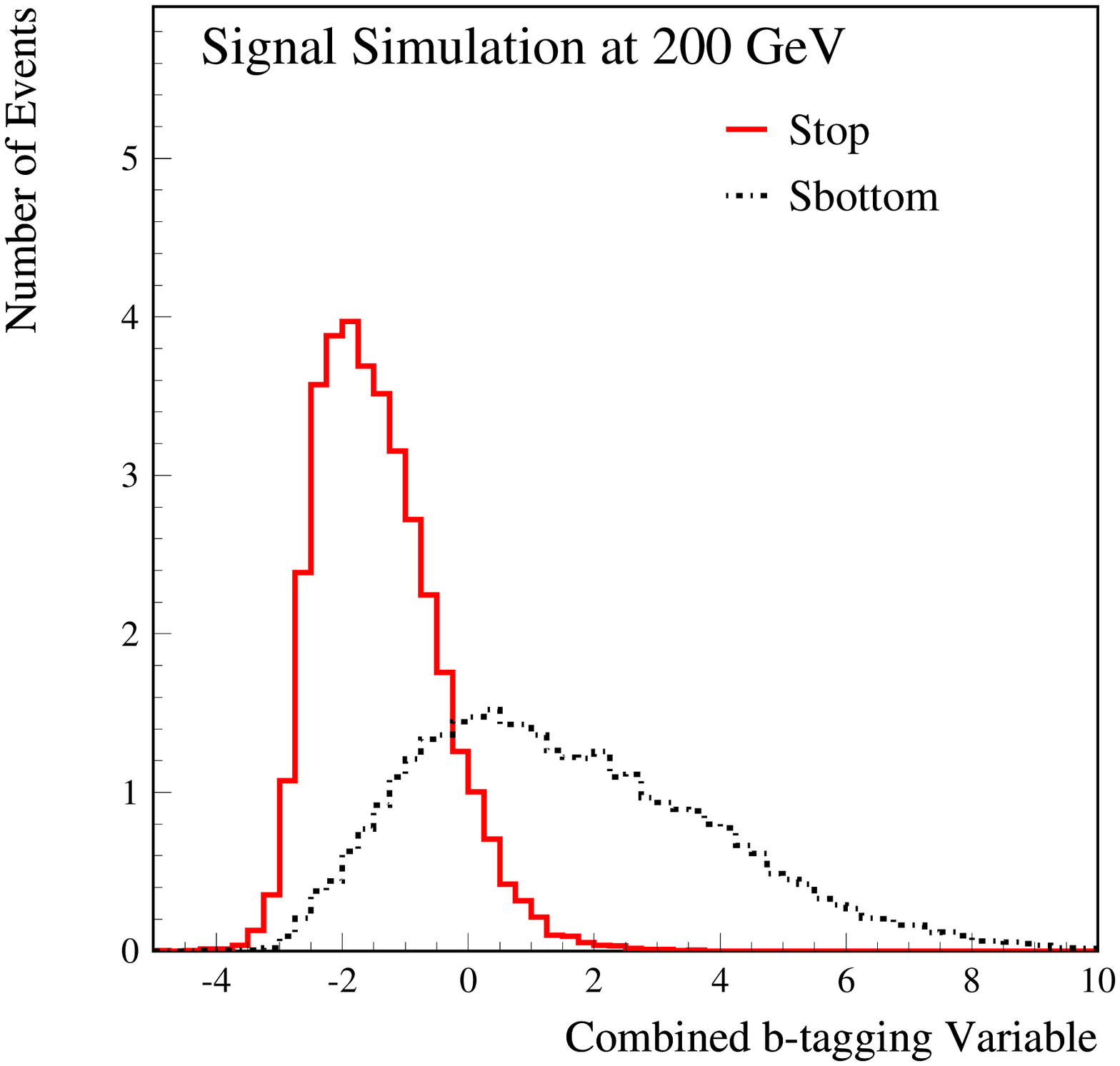} \\
\end{tabular}
\caption{Data-simulation comparison at the preselection level of the LEP2 
\qqbar\roro\ analysis. Data taken in the centre-of-mass energy 
range between 189 and 208~GeV were included. Right-hand side histograms 
show the expected distributions with arbitrary normalization  for the stop and
the sbottom signal at 200~GeV when all simulated samples are added together.}
\label{fi:rnpresel2}
\end{center}
\end{figure}

After the preselection described in section~7.1, the transverse missing momentum 
was required to be greater than 4~\GeVc, the angle of the missing momentum had 
to point in the polar angle region $[20^\circ,160^\circ]$, and the energy in a 
$40^\circ$ cone around the beam axis was required to be less than 40\% of the event energy. A 
veto algorithm was then applied based on the hermeticity taggers at polar angles
close to $40^\circ$ and $90^\circ$. Figures~\ref{fi:rnpresel} 
and~\ref{fi:rnpresel2} show data 
Monte Carlo comparisons following this selection and table~\ref{tab:rnpresel} 
gives the observed and expected event numbers at the different steps.\\
\indent
The stop and sbottom analyses were then separated for different ranges of the
mass difference $\Delta m$ between the squark and the gluino:
\begin{itemize}
\item{$\Delta m \leq 20~\GeVcc:$}\\
For high gluino masses, the energy deposited by the neutral R-hadrons is quite 
small. In this respect, the gluino is not so different from a neutralino, and
the $\Sqi \to \mathrm{q}\glui$ events resemble $\Sqi \to \mathrm{q}\xoi$ events.
\item{$\Delta m > 20~\GeVcc:$}\\
In this case, the gluino deposits more energy.
\end{itemize}

\begin{figure}[p!]
\begin{center}
\begin{tabular}{cc}
(a) & (b) \\
\includegraphics[width=7cm]{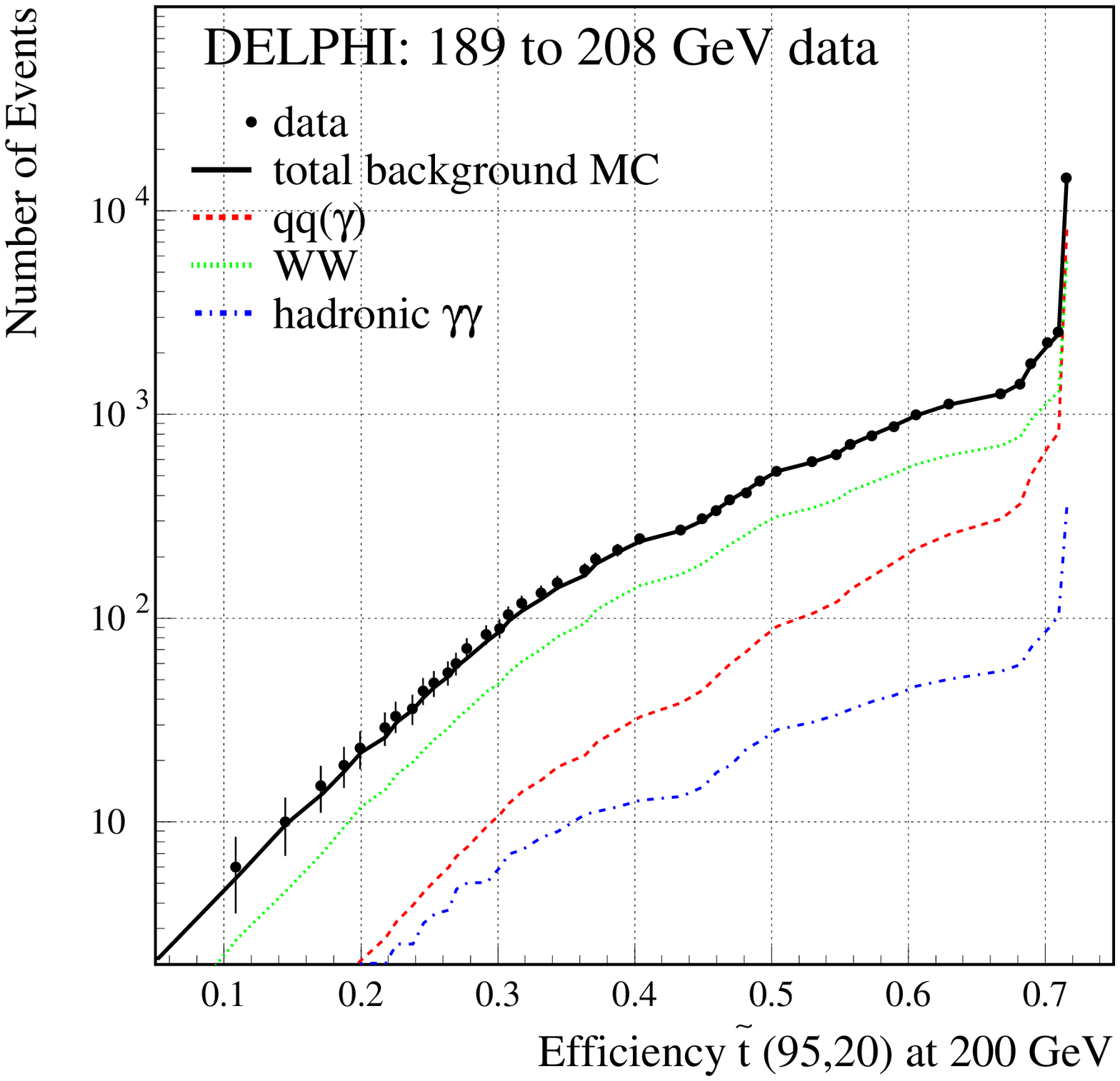} &
\includegraphics[width=7cm]{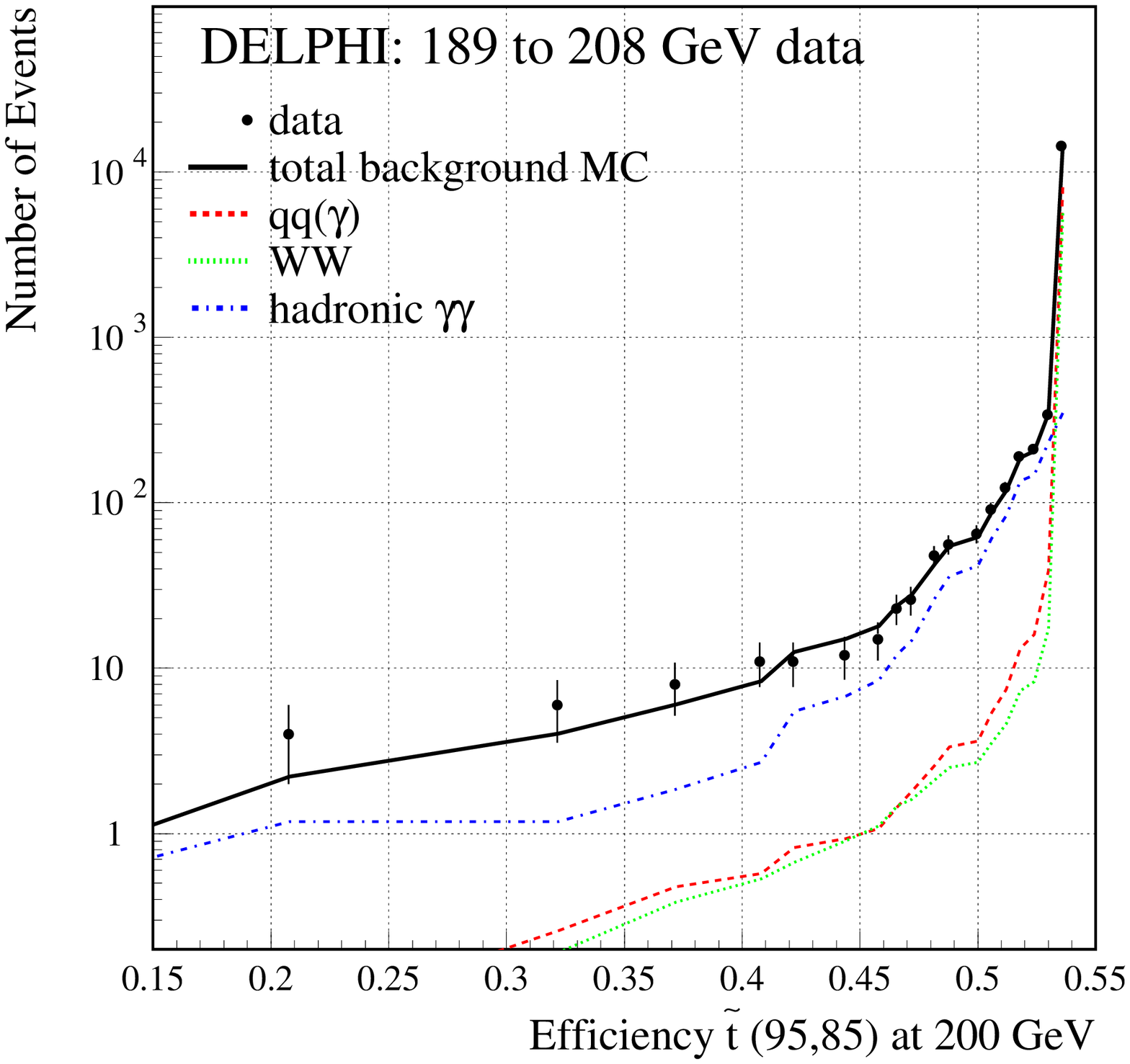} \\
(c) & (d) \\
\includegraphics[width=7cm]{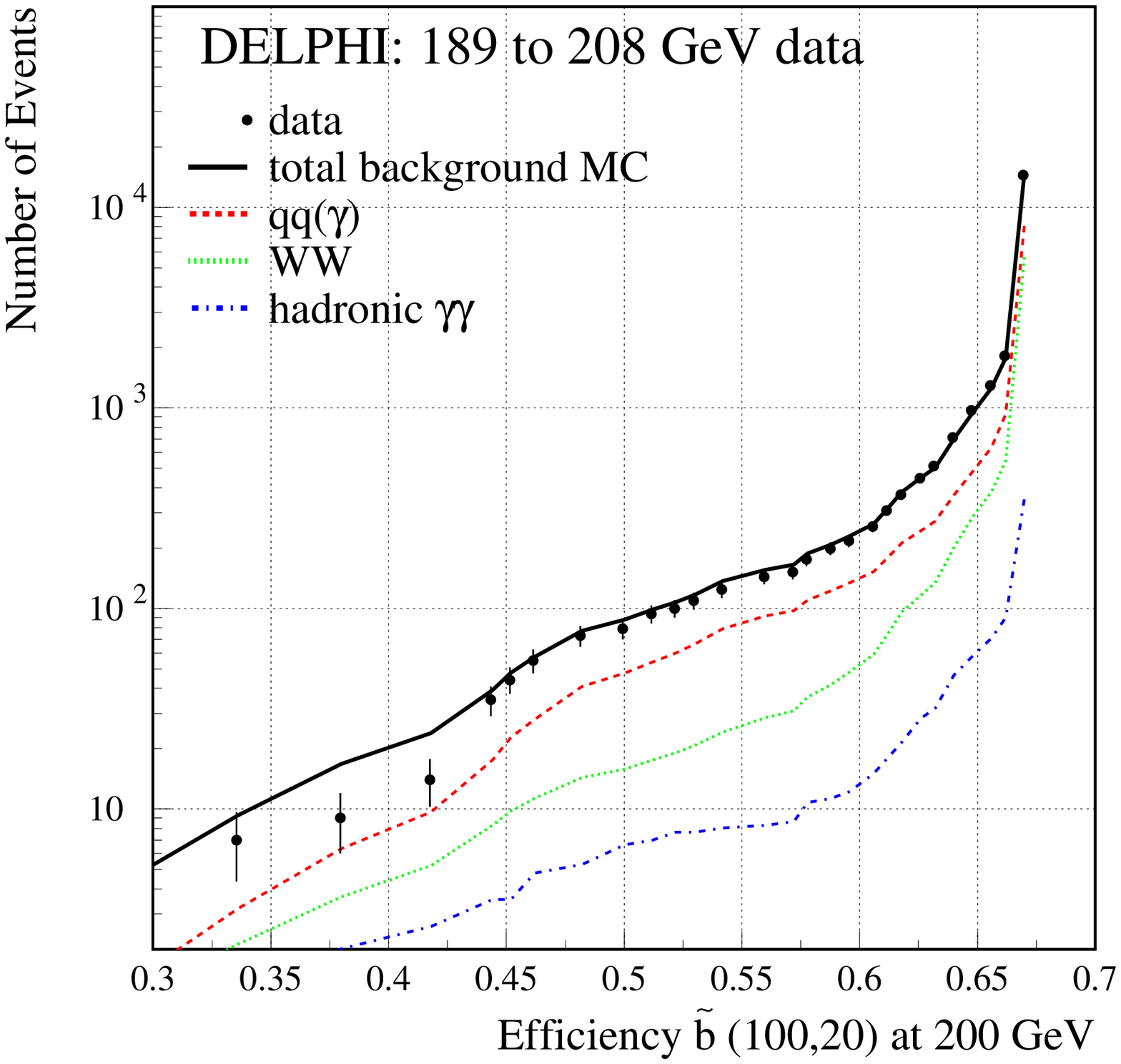} &
\includegraphics[width=7cm]{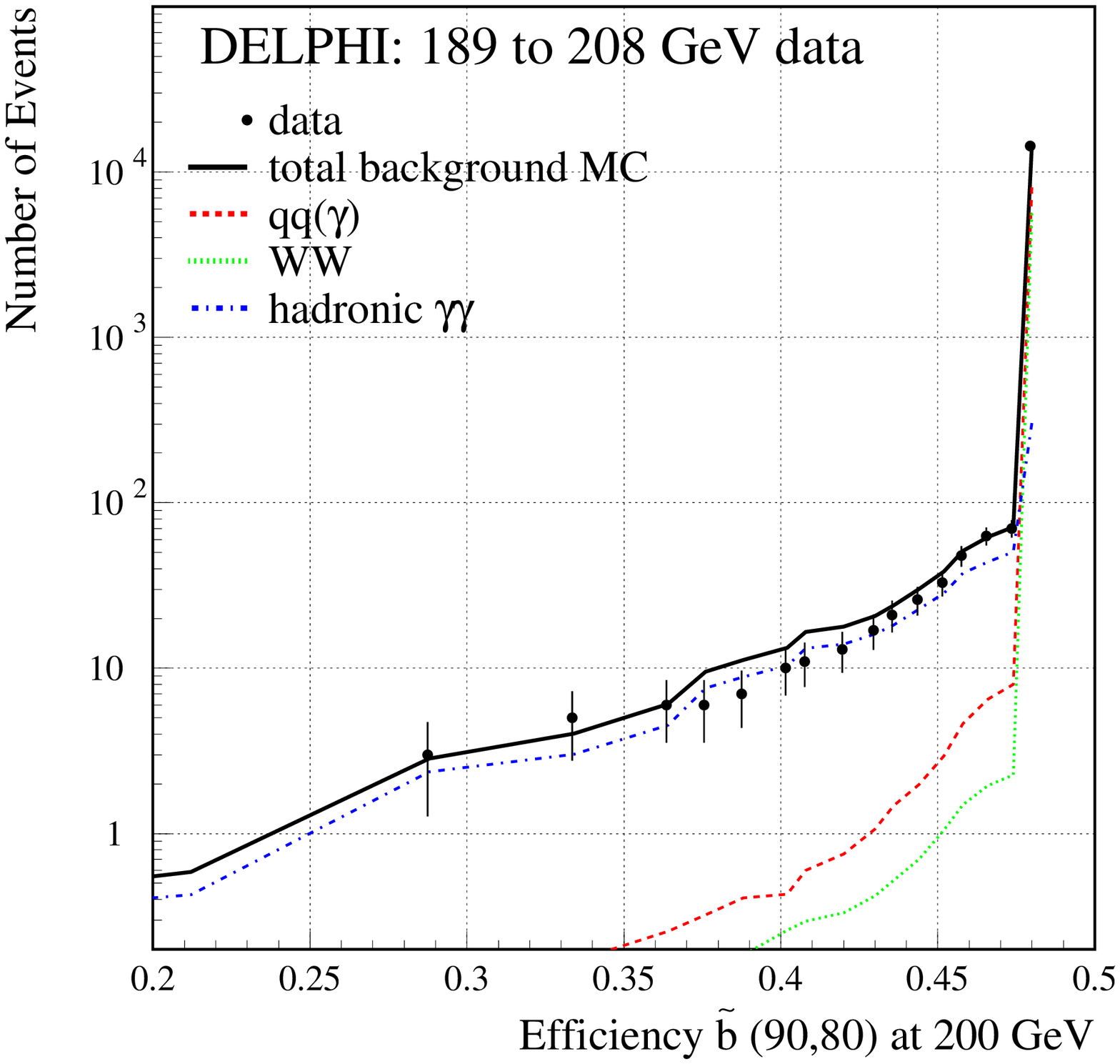} \\
\end{tabular}
\caption{Numbers of events as a function of the signal efficiencies for the 
stop and sbottom analysis. Data taken in the centre-of-mass energy range between
189 and 208~GeV were included.
(a) stop analysis for $\Delta m > 20~\GeVcc$ and (b) for $\Delta m \leq 20~\GeVcc$,
(c) sbottom analysis for $\Delta m > 20~\GeVcc$ and (b) for 
$\Delta m \leq 20~\GeVcc$. The squark and gluino mass values used for the signal
detection efficiencies are indicated on the $x$ axis, (\msq,\mglui).}
\label{fi:sqnneffi}
\end{center}
\end{figure}

\begin{figure}[p!]
\begin{center}
\begin{tabular}{cc}
(a) \\
\includegraphics[width=9cm]{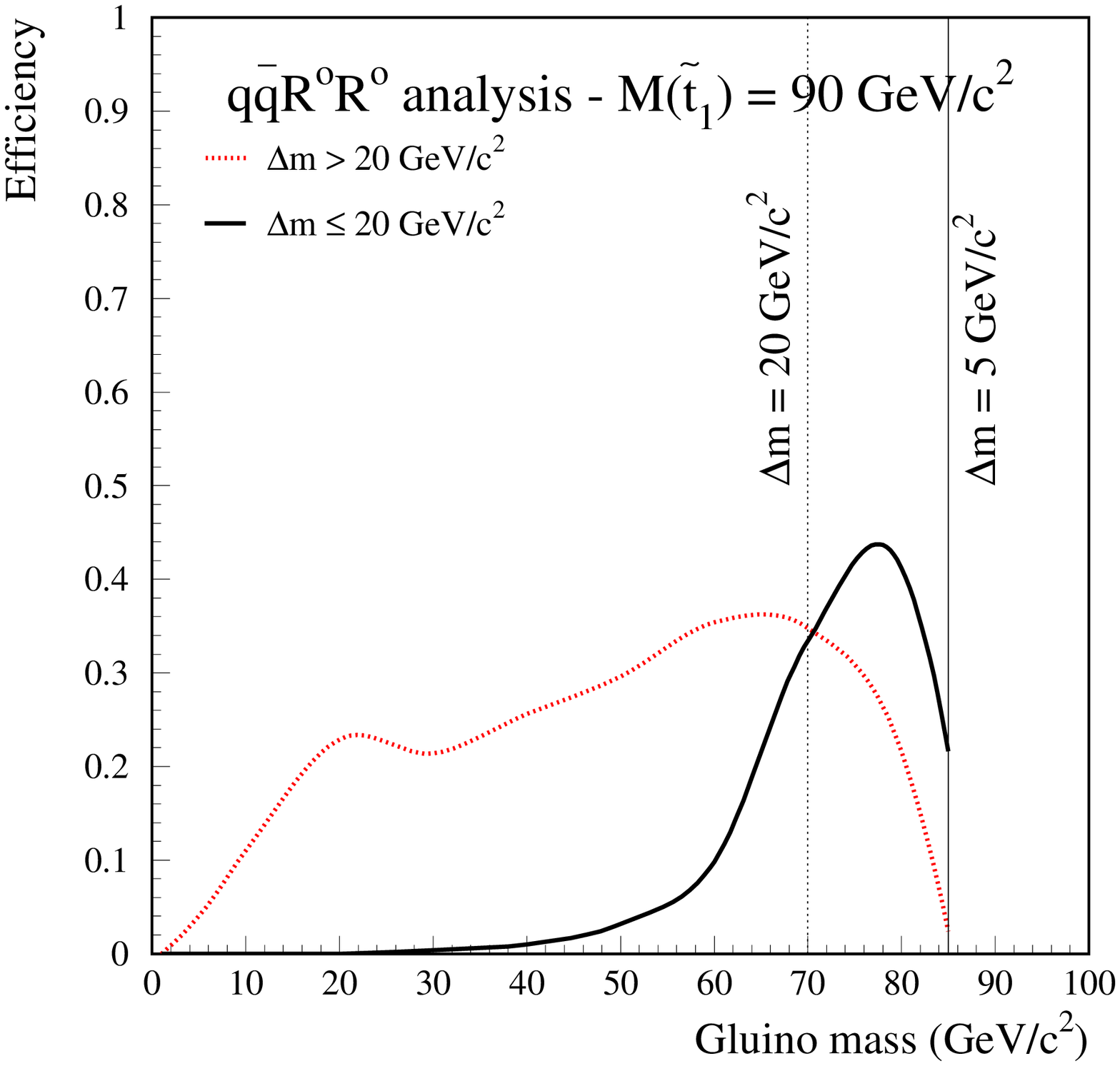} \\
(b) \\
\includegraphics[width=9cm]{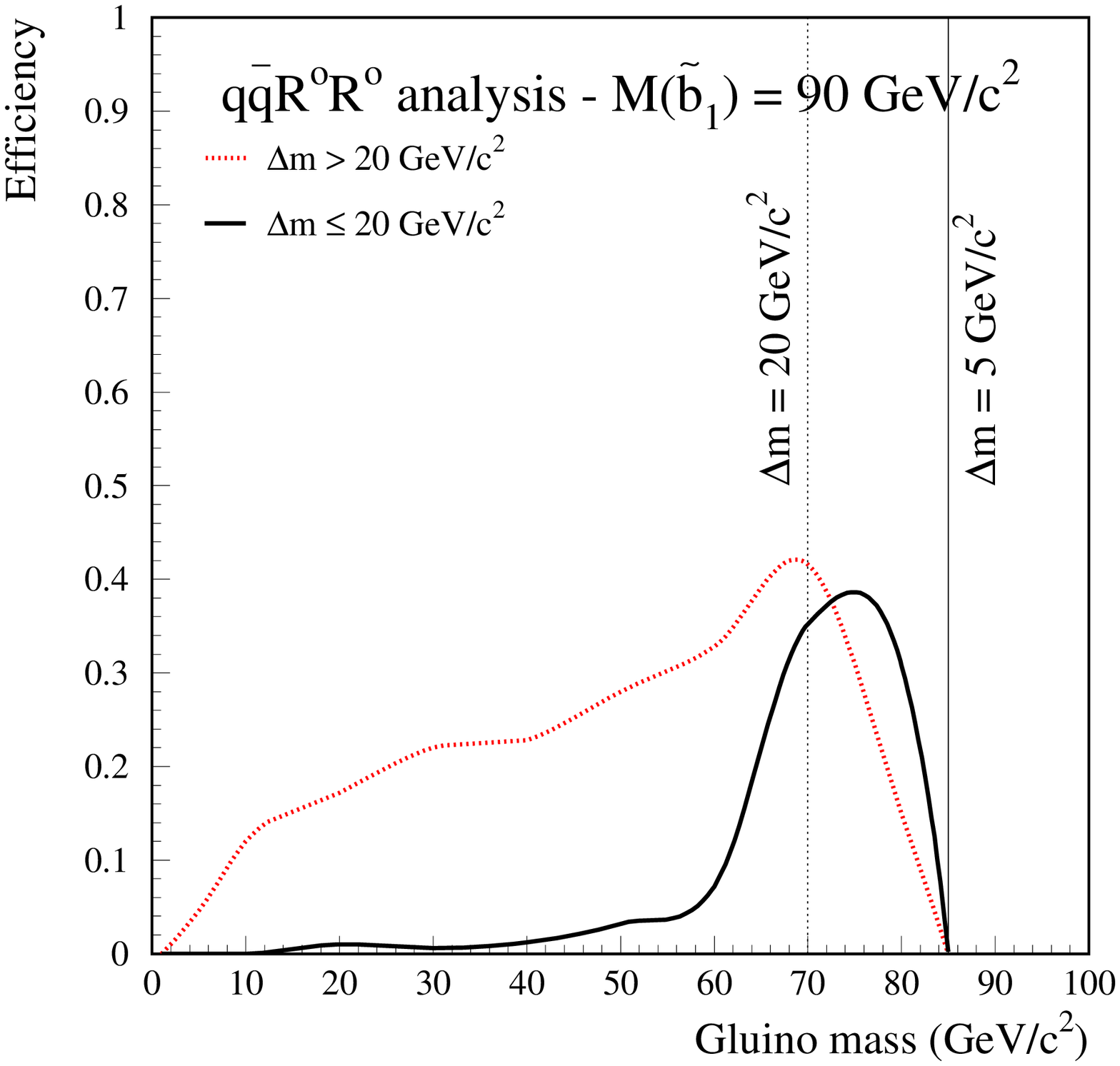} \\
\end{tabular}
\caption{Signal detection efficiencies at $\sqrt{s}=$200~GeV for the 
stop (a) and sbottom (b) \qqbar\roro\ analysis as a function of the gluino
mass ($\msqi=~90~\GeVcc$).}
\label{fi:rneffi}
\end{center}
\end{figure}

\indent
The neural networks were trained to isolate the \qqbar\roro\ 
signal in both $\Delta m$ windows. They were trained separately on stop
signals or sbottom signals.
The neural network structure was the same for 
the stop and the sbottom searches. It consisted of 10 input nodes, 
10 hidden nodes and 3 output nodes.\\
\indent
For $\Delta m \leq 20~\GeVcc$, the 10 input variables were: the ratio
between the transverse missing momentum and the visible energy, the transverse
energy, the visible mass, the softness defined as 
$M_{jet1}/E_{jet1}+M_{jet2}/E_{jet2}$, the acollinearity, the quadratic sum of
transverse momenta of the jets 
$\sqrt{(P_{t}^{jet1})^2+(P_t^{jet2})^2}$, the acoplanarity, 
the sum of the first and third Fox-Wolfram moments, the polar angle of the 
missing momentum and finally the combined b-tagging probability~\cite{BTAG}.\\
\indent
For $\Delta m > 20~\GeVcc$, the 10 input variables 
were: the charged energy, the transverse charged energy, 
the visible mass,
the thrust, the effective centre-of-mass energy~\cite{SPRIME}, the acollinearity,
the acoplanarity, the sum of the first and third Fox-Wolfram moments, the sum of the 
second and fourth Fox-Wolfram moments, and finally the combined b-tagging
probability.\\
\indent
The neural networks were trained to discriminate the 
signal from the combined two-fermion and four-fermion backgrounds, and from the 
$\gamma\gamma$ interactions leading to hadronic final states.
The first output node was trained to identify both $\Zo\to\qqbar$ and 
four-fermion events. The second node identified the $\gamma\gamma$ interactions 
leading to hadronic final states. And the third node was trained to select the
signal. The three output nodes were useful in the training of the network,
but the selection was made according to the output of the signal node only.
Figure~\ref{fi:sqnneffi} shows the 
number of events as a function of the signal efficiency for the two mass 
analysis windows of the stop and the sbottom analysis. The number of real 
events was in agreement with the Standard Model predictions over the full range
of the neural network outputs. The optimisations of the final cuts were 
performed by 
minimising the expected confidence level of the signal 
hypothesis~\cite{ALRMC}.\\
\indent
Tables~\ref{tab:resstn} and~\ref{tab:ressbn} contain the numbers of events
selected in the stop and sbottom analyses. Combining all data from 189 to
208~GeV, 32 and 11 events were selected in the stop analysis for 
$\Delta m > 20~\GeVcc$ and $\Delta m \leq 20~\GeVcc$, while the expected number 
of events were 30.1 and 11.1. In the sbottom analysis, no candidates were observed
for $\Delta m > 20~\GeVcc$ and five were selected for $\Delta m \leq 20~\GeVcc$.
The expected number of events were 3.0 and 5.3. Figure~\ref{fi:rneffi} shows 
the signal detection efficiencies for the stop and for the sbottom
($\msqi=~90~\GeVcc$). They are 
very low when the gluino mass is close to zero.  

\section{Results}

\begin{figure}[p!]
\begin{center}
\includegraphics[width=14cm]{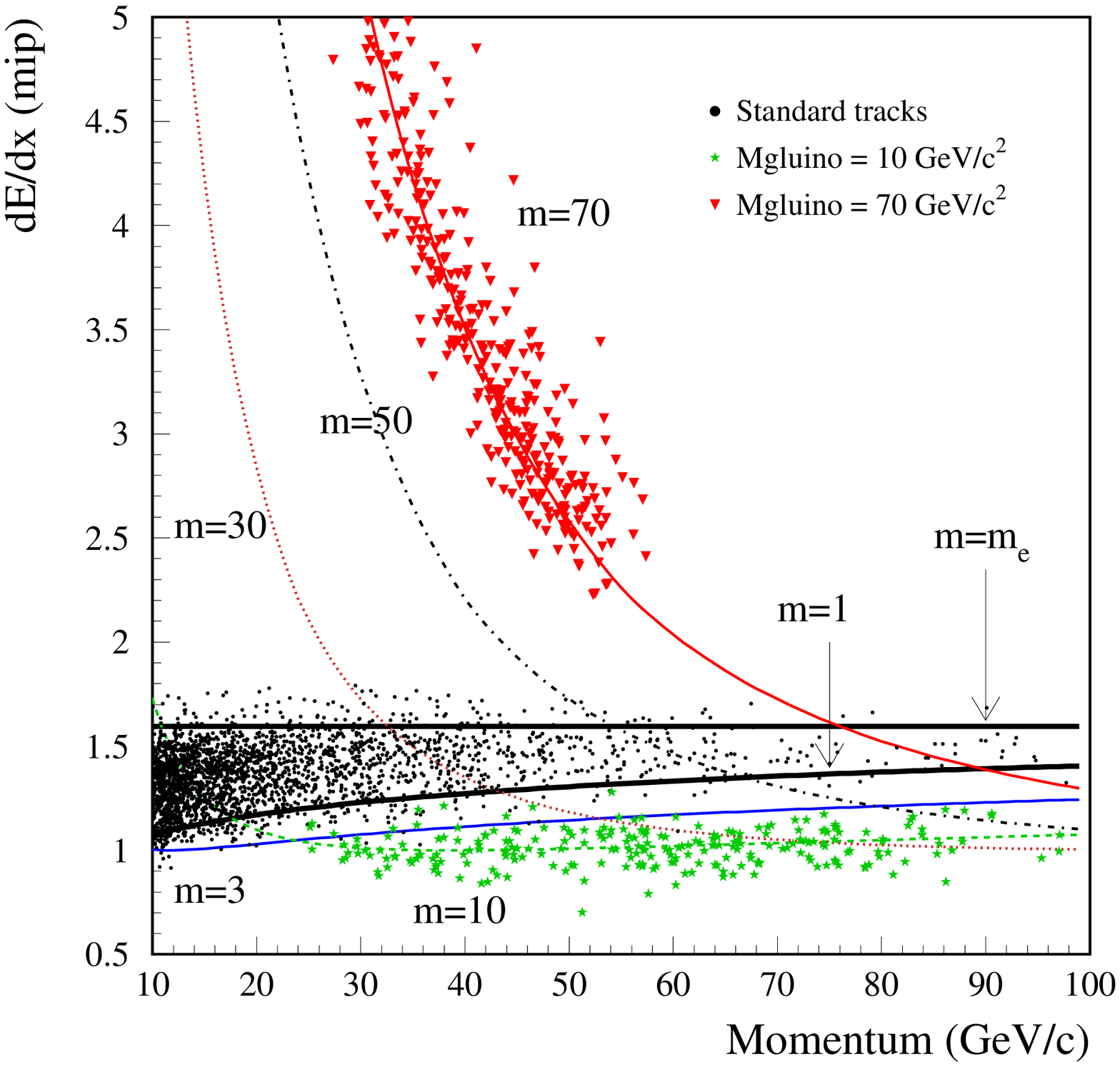}
\caption{The lines show the theoritical ionization energy loss as a function 
of the momentum of the particle for different mass cases (\GeVcc).
The black points are reconstructed tracks coming from $\Zo\to\qqbar$
events generated at $\sqrt{s}=$200~GeV, while the stars and the triangles 
correspond to $\rpm$ tracks with mass of 10 and 70~\GeVcc respectively. The 
dE/dx is expressed in units of energy loss for a minimum ionizing particle.}
\label{fi:dedx}
\end{center}
\end{figure}

\begin{figure}[p!]
\begin{center}
\includegraphics[width=14cm]{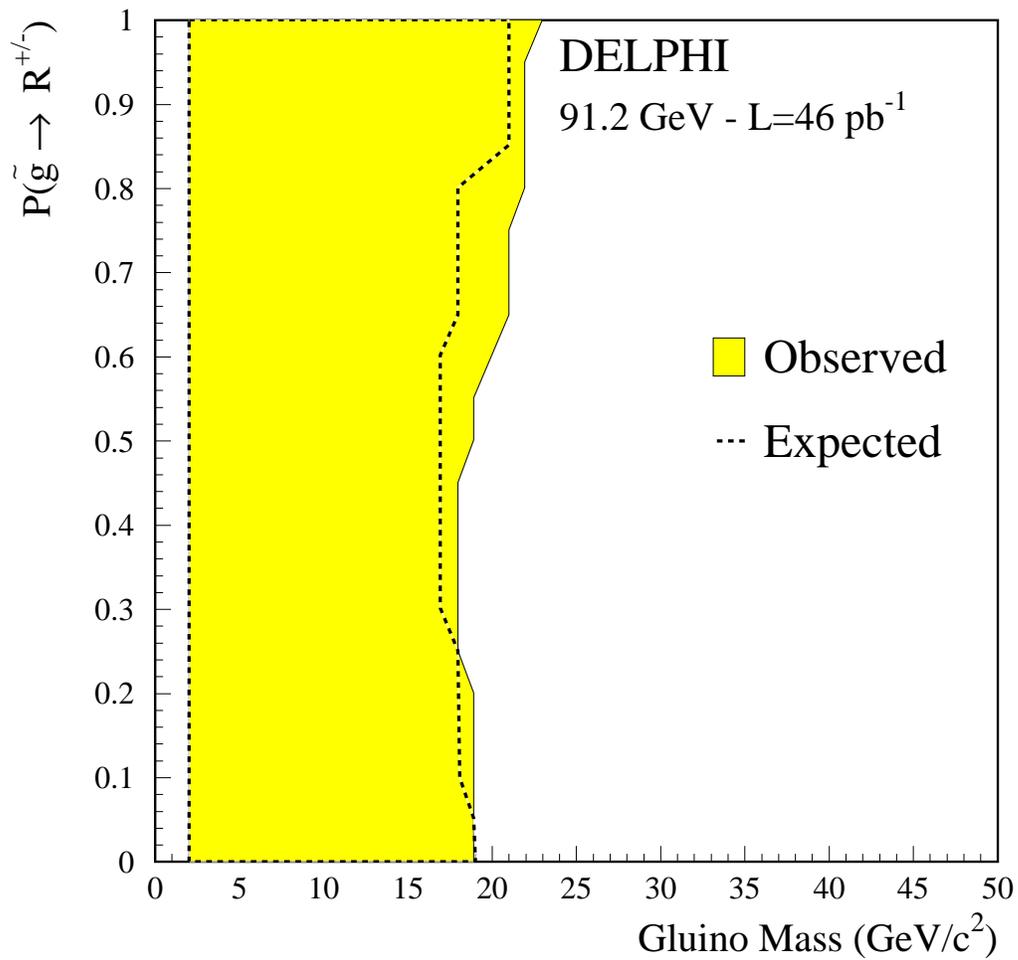}
\caption{Results of the LEP1 analysis: excluded region at 95\% confidence level 
in the plane (\mglui,P). P is the probability that the gluino hadronizes to
a charged R-hadron. The shaded region corresponds to the observed 
exclusion and the line to the expected one.}
\label{fi:exrhlep1}
\end{center}
\end{figure}

\begin{figure}[p!]
\begin{center}
\includegraphics[width=7cm]{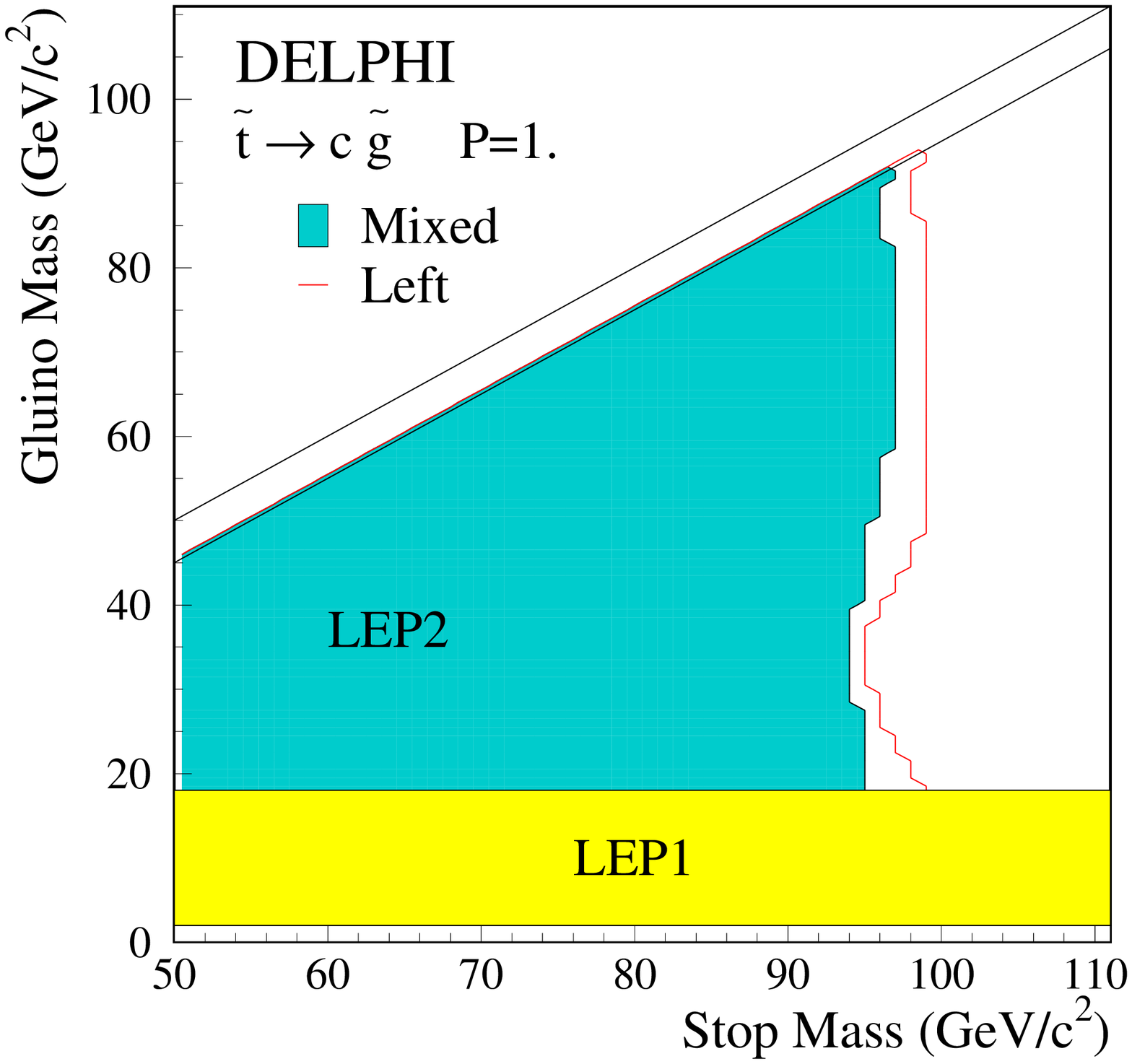}
\includegraphics[width=7cm]{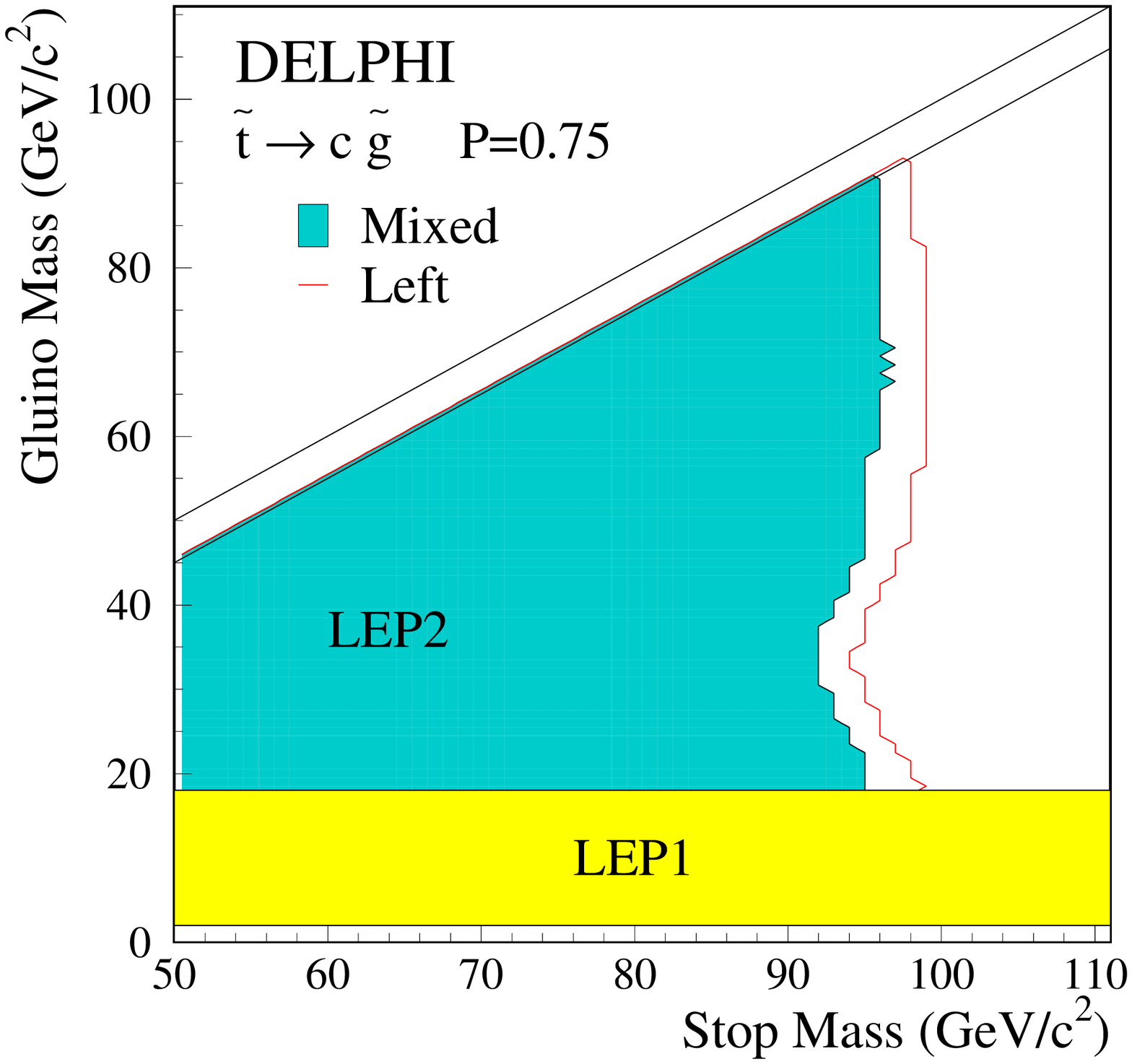}
\includegraphics[width=7cm]{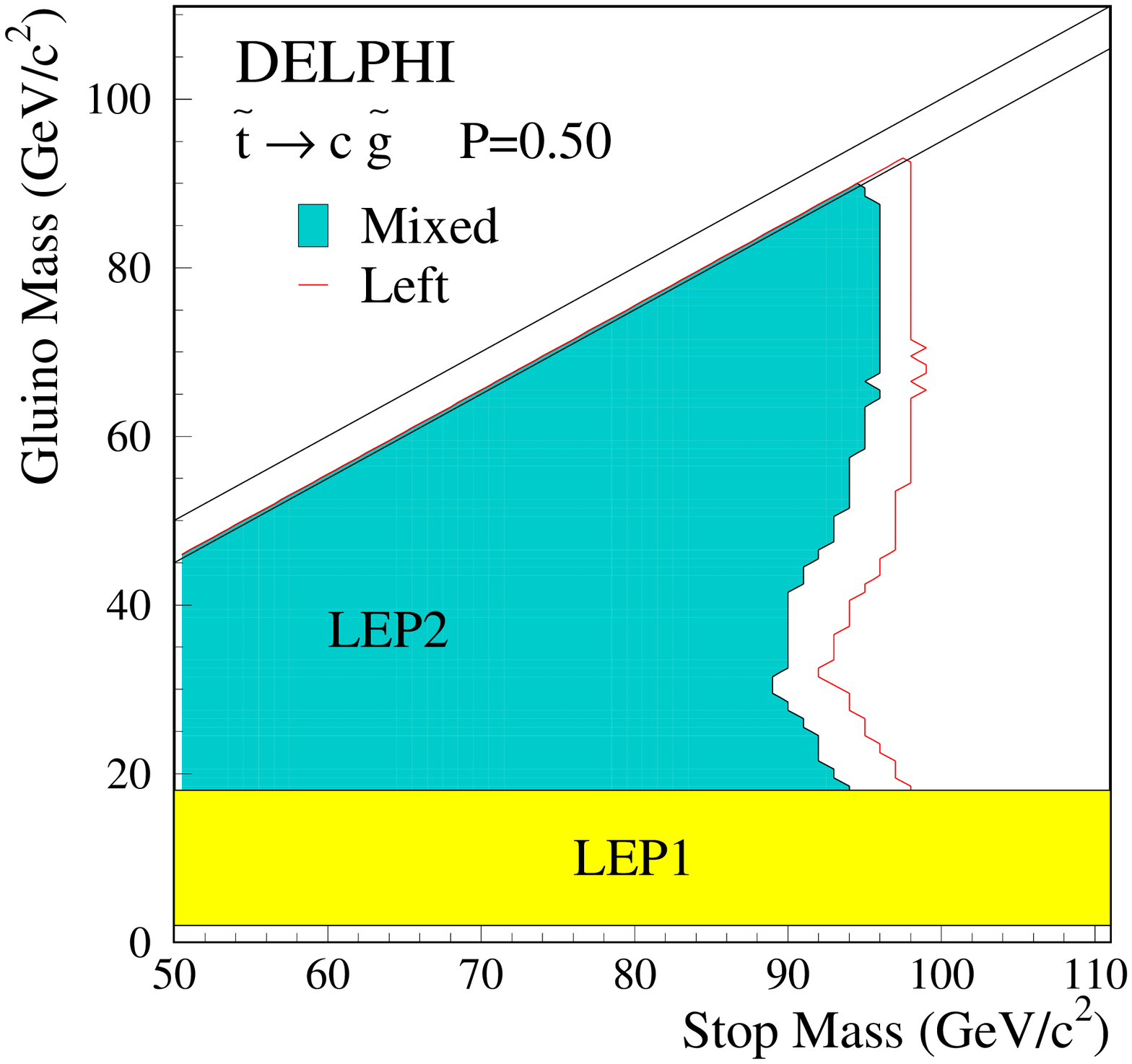}
\includegraphics[width=7cm]{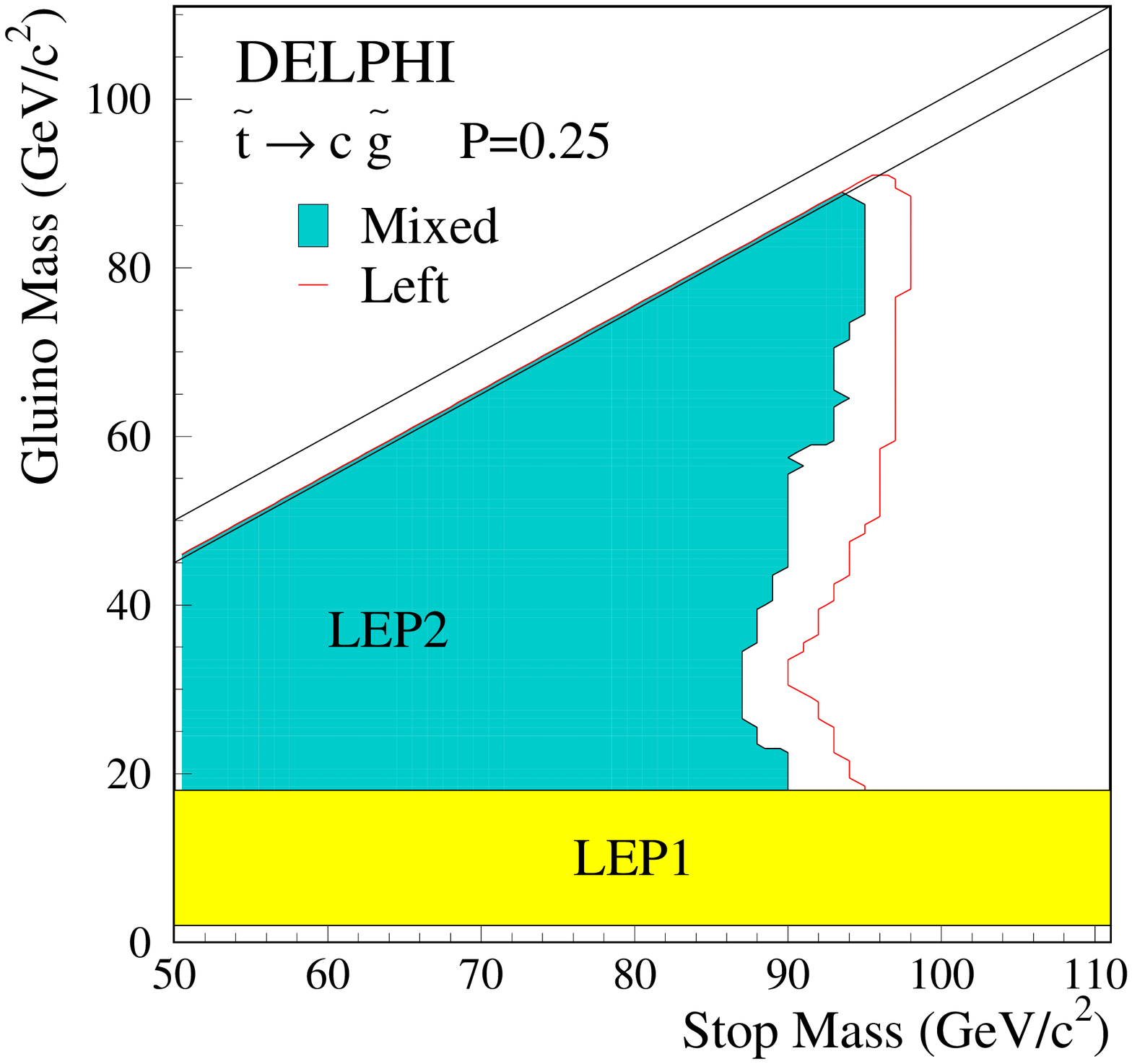}
\includegraphics[width=7cm]{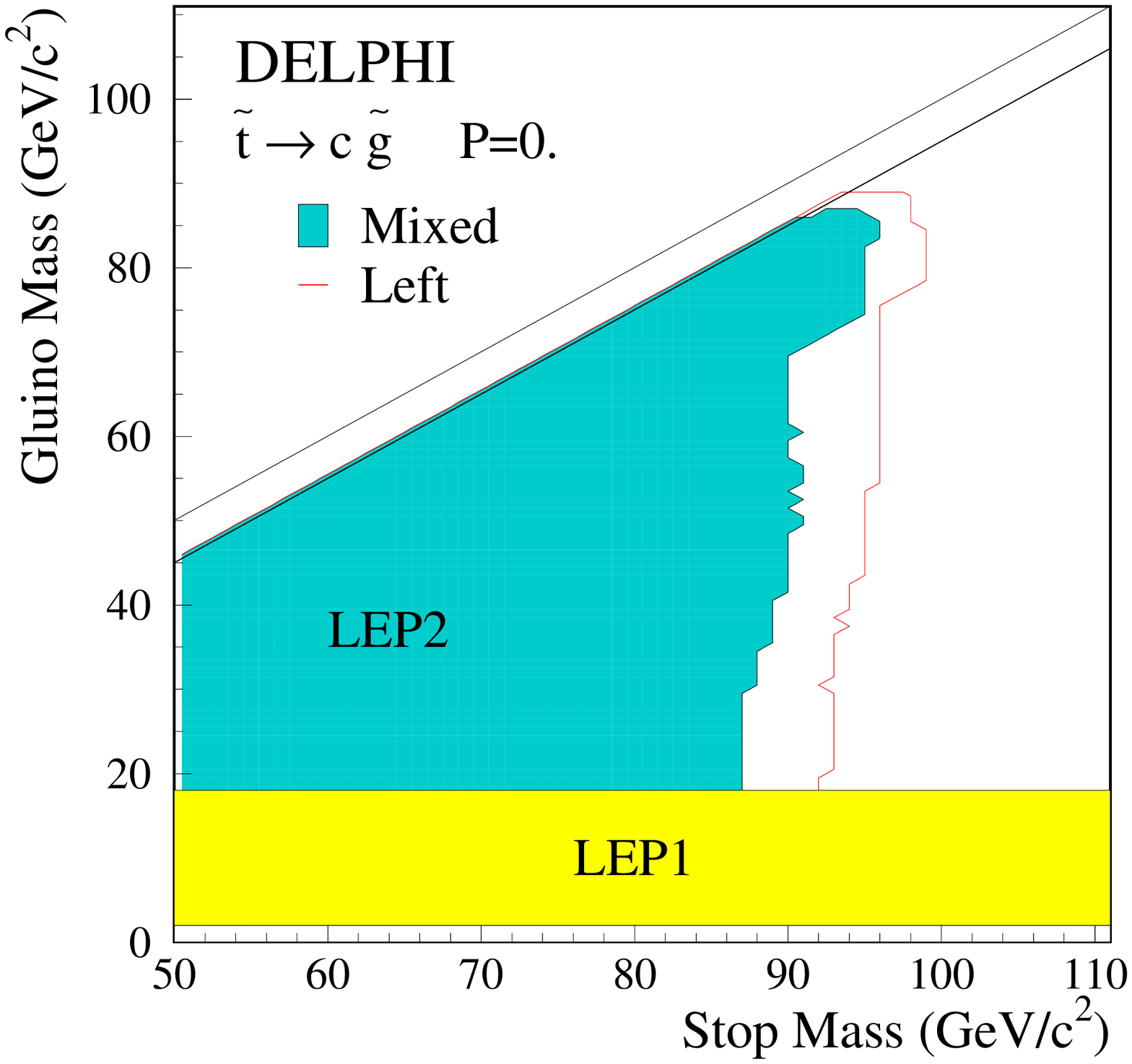}
\caption{Results of the LEP2+LEP1 stop analysis: excluded region at 95\% confidence level 
in the plane (\msti,\mglui). The line corresponds to the exclusion 
for purely left stop, and the shaded region to exclusion obtained 
for the mixing angle giving the minimal cross-section.
Excluded regions are given for different values of P, the 
probability that the gluino hadronizes to charged R-hadron: 0, 0.25, 0.5, 0.75 
and 1.}
\label{fi:exrhst}
\end{center}
\end{figure}

\begin{figure}[p!]
\begin{center}
\includegraphics[width=7cm]{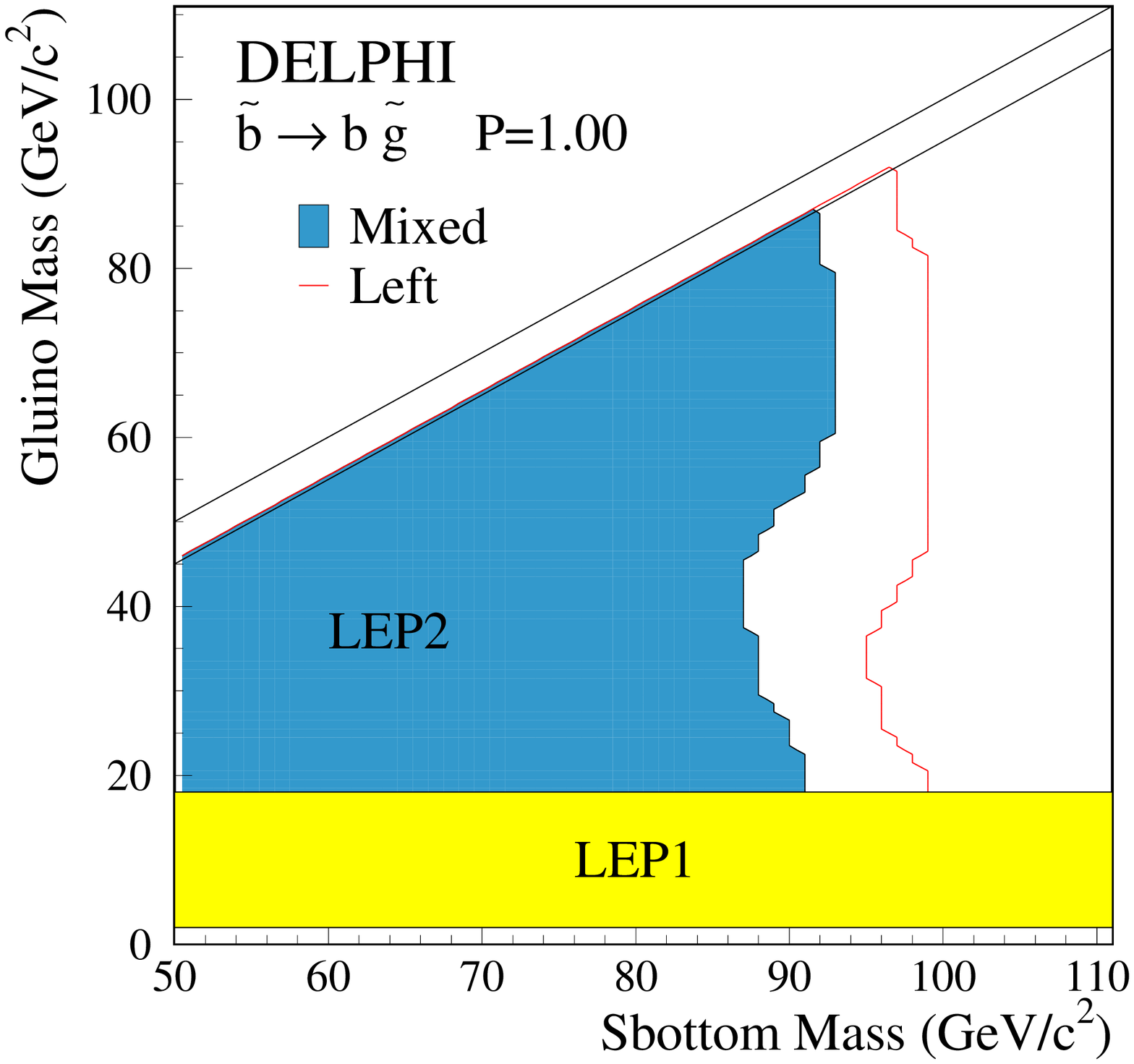}
\includegraphics[width=7cm]{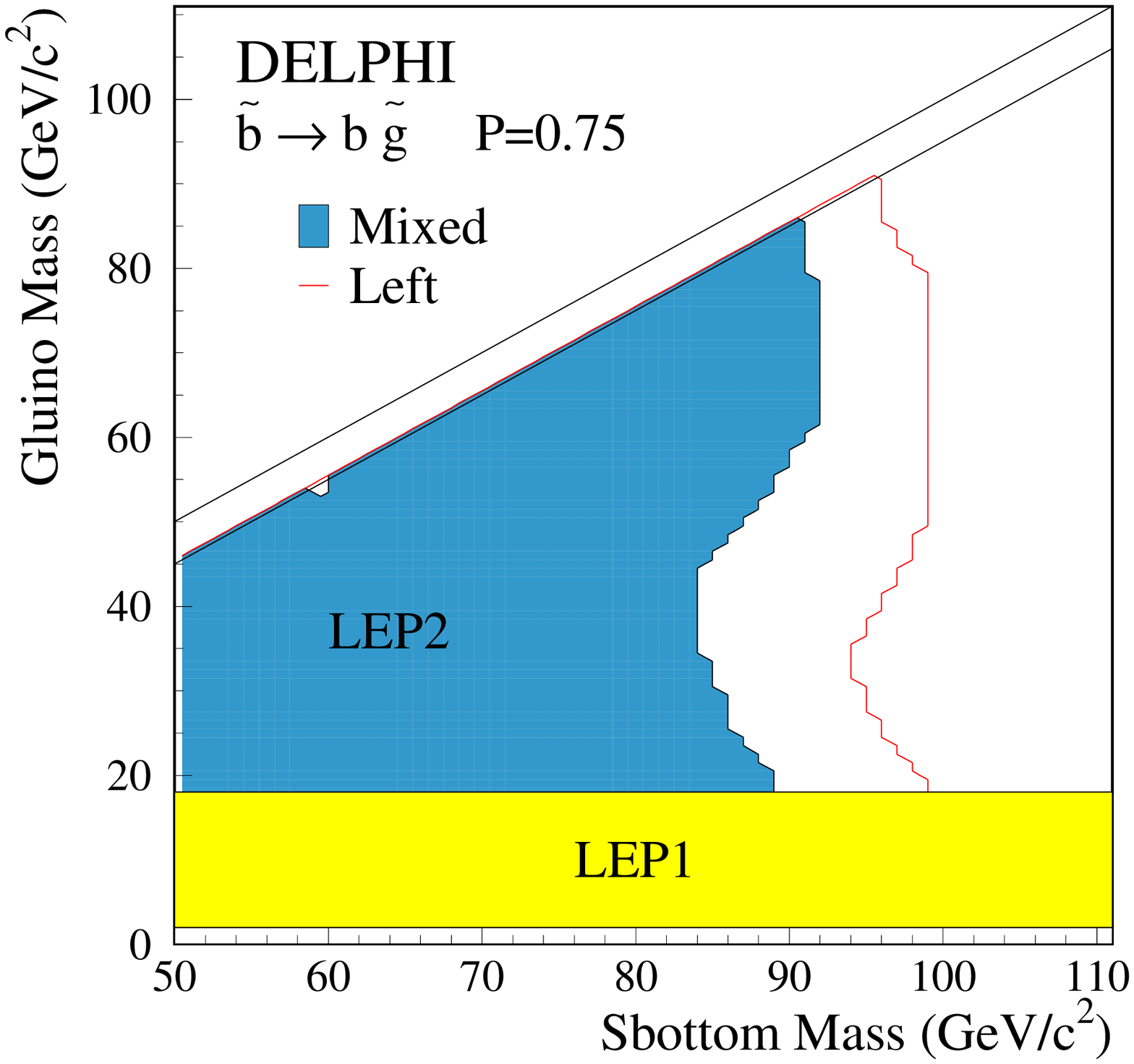}
\includegraphics[width=7cm]{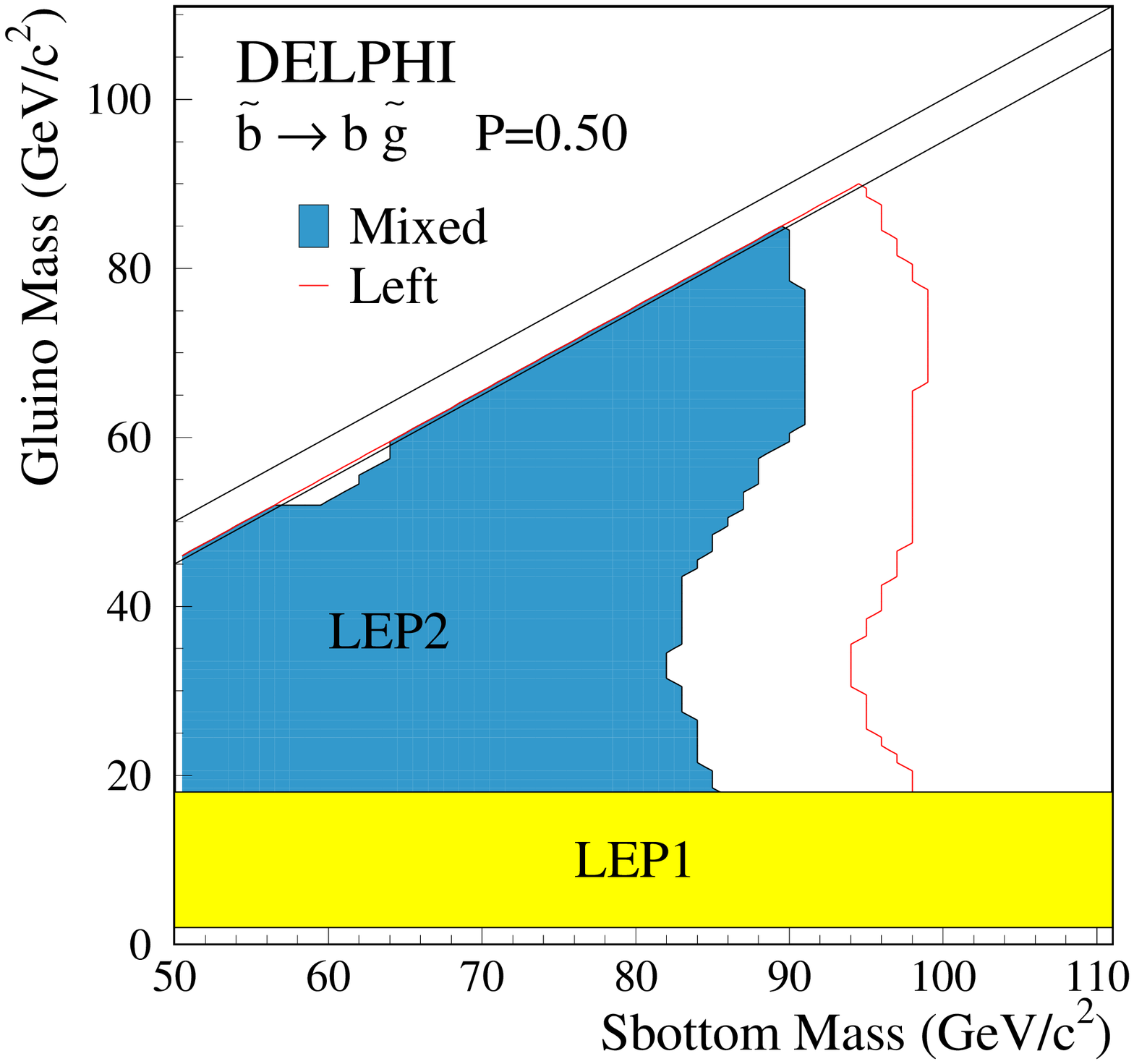}
\includegraphics[width=7cm]{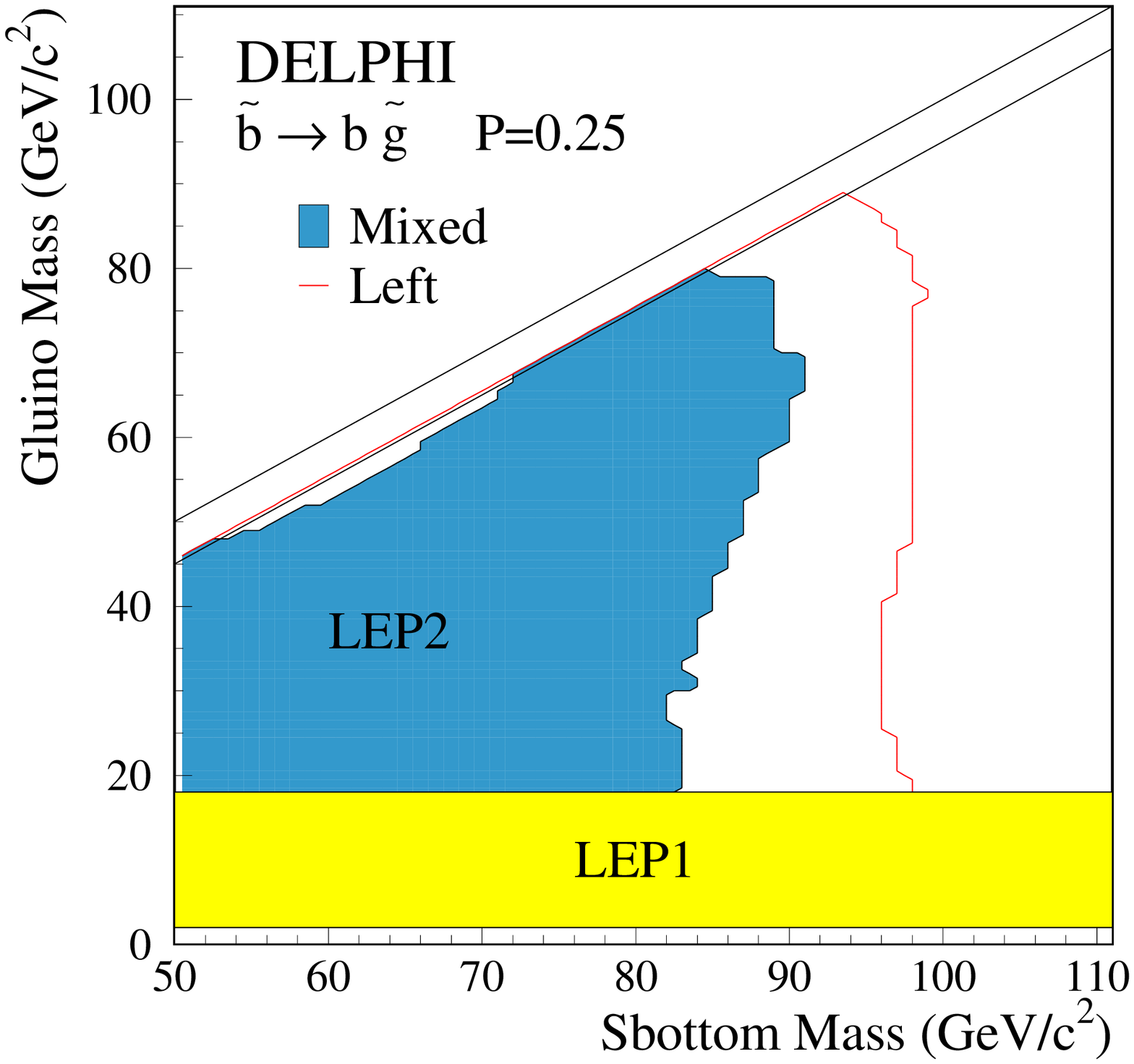}
\includegraphics[width=7cm]{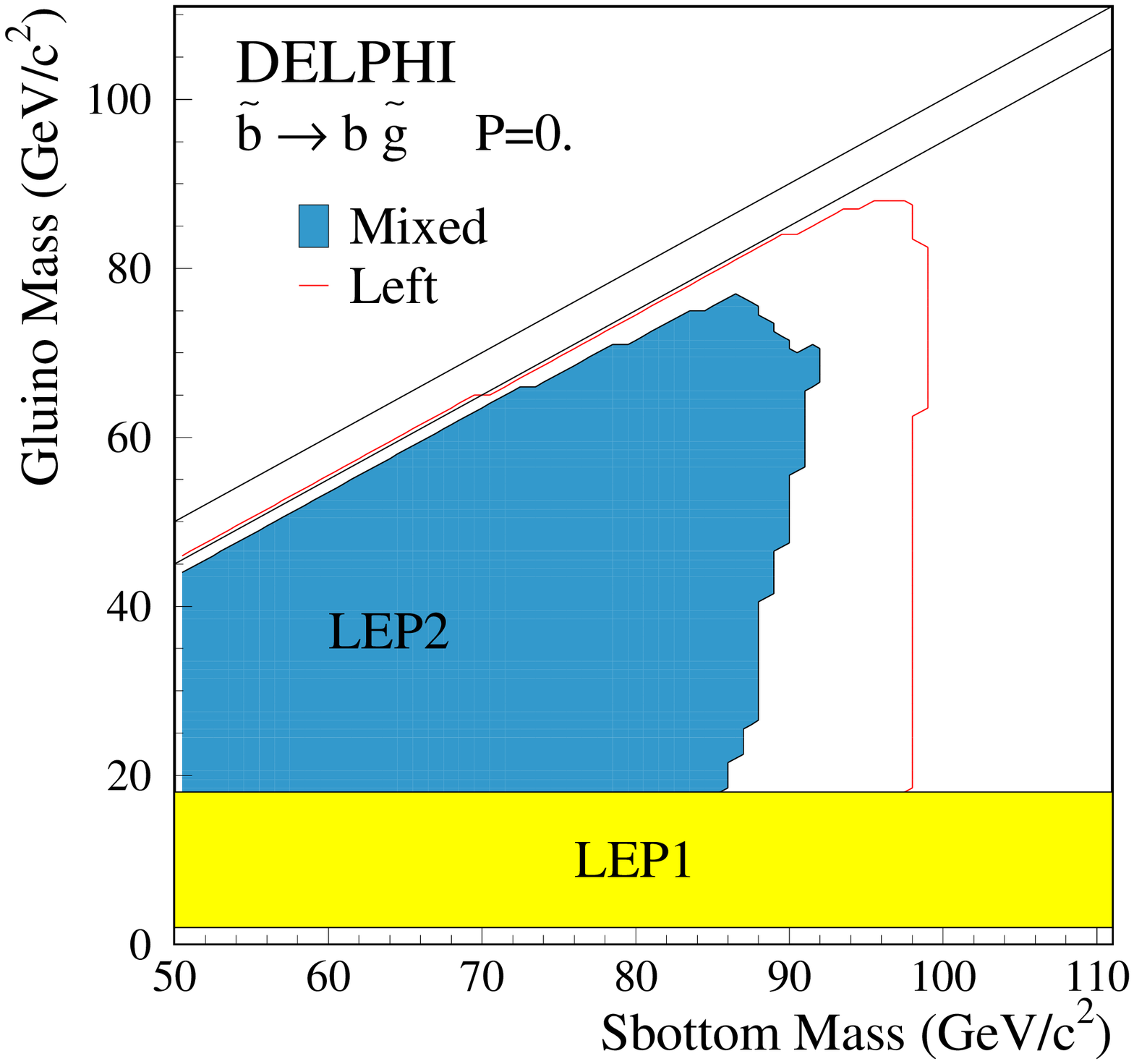}
\caption{Results of the LEP2+LEP1 sbottom analysis: excluded region at 95\% confidence level 
in the plane (\msbi,\mglui). The line corresponds to the exclusion 
for purely left sbottom, and the shaded region to exclusion obtained
for the mixing angle giving the minimal cross-section.
Excluded regions are given for different values of P, the 
probability that the gluino hadronizes to charged R-hadron: 0, 0.25, 0.5, 0.75 
and 1.}
\label{fi:exrhsb}
\end{center}
\end{figure}

No excess of events was observed in any analysis performed at LEP1 or 
at LEP2 in the stable gluino scenario. Results were therefore combined to 
obtain
excluded regions at 95\% confidence level in the parameter space.
The  limits were computed using the likelihood ratio method described 
in~\cite{ALRMC}. For different values of the parameter P describing the 
probability that the gluino hadronizes to a charged R-hadron, the relative
cross-sections for the different channels were given by:
\begin{eqnarray*}
\left\{ 
\begin{array}{lcl}
\sigma(\rpmrpm)   & = &   P^2	   \sigma \\
\sigma(\rpmro)    & = &   2P(1-P) \sigma \\
\sigma(\roro) 	  & = &   (1-P)^2 \sigma \\
\end{array}
\right.
\end{eqnarray*}
where $\sigma$ was either the $\ee \to \qqbar\glui\glui$ cross section at 
LEP1, the \Sti\aSti\ or the \Sbi\aSbi\ cross-section at LEP2.\\
\indent
For the LEP1 analysis, results were interpreted in terms of excluded gluino 
masses for different P. Figure~\ref{fi:exrhlep1} shows the excluded region at 
95\% confidence level. From this figure, a stable gluino with mass between 2 and
18~\GeVcc\ is excluded regardless of the charge of the R-hadrons. 
The minimum upper limit, 18~\GeVcc, is obtained for intermediate values of P
(between 0.2 and 0.45), while an upper limit of 23~\GeVcc\ is obtained for 
P=1.\\
\indent
For the LEP2 analysis, excluded regions in the planes (\msti,\mglui) and 
(\msbi,\mglui) were derived for five different values of P: 0., 0.25, 0.5, 
0.75 and 1. Moreover, the stop and sbottom
cross-sections were calculated for two cases. In the first case, the squark
mixing angle was set to zero which corresponds to the maximal cross-sections.
In the second case, the mixing angle was equal to $56^\circ$ for the stop and to
$68^\circ$ for the sbottom, which gives the minimal cross-sections. 
Squark masses below 50~\GeVcc\ were
not taken into account in these analyses.
Figures~\ref{fi:exrhst} and 
\ref{fi:exrhsb} show the excluded regions thus obtained.
In the case of minimal cross-sections,
a hole appears in the sbottom exclusion histograms around (60~\GeVcc,~50\GeVcc) 
in the plane (\msbi,\mglui) for intermediate values of P.
It can be explained by different effects:
\begin{itemize}
\item{}
In the analyses searching for charged R-hadrons, 
the signal detection efficiencies are small in this region of the 
(\msqi,\mglui) plane, because the dE/dx 
distribution of the charged R-hadrons as a function of the momentum crosses 
the band corresponding to Standard Model particles (cf Figure~\ref{fi:dedx}).
\item{}
When the $\Zo-\Sqi-\aSqi$ coupling is suppressed, the squark pair production 
cross-section is much more reduced for the sbottom than for the stop.
\item{}
The visible energy of \qqbar\roro\ events becomes small for low values 
of  $\Delta M$. At $\Delta M = 5~\GeVcc$, this has no effect on the stop 
exclusion results, because of the c-quark which is lighter than the b-quark.
\end{itemize}
Lower limits on the 
stop and sbottom masses are given in 
table~\ref{tab:resgluioana} for $\Delta m \geq 10~\GeVcc$ and for a gluino 
mass greater than 2~\GeVcc\ for different values of P.

\section{Conclusion}

The analysis of the LEP1 data collected in 1994 
excludes at 95\% confidence level a stable gluino with mass between 2 and 
18~\GeVcc. These limits are valid for any charge of the produced R-hadrons.\\
\indent
Stop and sbottom squarks have been searched for in the 609~\pb\ collected by 
DELPHI at centre-of-mass energies ranging from 189 to 208~GeV. In the stable 
gluino scenario, the dominant decays are $\Sti \to \qc\glui$ and 
$\Sbi \to \mathrm{b}\glui$.
No deviation from Standard Model predictions was observed. 
The observed limits at 95\% confidence level are:
\begin{itemize}
\item{}
$\msti>90~\GeVcc$, and $\msbi>96~\GeVcc$ for purely left squarks.
\item{}
$\msti>87~\GeVcc$, and $\msbi>82~\GeVcc$ independent of the mixing angle.
\end{itemize}

\section*{Acknowledgments}

We are greatly indebted to our technical
collaborators, to the members of the CERN-SL Division for the excellent
performance of the LEP collider, and to the funding agencies for their
support in building and operating the DELPHI detector.\\
\indent
We acknowledge in particular the support of \\
Austrian Federal Ministry of Science and Traffics, GZ 616.364/2-III/2a/98, \\
FNRS--FWO, Belgium,  \\
FINEP, CNPq, CAPES, FUJB and FAPERJ, Brazil, \\
Czech Ministry of Industry and Trade, GA CR 202/96/0450 and GA AVCR A1010521,\\
Danish Natural Research Council, \\
Commission of the European Communities (DG XII), \\
Direction des Sciences de la Mati$\grave{\mbox{\rm e}}$re, CEA, France, \\
Bundesministerium f$\ddot{\mbox{\rm u}}$r Bildung, Wissenschaft, Forschung
und Technologie, Germany,\\
General Secretariat for Research and Technology, Greece, \\
National Science Foundation (NWO) and Foundation for Research on Matter (FOM),
The Netherlands, \\
Norwegian Research Council,  \\
State Committee for Scientific Research, Poland, 2P03B06015, 2P03B1116 and
SPUB/P03/178/98, \\
JNICT--Junta Nacional de Investiga\c{c}\~{a}o Cient\'{\i}fica
e Tecnol$\acute{\mbox{\rm o}}$gica, Portugal, \\
Vedecka grantova agentura MS SR, Slovakia, Nr. 95/5195/134, \\
Ministry of Science and Technology of the Republic of Slovenia, \\
CICYT, Spain, AEN96--1661 and AEN96-1681,  \\
The Swedish Natural Science Research Council,      \\
Particle Physics and Astronomy Research Council, UK, \\
Department of Energy, USA, DE--FG02--94ER40817. \\


\newpage



\begin{table}
\begin{center}
\vspace{3cm}
\begin{tabular}{|l||r|rcr|}
\hline
                                 & Data & \multicolumn{3}{|c|}{$\Zo \to \qqbar$} \\
	   		         &      & \multicolumn{3}{|c|}{background}   \\
\hline
anomalous dE/dx                  &    99322  &  97170 & $\pm$ &  200 \\
$Y_{23}$                         &    24566  &  25794 & $\pm$ &  104 \\
1 \rpm\ candidate                &      421  &	  464 & $\pm$ &   14 \\
\hline
low dE/dx  (\mglui $<14$)        &        5  &     4.2 & $\pm$ &  1.3 \\      
high dE/dx (\mglui $\geq 14)$    &       12  &    13.5 & $\pm$ &  2.4 \\
\hline
\end{tabular}
\caption{Number of events selected after each cut of the charged R-hadron 
analysis at LEP1.}
\label{tab:prelep1rhadc}
\end{center}
\end{table}

\begin{table}
\begin{center}
\begin{tabular}{|l||r|rcr|}
\hline
                         & Data    & \multicolumn{3}{|c|}{$\Zo \to \qqbar$}	   \\
			 &         & \multicolumn{3}{|c|}{background}            \\
\hline
Acolinearity             &   41231 &   34853 & $\pm$ & 120     \\     
$N_{TPC}$                &   38977 &   33807 & $\pm$ & 120     \\ 
$\% E_{VD-ID}$           &   36877 &   32419 & $\pm$ & 120     \\ 
$E_{40}/E_{vis}$         &   19309 &   15311 & $\pm$ &  80     \\ 
$E_{20}/E_{vis}$         &   16664 &   13480 & $\pm$ &  75     \\ 
$E_t$                    &   16317 &   13453 & $\pm$ &  75     \\ 
$M_{vis}$                &    5932 &	6353 & $\pm$ &  52     \\ 
$|\cos \theta_{thrust}|$ &    5384 &	5725 & $\pm$ &  49     \\ 
$|\cos \theta_{Pmis}|$   &    2527 &	2294 & $\pm$ &  31     \\ 
$Y_{23}$                 &     214 &	 194 & $\pm$ &   9   \\ 
$N_{char.}$              &     183 &	 161 & $\pm$ &   8   \\  
Acoplanarity             &     134 &	 115 & $\pm$ &   7   \\  
Thrust                   &     105 &	81.7 & $\pm$ &   5.9   \\  
Acol. vs $M_{jet1}/E_{jet1}+M_{jet2}/E_{jet2}$        &      12 &	10.6 & $\pm$ &   2.1   \\ 
\hline
\end{tabular}
\caption{Number of events selected after each cut of the \qqbar\roro\
analysis at LEP1.}
\label{tab:prelep1rhadn}
\end{center}
\end{table}

\newpage

\begin{table}[h!]
\begin{center}
\begin{tabular}{|l||r|rcr||rcr|rcr|rcr|}
\hline
Cuts                & Data       & 
\multicolumn{3}{|c|}{Simulation} & \multicolumn{3}{|c|}{4-fermions} & 
\multicolumn{3}{|c|}{2-fermions} & \multicolumn{3}{|c|}{$\gamma\gamma$} \\
\hline\hline
$N_{TPC}$           & 175436 & 164146 & $\pm$ & 105 & 12418 & $\pm$ & 13 &  50391 & $\pm$ &  24 & 101338 & $\pm$ & 102 \\
$E_{STIC}/E_{vis.}$ & 145810 & 141362 & $\pm$ &  95 & 12062 & $\pm$ & 13 &  48170 & $\pm$ &  24 &  81131 & $\pm$ &  91 \\
$\theta_{thrust}$   &  54838 &  54933 & $\pm$ &  45 &  9739 & $\pm$ & 10 &  31510 & $\pm$ &  23 &  13685 & $\pm$ &  38 \\
$P_{TPC}$           &  48475 &  48846 & $\pm$ &  43 &  9141 & $\pm$ & 10 &  27580 & $\pm$ &  23 &  12126 & $\pm$ &  35 \\
$N_{cha.}$          &  45816 &  46227 & $\pm$ &  37 &  9040 & $\pm$ &  9 &  26969 & $\pm$ &  16 &  10219 & $\pm$ &  32 \\
$E_{neu.}^{max.}$   &  41880 &  42113 & $\pm$ &  37 &  8802 & $\pm$ &  9 &  23108 & $\pm$ &  16 &  10203 & $\pm$ &  32 \\
\hline
1 $R^{\pm}$ candidate   &  2187  & 2279.1 & $\pm$ & 6.3 & 1746.8 & $\pm$ & 4.6 & 470.3 & $\pm$ & 3.4 &  62.6 & $\pm$ & 2.6 \\
\hline
2 $R^{\pm}$ candidates  &     74 &   79.2 & $\pm$ & 0.8 &   75.2 & $\pm$ & 0.7 &   3.4 & $\pm$ & 0.3 &   0.5 & $\pm$ & 0.2 \\
\hline
\end{tabular}
\normalsize
\caption{Number of events after each cut of the LEP2 preselection. 189 to 208
GeV data are added.}
\label{tab:sqpresel} 
\end{center}
\end{table}

\newpage

\begin{table}[h!]
\begin{center}
\begin{tabular}{|l||c|c||c|c||c|c|}
\hline
& \multicolumn{2}{|c||}{$\mglui \leq 30~\GeVcc$}
	 & \multicolumn{2}{|c||}{$30~\GeVcc<\mglui< 60~\GeVcc$}
	 & \multicolumn{2}{|c|}{$\mglui\geq 60~\GeVcc$} \\
\hline
$\sqrt{s}$& Data & Simulation            & Data & Simulation            & Data & Simulation            \\ \hline
188.7    & 0    & 0.029$\pm$0.016 & 0    & 0.001$\pm$0.001 & 0    & 0.003$\pm$0.003 \\
191.6    & 0    & 0.005$\pm$0.004 & 0    & 0.001$\pm$0.001 & 0    & 0.001$\pm$0.001 \\ 
195.6    & 0    & 0.025$\pm$0.020 & 0    & 0.001$\pm$0.001 & 0    & 0.001$\pm$0.001 \\
199.6    & 0    & 0.007$\pm$0.007 & 0    & 0.001$\pm$0.001 & 0    & 0.001$\pm$0.001 \\
201.7    & 0    & 0.011$\pm$0.007 & 0    & 0.001$\pm$0.001 & 0    & 0.001$\pm$0.001 \\  
204.8    & 0    & 0.009$\pm$0.009 & 0    & 0.001$\pm$0.001 & 0    & 0.001$\pm$0.001 \\  
206.7    & 0    & 0.012$\pm$0.009 & 0    & 0.001$\pm$0.001 & 0    & 0.001$\pm$0.001 \\  
208.1    & 0    & 0.000$\pm$0.000 & 0    & 0.001$\pm$0.001 & 0    & 0.001$\pm$0.001 \\   
206.5(*) & 0    & 0.017$\pm$0.013 & 0    & 0.001$\pm$0.001 & 0    & 0.001$\pm$0.001 \\ \hline
total    & 0    & 0.115$\pm$0.033 & 0    & 0.009$\pm$0.003 & 0    & 0.011$\pm$0.004 \\
\hline
\end{tabular}
\caption{Number of events selected by the \qqbar\rpmrpm\ analysis at 
LEP2. (*) indicates 2000 data taken without sector 6 working.}
\label{tab:resrhadc}
\end{center}
\end{table}

\begin{table}[h!]
\begin{center}
\begin{tabular}{|l||c|c||c|c||c|c|}
\hline
         & \multicolumn{2}{|c||}{$\mglui \leq 30~\GeVcc$}
	 & \multicolumn{2}{|c||}{$30~\GeVcc<\mglui< 60~\GeVcc$}
	 & \multicolumn{2}{|c|}{$\mglui\geq 60~\GeVcc$} \\
\hline
$\sqrt{s}$& Data & Simulation            & Data & Simulation            & Data & Simulation            \\ \hline
188.7    & 0    & 0.577$\pm$0.101 & 0    & 2.514$\pm$0.264 & 0    & 1.937$\pm$0.243 \\
191.6    & 0    & 0.030$\pm$0.011 & 1    & 0.454$\pm$0.132 & 1    & 0.425$\pm$0.131 \\ 
195.6    & 2    & 0.135$\pm$0.042 & 2    & 0.783$\pm$0.097 & 0    & 0.648$\pm$0.088 \\
199.6    & 0    & 0.266$\pm$0.071 & 1    & 1.333$\pm$0.158 & 1    & 1.068$\pm$0.141 \\
201.7    & 0    & 0.097$\pm$0.025 & 2    & 0.489$\pm$0.056 & 2    & 0.392$\pm$0.050 \\  
204.8    & 1    & 0.208$\pm$0.051 & 1    & 0.961$\pm$0.106 & 0    & 0.753$\pm$0.093 \\  
206.6    & 0    & 0.187$\pm$0.040 & 1    & 0.848$\pm$0.089 & 1    & 0.661$\pm$0.079 \\  
208.1    & 0    & 0.011$\pm$0.005 & 0    & 0.095$\pm$0.014 & 0    & 0.085$\pm$0.013 \\   
206.5(*) & 0    & 0.083$\pm$0.025 & 1    & 0.679$\pm$0.074 & 1    & 0.596$\pm$0.070 \\ \hline
total    & 3    & 1.59$\pm$0.15 & 9    & 8.16$\pm$0.39 & 6    & 6.57$\pm$0.36 \\
\hline
\end{tabular}
\caption{Number of events selected by the \qqbar\rpmro\ analysis at 
LEP2. (*) indicates 2000 data taken without sector 6 working.}
\label{tab:resrhadm}
\end{center}
\end{table}

\newpage

\begin{table}[h!]
\begin{center}
\begin{tabular}{|l||r|r||r|r|r|}
\hline
Cuts     & Data & simulation  &  4-fermions & 2-fermions & $\gamma\gamma$ \\
\hline\hline
$P_t^{mis.}$        & 26423 & 26938 $\pm$ 20 & 8117 $\pm$ 8 & 18012 $\pm$ 15 & 809 $\pm$ 9 \\
$\theta_{P_{mis.}}$ & 16379 & 16821 $\pm$ 15 & 7191 $\pm$ 6 &  9088 $\pm$ 12 & 542 $\pm$ 8 \\
$E_{40^0}/E_{vis.}$ & 14694 & 15231 $\pm$ 14 & 6395 $\pm$ 6 &  8471 $\pm$ 12 & 364 $\pm$ 6 \\
Hermeticity         & 14422 & 14651 $\pm$ 14 & 6150 $\pm$ 6 &  8140 $\pm$ 12 & 361 $\pm$ 6 \\
\hline
\end{tabular}
\normalsize
\caption{\qqbar\roro\ analysis at LEP2: number of events after each
selection cut. Data with $\sqrt{s}$ in the range
189~GeV-208~GeV are included.}
\label{tab:rnpresel} 
\end{center}
\end{table}

\newpage

\begin{table}[h!]
\begin{center}
\begin{tabular}{|l||c|c||c|c|}
\hline
         & \multicolumn{2}{|c||}{$\Delta M > 20~\GeVcc$}
	 & \multicolumn{2}{|c|}{$\Delta M \leq 20~\GeVcc$} \\
\hline
$\sqrt{s}$ & Data & Simulation    & Data & Simulation	       \\ \hline
188.7    & 4    & 6.634$\pm$0.741 & 6    & 3.685$\pm$1.158 \\
191.6    & 4    & 1.054$\pm$0.115 & 0    & 0.482$\pm$0.097 \\ 
195.6    & 5    & 3.532$\pm$0.236 & 3    & 1.408$\pm$0.256 \\
199.6    & 7    & 4.324$\pm$0.270 & 0    & 1.617$\pm$0.280 \\
201.7    & 1    & 2.055$\pm$0.140 & 0    & 0.836$\pm$0.138 \\  
204.8    & 4    & 4.432$\pm$0.302 & 0    & 1.197$\pm$0.268 \\  
206.6    & 5    & 4.287$\pm$0.290 & 2    & 1.227$\pm$0.272 \\  
208.1    & 0    & 0.418$\pm$0.031 & 0    & 0.117$\pm$0.026 \\   
206.5(*) & 2    & 3.411$\pm$0.203 & 0    & 0.556$\pm$0.072 \\ \hline
total    & 32   & 30.2$\pm$1.0    & 11   & 11.1$\pm$1.3 \\
\hline
\end{tabular}
\caption{Number of events selected by the stop \qqbar\roro\
analysis. (*) indicates 2000 data taken without sector 6 working.}
\label{tab:resstn}
\end{center}
\end{table}

\begin{table}[h!]
\begin{center}
\begin{tabular}{|l||c|c||c|c|}
\hline
         & \multicolumn{2}{|c||}{$\Delta M > 20~\GeVcc$}
	 & \multicolumn{2}{|c|}{$\Delta M \leq 20~\GeVcc$} \\
\hline
$\sqrt{s}$& Data & Simulation     & Data  & Simulation       \\ \hline
188.7    & 0    & 1.038$\pm$0.657 & 3	& 2.475$\pm$1.139 \\
191.6    & 0    & 0.085$\pm$0.026 & 0	& 0.180$\pm$0.058 \\ 
195.6    & 0    & 0.314$\pm$0.088 & 1	& 0.571$\pm$0.178 \\
199.6    & 0    & 0.295$\pm$0.094 & 0	& 0.663$\pm$0.195 \\
201.7    & 0    & 0.225$\pm$0.049 & 1	& 0.319$\pm$0.096 \\  
204.8    & 0    & 0.295$\pm$0.040 & 0	& 0.428$\pm$0.174 \\  
206.6    & 0    & 0.312$\pm$0.05  & 0	& 0.379$\pm$0.175 \\  
208.1    & 0    & 0.022$\pm$0.005 & 0	& 0.037$\pm$0.017 \\	
206.5(*) & 0    & 0.417$\pm$0.054 & 0	& 0.281$\pm$0.160 \\ \hline
total    & 0    & 3.0$\pm$0.7     & 5	& 5.3$\pm$1.2 \\
\hline
\end{tabular}
\caption{Number of events selected by the sbottom \qqbar\roro\
analysis. (*) indicates 2000 data taken without sector 6 working.}
\label{tab:ressbn}
\end{center}
\end{table}

\newpage

\begin{table}[p!]
\begin{center}
\begin{tabular}{|l||c|c||c|c|}
\hline
       & \multicolumn{2}{|c||}{Stop} & \multicolumn{2}{|c|}{Sbottom} \\
\hline
$P$    & $\theta_{\tilde{t}}=0^\circ$ & $\theta_{\tilde{t}}=56^\circ$ & 
         $\theta_{\tilde{b}}=0^\circ$ & $\theta_{\tilde{b}}=68^\circ$ \\
\hline\hline
0.00   & 92 & 87 & 98 & 86 \\ 
0.25   & 90 & 87 & 96 & 82 \\  
0.50   & 92 & 89 & 94 & 82 \\  
0.75   & 94 & 92 & 94 & 84 \\ 
1.00   & 95 & 94 & 95 & 87 \\ 
\hline
\end{tabular}
\caption{Upper limits on the stop and sbottom masses as a function of the
probability P that the gluino hadronizes to charged R-hadrons. Limits are set
for $\Delta m \geq 10~\GeVcc$ and for a gluino mass greater than 2~\GeVcc. 
Mixing angles equal to zero corresponds to purely left-handed squarks, while 
$\theta_{\tilde{t}}=56^\circ$ and $\theta_{\tilde{b}}=68^\circ$ corresponds to
minimal cross-section case.}
\label{tab:resgluioana}
\end{center}
\end{table}

\clearpage


\newpage

\newpage

\newpage

\newpage

\newpage

\newpage

\newpage

\newpage

\newpage

\newpage

\newpage

\newpage

\newpage

\newpage

\end{document}